\documentclass[11pt,a4paper]{article}
\usepackage{jheppub}
\usepackage{amsmath}
\usepackage{multicol,bbm}
\usepackage{enumerate}
\usepackage{bibentry}
\usepackage{epstopdf}

\usepackage{amssymb}

\newcommand{\be}{\begin{eqnarray}}
\newcommand{\ee}{\end{eqnarray}}
\newcommand{\M}{\mathcal{M}}

\newcommand{\lb}{\left <}
\newcommand{\rb}{\right >}
\newcommand\remind[1]{{\color{red} #1}}

\def\M{{\mathcal{M}}}
\def\G{{\mathcal{G}}}
\def\T{{\mathcal{T}}}
\def\R{{\mathbb R}}
\def\A{{\mathbb A}}
\def\P{{\mathbb P}}

\def\GL{{\rm GL}}

\def\l{\langle}
\def\r{\rangle}
\def\nl{\nonumber\\}
\def\Z{\mathcal{Z}}
\def\ZZ{\mathbb{Z}}
\begin{document}

\title{The Amplituhedron and the One-loop Grassmannian Measure}
\author[a]{Yuntao Bai,}
\author[b]{Song He,}
\author[c]{Thomas Lam}
\affiliation[a]{Department of Physics, Princeton University, Princeton, NJ 08544 } 
\affiliation[b]{State Key Laboratory of Theoretical Physics and Kavli Institute for Theoretical Physics China, Institute of Theoretical Physics, Chinese Academy of Sciences, Beijing 100190, P. R. China}
\affiliation[c]{Department of Mathematics,
University of Michigan, 530 Church St, Ann Arbor, MI 48109, USA}
\emailAdd{ytbai@princeton.edu, she@pitp.ca, tfylam@umich.edu} 

\abstract{All-loop planar scattering amplitudes in maximally supersymmetric Yang-Mills theory can be formulated geometrically in terms of the  ``amplituhedron". We study the mathematical structures of the one-loop amplituhedron, and present a new formula for its canonical measure, or the one-loop Grassmannian measure formula. Using the recently proposed momentum-twistor diagrams, we show that there is a correspondence between the cells of one-loop amplituhedron, BCFW terms or equivalently on-shell diagrams, and residues of the one-loop Grassmannian formula. In particular, for the first non-trivial case of one-loop NMHV, these structures are naturally associated with a nice geometric picture as polygons in projective space, as we discuss in various illustrative examples. }

\maketitle

\section{Introduction}

Recent years have witnessed formidable progress in understanding hidden structures of scattering amplitudes in gauge theory and gravity~\cite{Elvang:2013cua}, especially in planar $\mathcal{N}=4$ super Yang-Mills (SYM).The most notable example is Witten's twistor string \cite{Witten:2003nn}. Another exciting example is the Grassmannian/on-shell-diagram program~\cite{Grassmannian, posGrassmannian}, which expresses scattering amplitudes in planar $\mathcal{N}=4$ SYM in terms of on-shell diagrams and Grassmannian integrals, and makes manifest~\cite{Mason:2009qx, dualGrassmannian} the full Yangian symmetry of the theory~\cite{Yangian}. This line of work led to the ``amplitudedron" proposal~\cite{amplituhedron, Arkani-Hamed:2013kca}, which provides a reformulation of amplitudes without referring to spacetime, Feynman diagrams, or even recursion relations~\cite{all loop}. As we will review shortly, the kinematic variables of planar amplitudes are most conveniently given in terms of momentum-twistor variables, introduced by Hodges~\cite{hodges}. These are twistors of dual momentum space coordinates that make manifest the dual superconformal symmetry
. The key idea of the amplituhedron construction is to consider both external and internal (loop) variables, expressed in momentum-twistor space, to be ``positive" in a geometric sense. It is conjectured that the planar integrand of $\mathcal{N}=4$ SYM at any loop order is given by a canonical top form on the amplituhedron that has logarithmic singularities on its boundary. 

Already at tree level and much more so at loop level, the geometry of the amplituhedron is highly non-trivial and understanding it is an important open problem. Without understanding the cell decomposition of the amplituhedron, the definition of the amplituhedron is still formal, and in practice it can be very difficult to find the canonical form. In~\cite{Bai:2014cna}, two of the authors have proposed a new diagrammatic formulation of the amplituhedron, dubbed ``momentum-twistor diagrams". The diagrams are basic building blocks, which are BCFW terms or equivalently amplituhedron cells, of the all-loop integrand written directly in momentum-twistor space. At tree-level, it is straightforward to relate the new diagrams to the original on-shell diagrams, although the interpretations are very different. To any loop order, the new diagrams have provided a very powerful tool for computing the planar integrand, and a systematic way for studying the amplituhedron geometry. 

One of the main motivations behind this paper is to use momentum-twistor diagrams to study the structure of the one-loop amplituhedron. Just as the standard Grassmannian $G(k,n)$ is crucial for the tree-level $n$-point N${}^k$ MHV amplituhedron, the key to understanding the geometry of the one-loop amplituhedron is a generalization which we refer to as the ``one-loop Grassmannian" $G(k,n;1)$.  Our main result is a description of a set of canonical measures on the one-loop Grassmannian, whose residues correspond to the one-loop momentum-twistor diagrams proposed in~\cite{Bai:2014cna} and refined in the current work. This correspondence is the one-loop generalization of the well-known correspondence between (reduced) on-shell diagrams, or BCFW of tree amplitudes, and residues of the Grassmannian contour integral~\cite{Grassmannian, posGrassmannian}. We will see that all the ideas familiar at tree-level, such as BCFW bridges, decorated permutations, positroid cells, etc., are generalized to the one-loop case in a natural way. More importantly, these residues exactly correspond to cells of the one-loop amplituhedron, thus the study of $G(k,n;1)$ and the associated diagrammatic construction turns out to be the key for understanding the cell decomposition of any one-loop amplituhedron. In particular, for $k=1$,  the geometric picture behind such one-loop cells can be easily visualized: each cell corresponds to a triangle on the projective plane. This provides, for the first time, a simple geometric meaning of the complete one-loop NMHV amplituhedron and its cell decompositions. 

Our paper is organized as follows. In section~\ref{pre} we review several important ingredients of our construction: the one-loop Grassmannian and the amplituhedron, as well as momentum-twistor diagrams. We present the most illustrative results, those for the one-loop $k=1$ amplituhedron, in section~\ref{k=1}: we first give a geometric interpretation of the one-loop positive Grassmannian $G_+(k,n;1)$, and then the construction of general cells, from $0$-dimensional cell to top cells, which are illustrated by detailed examples. In section~\ref{general}, we give the one-loop Grassmannian measure for general cases, and discuss how the residues are related to one-loop momentum-twistor diagrams, or the BCFW terms for one-loop integrand. A detailed example for the one-loop six-point NMHV amplitude is given in the Appendix. 

\section{Preliminaries}\label{pre}
We begin by reviewing the geometry of the Grassmannian, which is likely familiar to the reader. The Grassmannian is a simple but crucial building block in the construction of amplitudes, and it is important to understand it thoroughly. We will then discuss a generalization of the Grassmannian which we call the one-loop Grassmannian and denote as $G(k,n;1)$, with the latter index indicating the loop-level. While the Grassmannian is the basic building block of tree amplitudes, the one-loop Grassmannian is the building block of one-loop amplitudes. In fact, extensions to any loop level $L$ exist which we may denote as $G(k,n;L)$, but their geometry is much richer and we will not have occasion to discuss them in this article.

\subsection{The one-loop Grassmannian}
The Grassmannian $G(k,n)$ is the set of all $k$-dimensional subspaces of an $n$-dimensional linear (or vector) space. The linear space may be chosen to be $\mathbb{R}^n$ or $\mathbb{C}^n$. The real case is important for defining the positive Grassmannian, while the complex case is important for taking multi-variable residues in the Grassmannian. It has been shown, for instance, that leading singularities of all loop amplitudes are residues in the complex Grassmannian. We will need both for our discussion, and the context should make it clear which we are referring to.

Now consider the set $M(k,n)$ of all $k\times n$ matrices of rank $k$, and let $M\in M(k,n)$. The rows of $M$ thus span a $k$-dimensional subspace of $n$-dimensional linear space. Now if we let $M'=G.M$ for some $G\in \GL(k)$, then the rows of $M'$ are a linear combination of the rows of $M$, and conversely the rows of $M$ are a linear combination of rows of $M'$. Thus, the rows of $M'$ and $M$ span the same subspace, which is a point in $G(k,n)$.  Define two matrices $M,M'\in M(k,n)$ to be equivalent whenever $M'=G.M$ for some $G\in \GL(k)$.  Then we find that the Grassmannian $G(k,n)$ is precisely the set of all equivalence classes of $M(k,n)$. That is, $G(k,n)=\GL(k) \backslash M(k,n)$. As a manifold, the dimension of the Grassmannian is $k\times(n-k)$. This discussion holds for both the real and complex cases.

We also wish to define the positive Grassmannian $G_+(k,n)$, which we can think of as the positive part of the Grassmannian $G(k,n)$. See \cite{Postnikov:2006kva} for a comprehensive discussion. For a matrix $M\in M(k,n)$, we say that $M$ is a positive matrix if and only if all of its ordered $k\times k$ minors have the same sign (whenever the minors are non-zero). That is, $\epsilon^{i_1,...,i_k}M_{i_1a_1}...M_{i_ka_k}$ has the same sign for all $1\le a_1<a_2<...<a_k\le n$. Here we are implicitly summing over repeated indices, and $\epsilon$ denotes the Levi-Civita symbol. The set of all positive matrices is denoted $M_+(k,n)$. Clearly, if $M\in M_+(k,n)$, then $G.M\in M_+(k,n)$ for any $G\in \GL(k)$. We define the positive Grassmannian $G_+(k,n)$ to be the set of all equivalence classes of $M_+(k,n)$.

We now define the one-loop Grassmannian. We consider the set $M(k+2,n)$ with a new equivalence relation. We say that two matrices $M,M'\in M(k+2,n)$ are equivalent if and only if $M'=G.M$ for some $G\in \GL(k;1)$, where $\GL(k;1)$ is the group of all invertible $(k+2)\times (k+2)$ matrices of the form
\be\label{gauge}
G \equiv \begin{pmatrix}
K_{11}& K_{12}& \mu_{11} & ... & \mu_{1k}\\
K_{21}& K_{22}& \mu_{21} & ... & \mu_{2k}\\
0     & 0     & J_{11} & ... & J_{1k}\\
0     & 0     & J_{21} & ... & J_{2k}\\
...\\
0     & 0     & J_{k1} & ... & J_{kk}\\
\end{pmatrix}
\equiv
\begin{pmatrix}
K & \mu \\
0 & J \\
\end{pmatrix}
\ee

The group $\GL(k;1)$ is known as a ``parabolic subgroup'' of $\GL(k+2)$. The lower left $k\times 2$ block of the matrix is zero. The set of all equivalence classes is denoted $G(k,n;1)=\GL(k;1) \backslash M(k+2,n)$. The dimension of the group $\GL(k;1)$ is $k^2+2k+4$ so that $\dim G(k,n;1)=\dim M(k+2,n)-\dim \GL(k;1) = k(n-k-2)-4$. We will refer to the subgroup corresponding to the upper left $2\times 2$ matrix as the $\GL(2)$ symmetry, the subgroup corresponding to the lower right $k\times k$ matrix as the $\GL(k)$ symmetry, and the subgroup corresponding to the upper right $2\times k$ matrix as the translation symmetry $T$.

Furthermore, we wish to define the positive one-loop Grassmannian $G_+(k,n;1)$. We say that a matrix $M\in M(k+2,n)$ is positive in the one-loop sense if $M\in M_+(k+2,n)$ and $M_0\in M_+(k,n)$, where $M_0$ denotes the matrix $M$ with the first two rows removed. It is not hard to see that both positivity conditions are invariant under $\GL(k;1)$ transformations. Hence, we define the positive one-loop Grassmannian $G_+(k,n;1)$ to be the set of equivalence classes of matrices which are positive in the one-loop sense.

For the purpose of using uniform notation, we note that $G(k,n)=G(k,n;0)$ and $\GL(k)=\GL(k;0)$, and similarly for their positive parts. 



\subsection{The one-loop amplituhedron and the amplituhedron form}

We now describe how Grassmannians are used to construct the amplituhedron, the central geometry of planar loop integrands for $\mathcal{N}=4$ SYM. Again, we will restrict our attention to the one-loop story. The all loop story was first described in the papers \cite{amplituhedron, Arkani-Hamed:2013kca}, to which we refer the interested reader. For a more mathematical discussion of the amplituhedron as Grassmann polytopes, see \cite{Lam:2015uma}. The relation between Grassmannians and amplitudes, especially for ABJM theory, is also discussed in \cite{Elvang:2014fja}.

We define the one-loop amplituhedron $\mathcal{A}(k,n;1)$, which is a subset of $G(k,k+4;1)$, for each $k,n$, as follows: 
\be
\mathcal{A}(k,n;1) = \left\{\mathcal{Y} = P.Z\; : \; P\in G_+(k,n;1), Z\in M_+(n,4+k)\right\}.
\ee

Here we will write
\be
P = \begin{pmatrix}
D \\
C
\end{pmatrix}
\ee
where the $C$ matrix contains the bottom $k$ rows of $P$ while the $D$ matrix contains the top two rows. The $C$ matrix corresponds to the tree part while the $D$ matrix the loop part of the amplituhedron.
	Furthermore, we define
\be
\mathcal{Y} = \begin{pmatrix}
A \\
B\\
Y
\end{pmatrix}
\ee
where $A,B$ are loop variables living in the top two rows while $Y$ is an auxiliary variable living in the bottom $k$ rows.
	The one-loop amplituhedron has dimension $4k+4$. In general, the $L$-loop amplituhedron $\mathcal{A}(k,n,L)$ has dimension $4k+4L$, where we have four extra dimensions for each loop level corresponding to the 4-dimensional loop integral. The leftover $4k$ degrees of freedom are integrated out by localizing $Y$ to a point, as described in the original reference. We will not need the all-loop amplituhedron for our purposes, hence we will avoid defining it. We move on to the definition of the amplituhedron form, which is where we make contact with the integrand of the amplitude.

There is a canonical top form on the amplituhedron space which is defined as the unique top form that has logarithmic unit singularities on the boundary of the amplituhedron. This definition is rather subtle and requires some clarification. The discussion in this part is completely general and applies to the amplituhedron $\mathcal{A}(k,n;L)$ for arbitrary quantum numbers $k,n,L$. We will let $d$ denote the dimension of the amplituhedron.

First we pick an orientation on the positive part of the amplituhedron and refer to that as the positive orientation. Then, we tile the amplituhedron by simplices $S_i$. A simplex $S_i$ in the amplituhedron is an open subset parametrized by some positive variables $\alpha_{ij} \;(j=1,...,d)$ which we assume to be positively oriented. We assign a top form to each simplex, which is simply given by the logarithmic form as follows.
\be
\omega(S_i) \equiv \frac{d\alpha_{i1}...d\alpha_{id}}{\alpha_{i1}...\alpha_{id}}
\ee
The amplituhedron form is the sum of all these logarithmic forms.
\be
\omega(\mathcal{A}) = \sum_{i}\omega(S_i)
\ee
For example, each BCFW term is a simplex whose parameters are given by a (non-unique) subset of bridge variables on its associated momentum-twistor diagram, as we will discuss in the next section. The positive orientation of the simplices is important for getting the correct relative signs between the logarithmic forms when summing over $i$.

There is an important and beautiful subtlety of the amplituhedron form, which is that it is triangulation independent. Roughly speaking, the logarithmic singularities that appear on the boundary between two adjacent simplices cancel each other out. These are called spurious singularities, and they reflect the fact that spurious poles cancel between BCFW terms. The singularities appearing at the boundary of the amplituhedron, however, are true singularities of the amplituhedron form, and they reflect the singularities of the integrand that appear when a Feynman propagator goes on-shell.

The cancellation of spurious singularities, and hence the existence and uniqueness of the logarithmic form, is a very general property of positive spaces that is worth studying even without physics motivations. It seems to be something that is still not very well understood, or even well known, in the mathematics literature. 

There is also the possibility of obtaining the top form without using triangulations. This is a deep and difficult problem on which some progress has been made. It is related to finding an "amplituhedron dual", which is discussed in \cite{Arkani-Hamed:2014dca}. We will leave discussions of this to future work.

Finally, the amplituhedron is a purely bosonic object while the super amplitude is fermionic of degree $4k$. Obtaining a fermionic object from a bosonic form involves a simple procedure outlined in~\cite{amplituhedron}, which involves integrating over $4k$ auxiliary fermonic variables first done in \cite{ArkaniHamed:2010gg}. This provides a bosonic way of studying supersymmetry. The implications for general supersymmetric theories is still mysterious.

\subsection{Momentum-twistors and momentum-twistor diagrams}
We now provide a brief review of amplitudes/Wilson loops in momentum-twistor space. See \cite{Elvang:2013cua} for a more comprehensive review. Denote the $n$-point N${}^k$MHV $L$-loop amplitude as $\mathcal{A}_{n,k}^{(L)}$, and throughout the paper we will consider the MHV-tree-stripped amplitude $A_{n,k}^{(L)}$. We have
\be
\mathcal{A}_{n,k}^{(L)}= \frac{\delta^{2\times (2|4)} (\sum_{i=1}^n \lambda^{\alpha}_i (\tilde\lambda^{\dot{\alpha}}_i |\tilde\eta^A_i) )}{\l 1 2\r \ldots\l n{-}1 n\r\l n 1\r} A_{n,k}^{(L)}\,, 
\ee
where $\alpha, \dot\alpha=1,2$ are SU(2) indices of spinors $\lambda_i$ and their conjugates $\tilde\lambda_i$ encoding the null momenta of $n$
particles,  and $A=1,...,4$ is the SU(4) index of Grassmann variables $\tilde\eta_i$ describing their helicity states. By going to the dual (super)space, one defines (super) momentum-twistors in the fundamental representation of its superconformal group 
\be \Z_i=
(Z_i^a | \eta_i^A)=(\lambda_{i\alpha}, \mu_i^{\dot\alpha} |\eta_i^A)\,,
\ee 
which are unconstrained variables and they determine $\tilde\lambda,\eta$ via,
\be
(\tilde\lambda |\tilde\eta)_i=\frac{\l i{-}1\,i\r (\mu|\eta)_{i{+}1}+\l i{+}1\,i{-}1\r (\mu|\eta)_{i}+\l i\,i{+}1\r (\mu|\eta)_{i{-}1}}{\l i{-}1\, i\r\l i\,i{+}1\r}.
\ee

The central object we will study in this paper is the {\it integrand} of amplitudes/Wilson loops in momentum-twistor space. We denote the four-dimensional integrand of $A_{n,k}^{(L)}$ as $Y_{n,k}^{(L)}$, which is a form of degree $4L$ in the $L$ loop variables denoted as $\ell$'s. Formally it reads
\be
A^{(L)}_{n,k}=\int_{\rm reg} Y^{(L)}_{n,k}(\Z_1,\ldots,\Z_n; \{\ell_1,\ldots,\ell_L\})=\int_{\rm reg} \prod_{m=1}^L d^4 \ell_m\,I_{n,k}^{(L)} (\Z_1,\ldots,\Z_n; \{\ell_1,\ldots,\ell_L\})\,,\nl
\ee
where ``reg" means regularizations needed for the loop integrals; by pulling out the integral measure explicitly, the remaining part of $Y^{(L)}_{n,k}$, as a rational function, is denoted as $I^{(L)}_{n,k}$.

The loop variables $\ell$'s are lines in momentum-twistor space, represented by bi-twistors: $\ell_m\equiv (A_m B_m)\equiv(A B)_m$, for $m=1,\ldots,L$, and the integral measure is defined as
\be\label{loopmeas}
d^4 \ell  \equiv \l A B d^2 A\r\l A B d^2 B\r = \frac{d^4 A d^4 B}{{\rm vol}\; \GL(2)}\,.
\ee
It is an integral over the space of lines $(AB)$, written as a pair of points (twistors) $A$ and $B$, divided by the $\GL(2)$ redundancies labeling their positions on the line~\cite{all loop}. 

Now we turn to the momentum-twistor diagrams, which will help us better understand the top cell measures. Momentum-twistor diagrams were first introduced in \cite{Bai:2014cna}, and are based on the original on-shell diagrams introduced in \cite{posGrassmannian}. We will not be able to go into much detail about how to compute them. We will simply remind the reader of the definition and briefly describe some important properties.

Momentum-twistor diagrams are graphs on a disc with points $1,...,n$ labelled in cyclic order along the boundary of the disc. The interior of the disc consists of black and white vertices with lines joining pairs of vertices together. To each external label $i$, there is a line connecting the label to one of the internal vertices. The vertices are trivalent when all the lines are included.

The white vertices are evaluated as an integral over $C\in G(1,3)$ while the black vertices are evaluated over $C\in G(2,3)$, as follows.
 
\be
\vcenter{\hbox{\includegraphics[width=3cm]{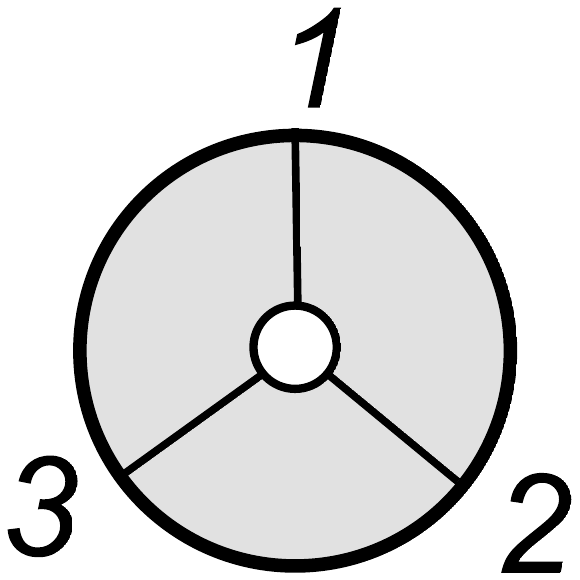}}}&&\int \frac{1}{\text{Vol}\;\GL(1)}\frac{d^{1\times3}C}{(1)(2)(3)}\delta^{4|4}\left(\sum_{a=1}^3 C_a \mathcal{Z}_a\right)\\
\vcenter{\hbox{\includegraphics[width=3cm]{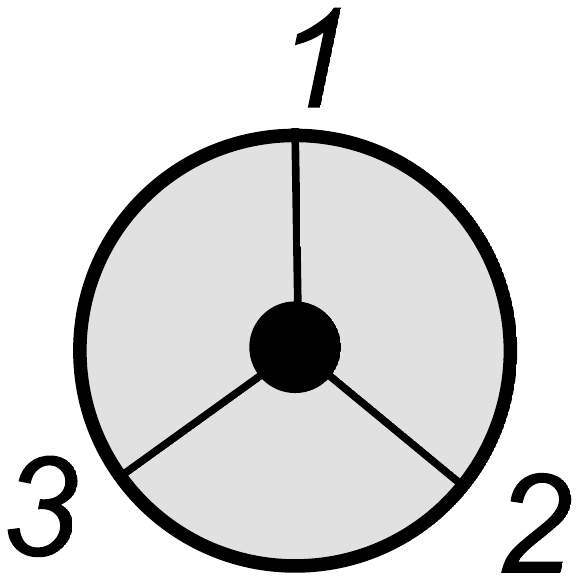}}}&&\int \frac{1}{\text{Vol}\;\GL(2)}\frac{d^{2\times3}C}{(12)(23)(34)}\prod_{\alpha=1}^2\delta^{4|4}\left(\sum_{a=1}^3C_{\alpha,a}\mathcal{Z}_a\right)
\ee
where in the first case $(a)$ denotes the $a^\text{th}$ component of $C$ for $a=1,2,3$, and in the second case $(ab)$ denotes the minor for columns $a,b$ of $C$ for $a,b=1,2,3$.

For each bridge between two vertices, we integrate over the momentum-twistor along the bridge via
\be
\frac{d^{4|4}\mathcal{Z}}{\text{Vol} \;\GL(1)}
\ee

Once all the vertices are joined together and all the bridge integrals are computed, the full diagram becomes an integral over a cell of the Grassmannian $G(k,n)$. The value of $k$ is determined by the Grassmann degree of the diagram, which must be $4k$. It can also be read off as the number of arrows pointing into the diagram from the boundary in any perfect orientation, which is a choice of direction along each bridge so that each white vertex has one sink and two sources, and each black vertex has two sinks and one source. The cell is parametrized by some variables in the space of $C$ matrices. For planar diagrams, the number of variables, and hence the dimension of the diagram, is $F{-}1$ at tree level, where $F$ is the number of faces appearing in the diagram. At loop level, the number of variables is $F{-}L{-}1$, where $L$ is the number of loops. For non-planar diagrams, the counting is slightly more involved.

Recall that two diagrams are equivalent if they are related by an equivalence move. At tree level, the only equivalence move is the square move \cite{posGrassmannian}. At loop level, there are new equivalence moves, making the loop level story even richer. We will discuss these features in future work.

We will assume that all our diagrams are reduced, meaning that sub-diagrams of the following form do not exist in any equivalent diagram. 
\be
\vcenter{\hbox{\includegraphics[width=2.6cm]{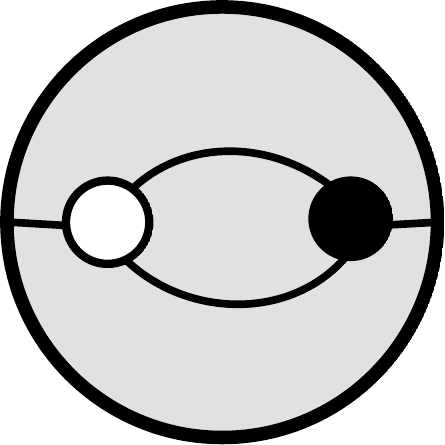}}}
\ee

In the momentum space on-shell diagrams, these sub-diagrams are interpreted as loop integration variables. This interpretation does not carry over to the analysis of momentum-twistor diagrams where the loop variables take on a totally different form.

At loop level, there is a new structure called a "bubble", which looks like the following.
\be
\vcenter{\hbox{\includegraphics[width=2.6cm]{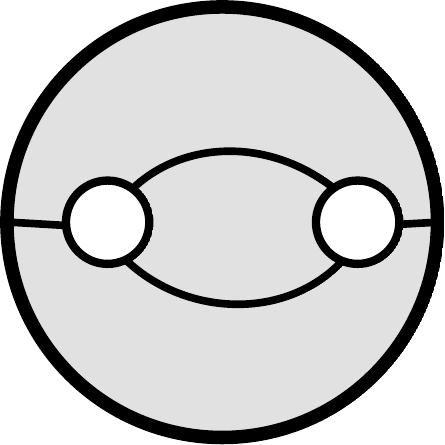}}}
\ee

We have one of these for each loop, with each bubble corresponding to one loop variable. In any perfect orientation, the arrows along the two external lines of the bubble point outward. Treating these two lines as though they were external particles, we can read off its boundary measurement giving us a cell in the space of $D$ matrices rather than $C$.

The all-loop BCFW recursion~\cite{all loop} can be translated into the language of momentum-twistor diagrams, where each BCFW term corresponds to a diagram, and the total integrand is the sum of a sequence of such diagrams. The diagrams capture all aspects of the computation, and in fact simplifies the computation in some cases. One particular advantage of the diagram is that it makes manifest that each BCFW term is a product of the $d\log$ of some variables, since the same is obviously true for each diagram. Furthermore, it is trivial to read off what those variables are by following the boundary measurement technique of Postnikov~\cite{Postnikov:2006kva}. For a detailed example, see the Appendix, which shows the 16 BCFW terms for the 6-point NMHV 1-loop integrand, their momentum-twistor diagrams and their coordinates in $P$ space.

The diagrams do NOT have to be planar. This is a novel feature of momentum-twistor diagrams. As explained in the original paper, one degree of non-planarity appears naturally when taking the forward limit in BCFW recursion. This may appear worrisome since we are describing the planar sector. However, this is not a problem for two reasons. One, the computation of the momentum-twistor diagrams does not require planarity. Two, there is a procedure, very similar to a square move but only appearing at loop level, that gets rid of all the non-planarities generated by BCFW. This is a rather deep fact which will be explored in a future paper.

\section{The one-loop Grassmannian for $k=1$}\label{k=1}

We now provide a detailed analysis of the Grassmannian geometry at one-loop. We will focus on the $k=1$ case because the geometry is easily visualizable as a polygon in two dimensions. The geometry of higher $k$ cases is harder to interpret and we will leave their exploration to future work.

\subsection{The geometry of $G_+(1,n;1)$}
Recall that for the one-loop NMHV Grassmannian, we think of a point $P\in G_+(1,n;1)$ as a $3 \times n$ matrix where the first two rows form the $D$ matrix and the last row is the $C$ matrix, modded out by the parabolic group $\GL(1;1)$.  There is a torus $T _+= \GL_+(1)^n$ that acts on $G_+(1,n;1)$ by scaling the columns independently by positive scalars.

We describe $X = G_+(1,n;1)/T_+$ explicitly in terms of spaces of $n$ points in the projective plane $\P_+^2\equiv(\mathbb{R}^2\times\mathbb{R}_{\ge 0})/\mathbb{R}_+$, which is the space of all vectors of the form $(d_1,d_2,c)^T$ with $c\ge 0$ modded out by overall multiplication by positive scalars. Fix a line at infinity $L_\infty \subset \P_+^2$ which we can take to be the set of all non-zero vectors with $c=0$, and let $\A^2 = \P_+^2 \setminus L_\infty$ denote the affine plane.
The space $X$ consists of collections $P\equiv(P_1,P_2,\ldots,P_n)$ of $n$ points where each $P_i$ is either
\begin{enumerate}
\item
a point $P_i \in \A^2$, or
\item
a point $P_i \in L_\infty$, or
\item
``zero" (equivalently, a point not on $\P_+^2$).
\end{enumerate}
Case (1) occurs if the third entry of the $i$-th column of $C$ is non-zero.  Then the $i$-th column of $P$ can be scaled to the column vector  $(d_1,d_2,1)^T$, representing the point $(d_1,d_2)^T\sim (d_1,d_2,1)^T \in \A^2$.  

\noindent
Case (2) occurs if the third entry of the $i$-th column of $C$ is zero, but the $i$-th column of $P$ is non-zero.  Then we have the column vector $(d_1,d_2,0)^T$, representing the point $(d_1,d_2)^T\sim (d_1,d_2,0)^T \in L_\infty$.  

\noindent
Case (3) occurs if the $i$-th column of $P$ is the zero vector. 

The positivity of the collection $P=(P_1,P_2,\ldots,P_n)$ implies the following. 
\begin{enumerate}
\item the points in $P$ span $\P_+^2$ -- in particular, there are at least 3 of them, and they cannot all lie on the line at infinity, and
\item
the points in $\P_+^2$ are the boundary points of a convex polygon which we will also refer to as $P$ (with some points allowed to be at infinity).
\end{enumerate}

Two matrices $P$ and $P'$ represent the same point in $G_+(1,n;1)$ if they are related by one of the three operations:
\begin{enumerate}
\item
the $\GL_+(1)$-action on the row $C$, which rescales the polygon relative to the origin of the affine plane, or
\item
the $\GL_+(2)$-action on the two rows $D$, which linearly transforms the polygon on the affine plane while preserving convexity, or
\item
the $T$ action that adds a multiple of $C$ to one of the rows of $D$, which translates the polygon.
\end{enumerate}
We note that the $\GL_+(1)$ action that multiplies the $C$ matrix by a scalar $\alpha$ can be thought of also as a $GL_+(2)$ action that rescales the $D$ matrix by $1/\alpha$, since the action of the torus $T_+$ has been modded out. In particular, two collections $P$ and $P'$ represent the same point of $G_+(1,n;1)$ if they are related by the action of $\GL_+(2) \ltimes \A^2$, where $\A^2$ acts as translations on $\A^2$ (fixing the line at infinity).  In other words, $X$ is the space of $n$-points in $\P_+^2$ in convex position (with a distinguished line at infinity), modulo the action of affine transformations.

A simple example of a polygon $P$ is given by the following, where $\theta_s = 2\pi s/6$ for $s=0,...,5$.
\be
\vcenter{\hbox{\includegraphics[width=5cm]{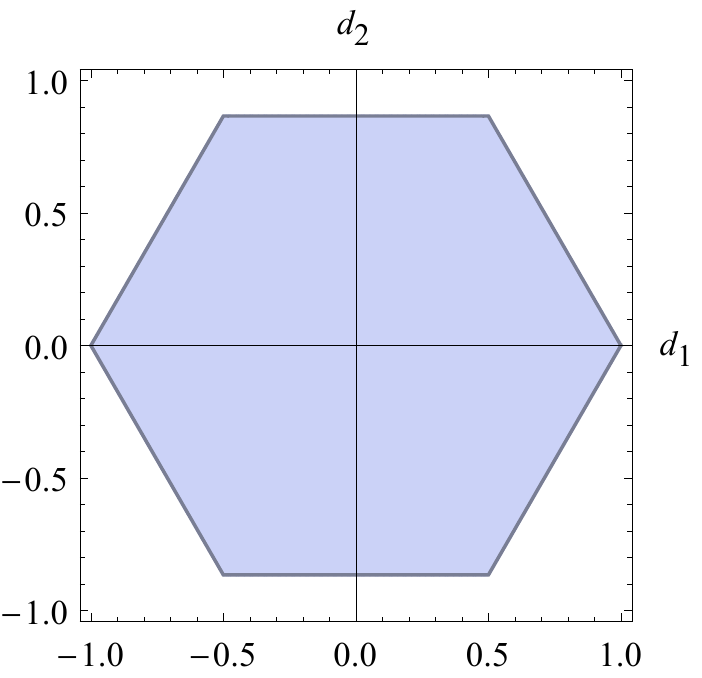}}}\;\;\;\;
P=\begin{pmatrix}
\cos\theta_0 & \cos\theta_1 & \cos\theta_2 & \cos\theta_3& \cos\theta_4& \cos\theta_5\\
\sin\theta_0 & \sin\theta_1 & \sin\theta_2 & \sin\theta_3& \sin\theta_4& \sin\theta_5\\
1 & 1 & 1 & 1 & 1 & 1
\end{pmatrix}
\ee

\subsection{The 0-dimensional cells}
We now wish to discuss the 0-dimensional cells of $G_+(1,n;1)$. These innocuous parts of the geometry are actually crucial, because they are the seeds from which all other diagrams can be constructed.

We first remind ourselves of the 0-dimensional cells appearing in the tree level Grassmannian $G_+(k,n)$. We pick $n$ external states $a_1,...,a_k$ and gauge fix the Grassmannian to be identity in the $k\times k$ block corresponding to columns $a_1,...,a_k$, and set all other components to zero. For instance, for $G_+(2,5)$ and $(a_1,a_2)=(2,3)$, the Grassmannian point is given by the following.
\be
\vcenter{\hbox{\includegraphics[width=4cm]{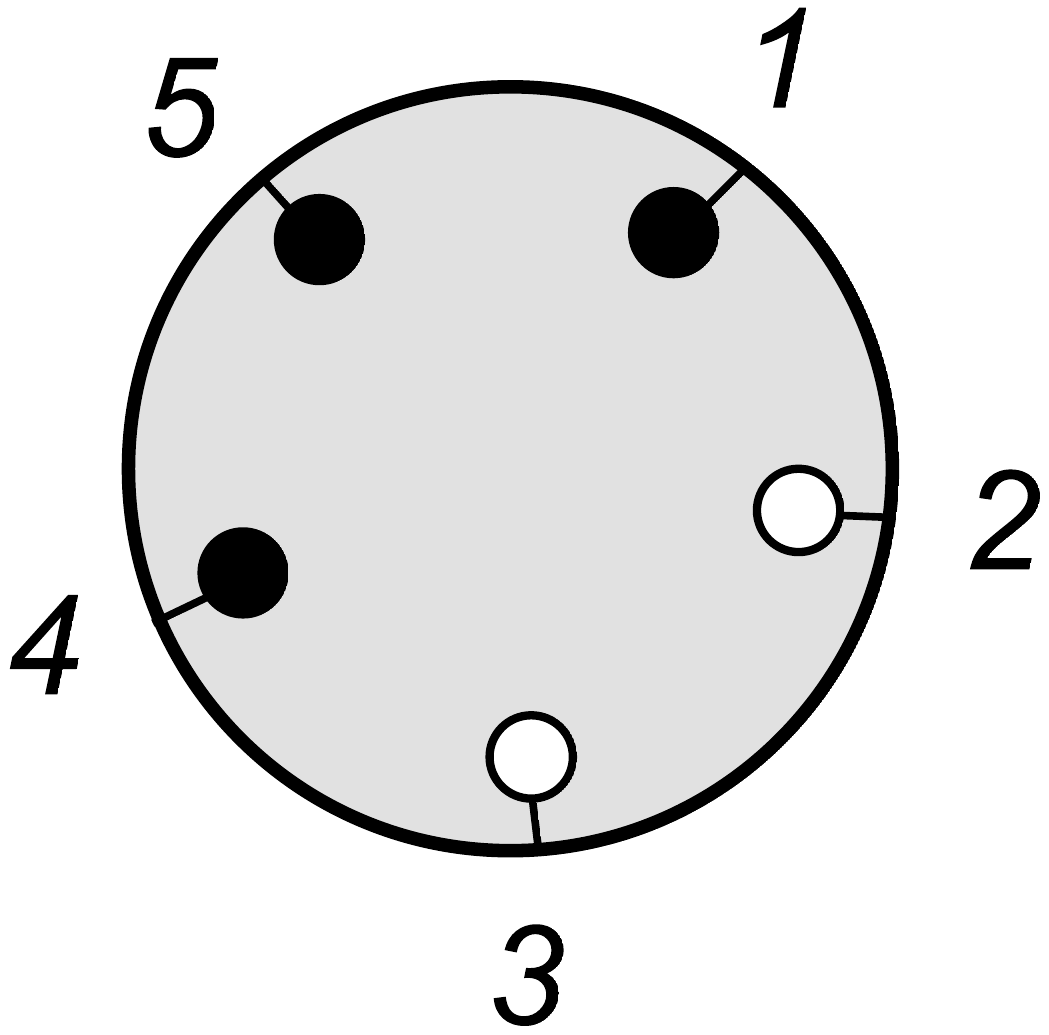}}}\;\;\;\;
C=\begin{pmatrix}
0 & 1 & 0 & 0 & 0 \\
0 & 0 & 1 & 0 & 0
\end{pmatrix}
\ee
Here we have also included the corresponding momentum-twistor diagram. In general, a 0-dimensional cell consists of black and white ``lollipops" attached to the external legs of the diagram. The white lollipops are positioned precisely at the indices $a_1,...,a_k$. The number of white lollipops is therefore the $k$ charge of the diagram. In particular, the MHV tree diagram is just a ring of black lollipops. 

We now return to the one loop $k=1$ case. The $0$-dimensional cells $\Pi_{a;b,c}$ of $G_+(1,n;1)$ are exactly the torus-invariant points of $G_+(1,n;1)$.  They are determined by picking $a \in \{1,2,\ldots,n\}$ and another unordered pair $(b,c) \in \{1,2,\ldots,n\} - \{a\}$.  The points $\Pi_{a;b,c}$ consists of a point $P_a \in \A^2$ and distinct points $P_b,P_c \in L_\infty$.  By an affine transformation, we can assume that $P_a, P_b, P_c$ are precisely the standard unit vector basis in $\P_+^2$.  For example, if $n = 5$, we have that $\Pi_{1;3,4}$ is represented by the following matrix.
\be
\vcenter{\hbox{\includegraphics[width=4cm]{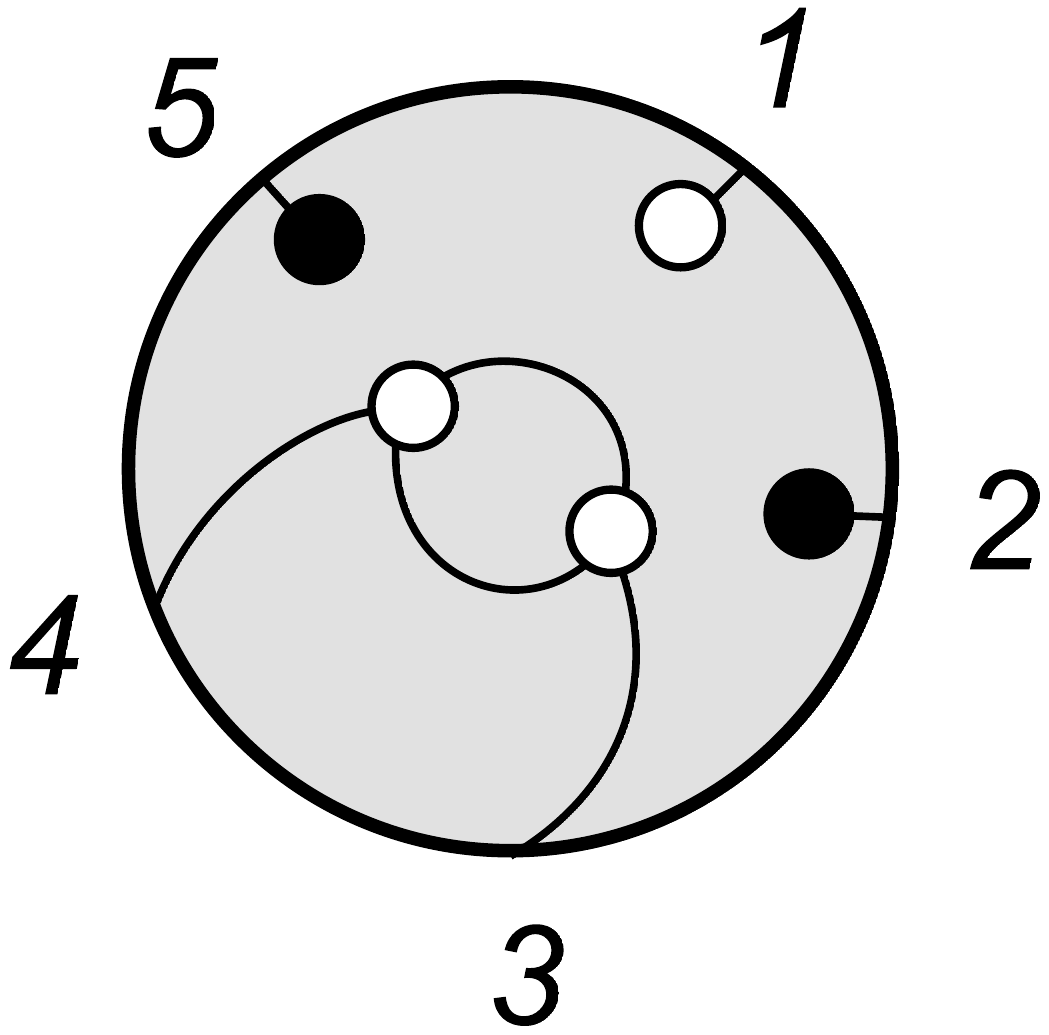}}}\;\;\;\;
P=\begin{pmatrix}0&0&1 & 0 & 0\\
0&0&0&1&0 \\
1& 0&0&0&0\end{pmatrix}
\ee
Again we have included the corresponding diagram. As in the tree level case, we must have one white lollipop at particle $a$ corresponding to the $k=1$ charge. The novelty at one loop is that we must have one bubble consisting of two external legs. The bubble represents the loop variable while the legs reflect the corresponding $GL(2)$ symmetry. For 0-dimensional cells, these two legs must be connected to the two external particles labelled $b,c$. At one loop, this is the only type of 0-dimensional cell that can appear. At higher loops, the classification of 0-dimensional cells is still unclear. For instance, we can imagine connecting two bubbles together through their external legs. Whether or not this type of diagram should be classified as a 0-dimensional cell, and what other cells are possible, are still open questions.

\subsection{Top cells}

We now describe the top cells of $G_+(1,n;1)$; more specifically we describe the corresponding top cells of $X = G_+(1,n;1)/T_+$.  Suppose that $P=(P_1,\ldots,P_n)$ is a convex $n$-gon in $\A^2$ with points occuring in cyclic counterclockwise order.  Suppose that $a, a+1, b, b+1$ are in cyclic order, where $a+1$ can equal $b$, but $a \neq b+1$.  Let us say that the edges $(a,a+1)$ and $(b,b+1)$ intersect {\it positively} if the intersection point $(a,a{+}1)\cap(b,b{+}1)\equiv P_b(a,a{+}1,b{+}1)-P_{b{+}1}(a,a{+}1,b)$ of the two lines is on the opposite side of the line $(a+1,b)$ to both $a$ and $b+1$.  If the intersection point $X$ is on the other side of $(a+1,b)$, then we say that $(a,a+1)$ and $(b,b+1)$ intersect negatively.  Note that $(a,a+1)$ and $(b,b+1)$ intersect positively if and only if $(b,b+1)$ and $(a,a+1)$ intersect negatively.  For short we say that $[a,a+1;b,b+1]$ is positive.  If $a+1 = b$, then $[a,a+1;a+1,a+2]$ is always positive.

The condition that $[a,a+1;b,b+1]$ is positive is equivalent to the algebraic condition
\begin{equation}\label{eq:can}
[(a,a{+}1)\cap(b,b{+}1)]=[b](a,a{+}1,b{+}1)-[b{+}1](a,a{+}1,b) > 0.
\end{equation}
Here $[a]$ denotes a minor of the row $C$, and $(a,b,c)$ denotes a minor of the $3 \times n$ matrix $P$.
For example, suppose $n = 4$ and $a = 1$ and $b = 3$.  By an affine transformation, a convex quadrilateral in $\A^2$ can be made to have vertices $(0,0),(1,0),(x,y),(0,1)$, so that the corresponding $P$ matrix is
\be
\underset{y<1}{
\vcenter{\hbox{\includegraphics[width=4.5cm]{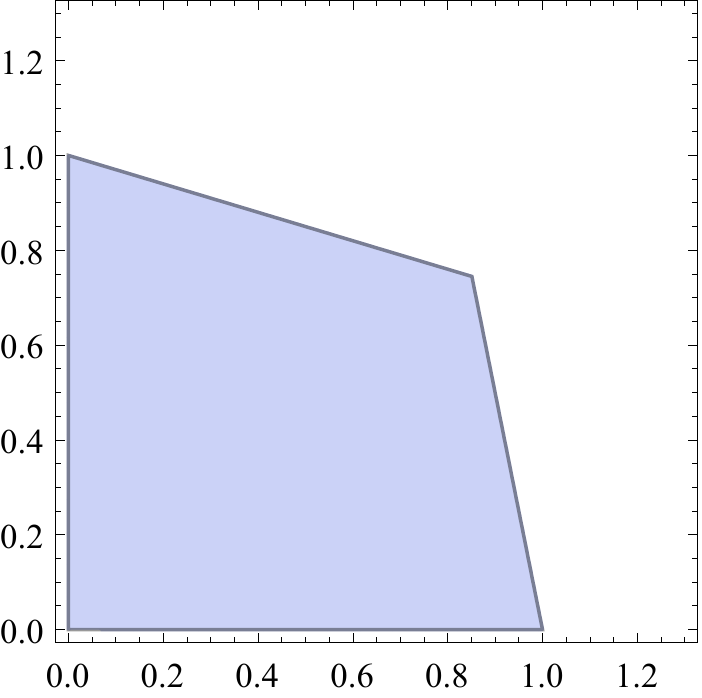}}}}\;\;\;\;
\underset{y>1}{
\vcenter{\hbox{\includegraphics[width=4.5cm]{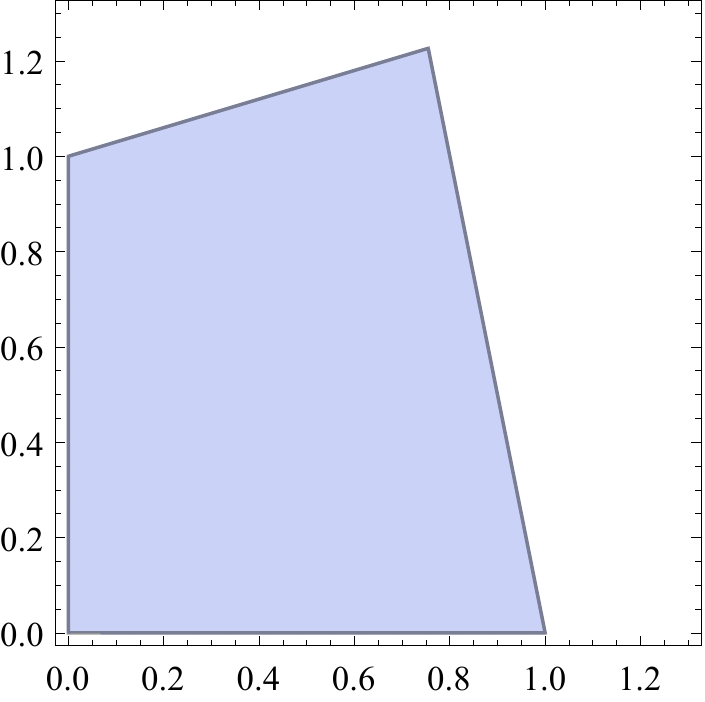}}}}\;\;\;\;
P = 
\begin{pmatrix}
0 & 1 & x & 0 \\
0 & 0 & y & 1 \\ 
1 & 1 &1 &1 
\end{pmatrix}.
\ee
The convexity of the quadrilateral is equivalent to the conditions $(123) = y >0 $ and $(134)= x > 0$ and $(234) = x+y-1 > 0$.  The condition that $(12)$ and $(34)$ intersect positively is the condition $y < 1$, or equivalently, $[(12)\cap(34)]=[3](124)-[4](123) > 0$.

A top cell can be described in the following way: it is the space of convex $n$-gons in $\A^2$, considered up to affine transformations, satisfying additional constraints specifying that certain pairs of edges $(a,a+1)$ and $(b,b+1)$ intersect positively.  These constraints are only specified for some pairs of edges.  Other pairs of edges are allowed to intersect either positively or negatively.  

There are $\binom{n}{3}$ top cells in total.  Given a $3$-element subset $S \subset \{1,2,\ldots,n\}$, we now describe the corresponding top cell $\Pi_S$.  Let $S = \{a< b < c\} \subset \{1,2,\ldots,n\}$.   Then the intersection constraints for $\Pi_S$ are that $[a,a+1;b,b+1]$, $[b,b+1;c,c+1]$, and $[c,c+1;a,a+1]$ are positive.  Note that $[a,a+1;b,b+1]$ being positive implies that $[a',a'+1;b,b+1]$ is positive for all $a < a' < b$.  Thus $\Pi_S$ is cut out of $G_+(1,n;1)$ by three quadratic equations of the form \eqref{eq:can}.

Another way to describe $\Pi_{\{a,b,c\}}$ is as follows.  Start with the intersection points 
\be\label{eq:triangle}
X_1 = (a,a+1) \cap (b,b+1), \; X_2 = (b,b+1) \cap (c,c+1), \; X_3 = (c,c+1) \cap (a,a+1).
\ee
The convex hull of these three points is a big triangle $\Delta$.  Then $\Pi_{\{a,b,c\}}$ is the space of $n$-gons that can be inscribed inside $\Delta$ so that the edges $(a,a+1)$, $(b,b+1)$ and $(c,c+1)$ lie on the three edges of $\Delta$, as shown in the following picture.

\be
\vcenter{\hbox{\includegraphics[width=4.5cm]{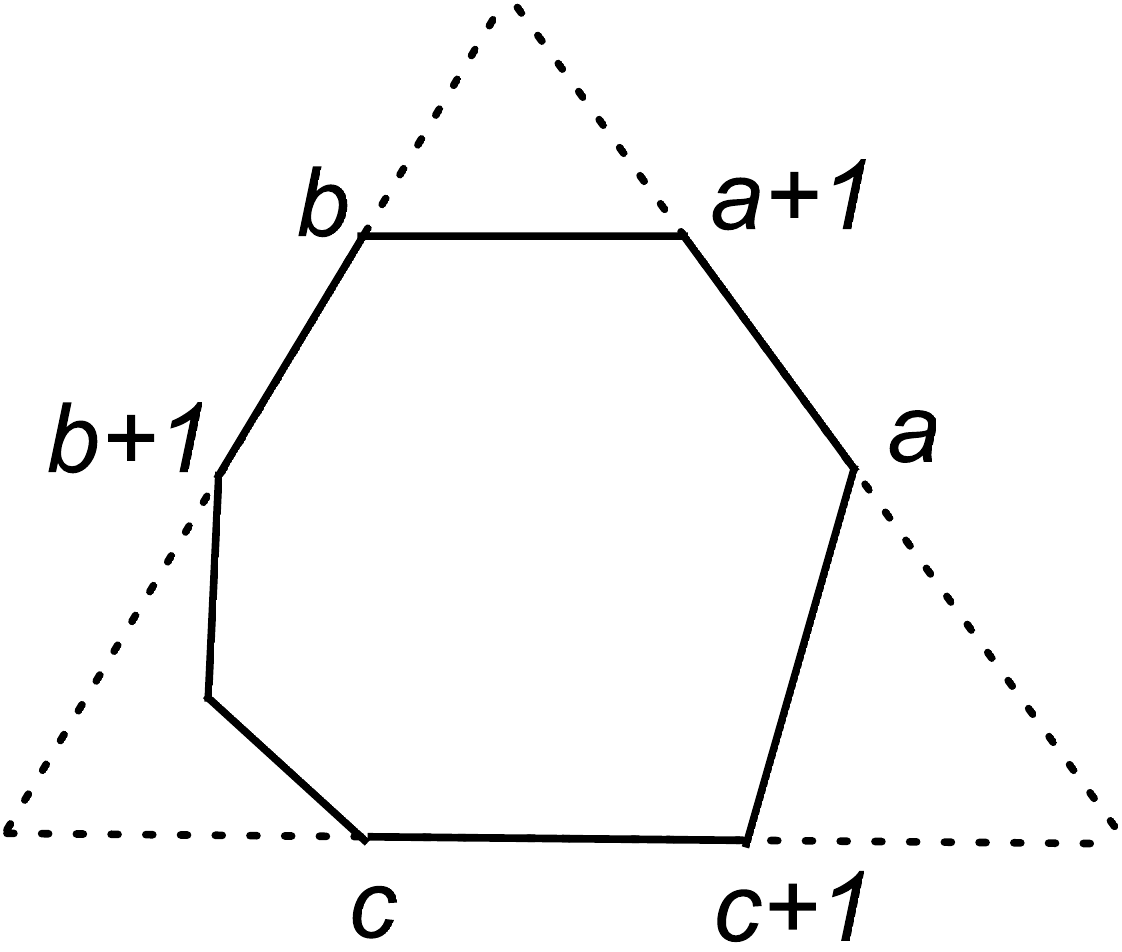}}}\;\;\;\;
\ee

Another way to think about the inscribed $n$-gon is that it is obtained from $\Delta$ by slicing it a number of times.  If $a+1 = b$, then the intersection point $X_1 = (a,a+1) \cap (a+1,a+2)$ is simply $a+1$, so sometimes $\Delta$ shares some of its vertices with the $n$-gon $P$. 

\subsection{Shifts and cells of all dimension}

The top cells $\Pi_S$ can be built up from $0$-dimensional cells via shifts of columns of the $P$ matrix.  The geometry of the intersection conditions on edges arises naturally in this way.

We consider two kinds of shifts: The shift ($i+1 \to i$) acts by adding $a P_i$ to $P_{i+1}$ while ($i \to i+1$) acts by adding $a P_{i+1}$ to $P_i$, where $a$ is positive and will later be interpreted as a bridge variable.  These shifts send points in $G_+(1,n;1)$ to points in $G_+(1,n;1)$.  In terms of point configurations, ($i+1 \to i$)  moves $P_{i+1}$ to a point on the interval connecting $P_i$ and $P_{i+1}$, while ($i \to i+1$) moves $P_i$ to a point on the same interval.  Note that if $P_{i+1} = 0$ is not on $\P^2$, then ($i+1 \to i$)  just places $P_{i+1}$ at the location of $P_i$.

\be
\vcenter{\hbox{\includegraphics[width=4cm]{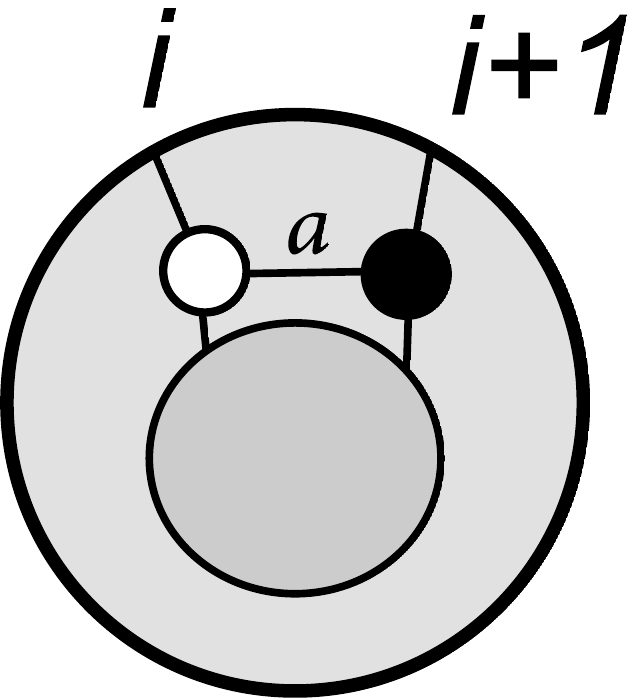}}}\;\;\;
\vcenter{\hbox{\includegraphics[width=4cm]{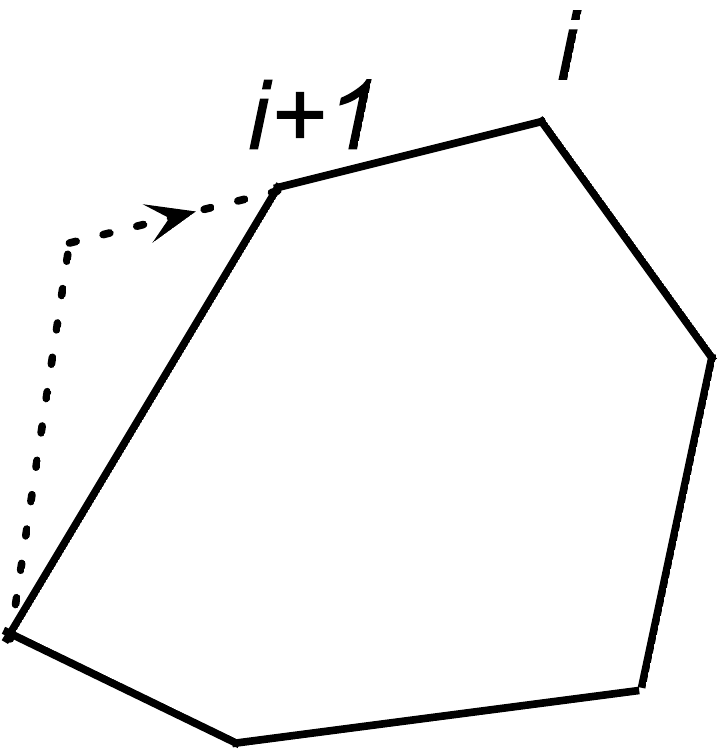}}}\\
\text{The shift $(i+1\rightarrow i)$ where $P_{i{+}1} \rightarrow P_{i{+}1}{+}a P_i$}
\ee

As we are showing in the diagram above, the shifts have a simple interpretation as a BCFW bridge. For instance, by doing the shift $(i+1\rightarrow i)$ we are simply attaching a new pair of black and white vertices at positions $i$ and $i+1$. The two ways of coloring the vertices reflects the two possible shifts $(i+1\rightarrow i)$ and $(i\rightarrow i+1)$, of which we have drawn the former. The variable $a$ appears as a bridge variable on the diagram, which is consistent with the fact that adding a bridge increases the number of faces, and hence the dimension, of the diagram by 1.

Using shifts where $a$ varies over $\R_{>0}$, we can build up higher-dimensional cells from 0-dimensional cells, at least at one loop. We can obtain all top cells of $G_+(1,n;1)$ in this way, though in general we need to allow non-adjacent shifts, and possibly negative values for $a$, in which case we use $-a$ so that $a>0$. The same is probably true to all loops, though it is not clear what all the 0-dimensional cells are in those cases, as discussed earlier.

Any momentum-twistor diagram can be decomposed into a sequence of bridges, or shifts. In other words, there exists a sequence of shifts that takes a 0-dimensional cell to the full diagram. In practice, this sequence is obtained by starting with the full diagram and removing one bridge at a time. This is related to the fact that each permutation in the symmetric group $S_n$ can be written as a composition of transpositions, as explained in the original reference~\cite{posGrassmannian}. Of course, the decomposition is not unique, and neither is the 0-dimensional cell.

\subsection{A 5-point example}\label{ssec:ex1}

 
Let $n= 5$.  As an example, we now illustrate how the cell B-1 in the Appendix is built up via shifts. Recall that B-1 is one of three BCFW terms (B-1, B-2, B-3) for the 5-point 1-loop NMHV integrand. In the Appendix, these three terms are given as part of the 16 terms for the 6-point case. In what follows, we have allowed ourselves to use non-adjacent shifts, which work in the obvious way.\\

\noindent
(0) Start with the 0-dimensional cell $\Pi_{1;3,5}$.  Thus we have a point $P_1 \in \A^2$, and two points $P_3,P_5 \in L_\infty$.

$$
P=\begin{pmatrix}0&0&1& 0 & 0\\
0&0&0&0&1 \\
1& 0&0&0&0\end{pmatrix}
\;\;\;\;\;
\vcenter{\hbox{\includegraphics[width=3.5cm]{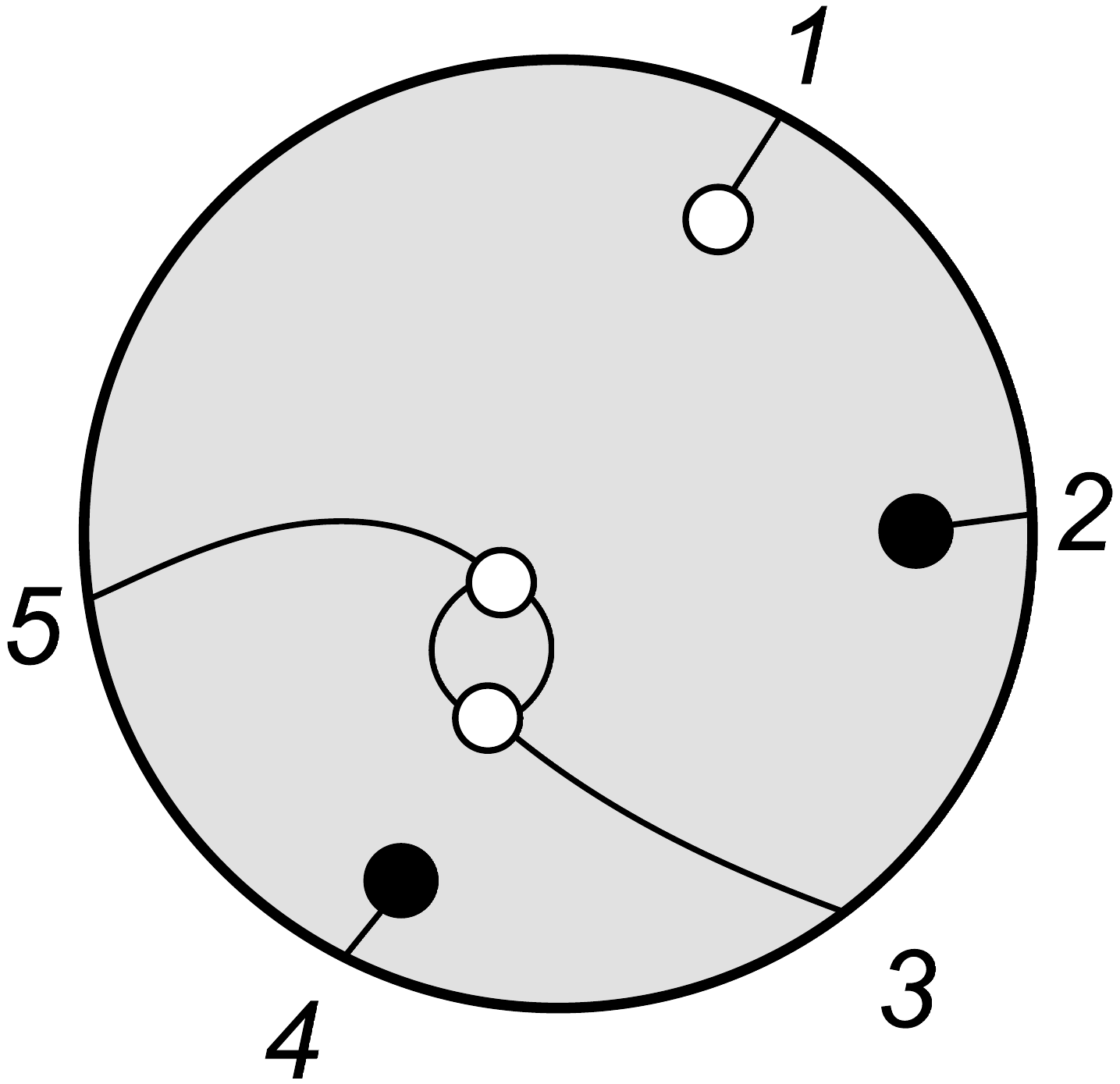}\;\;\;\;\;\includegraphics[width=3.5cm]{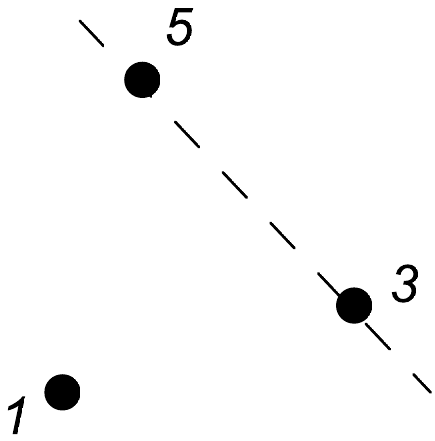}}}
$$

\noindent
(1) Apply the shift $(2 \to 3)$.  This places $P_2$ at the same location as $P_3$.  

$$
P=\begin{pmatrix}0&a_1&1& 0 & 0\\
0&0&0&0&1 \\
1& 0&0&0&0\end{pmatrix}
\;\;\;\;\;
\vcenter{\hbox{\includegraphics[width=3.5cm]{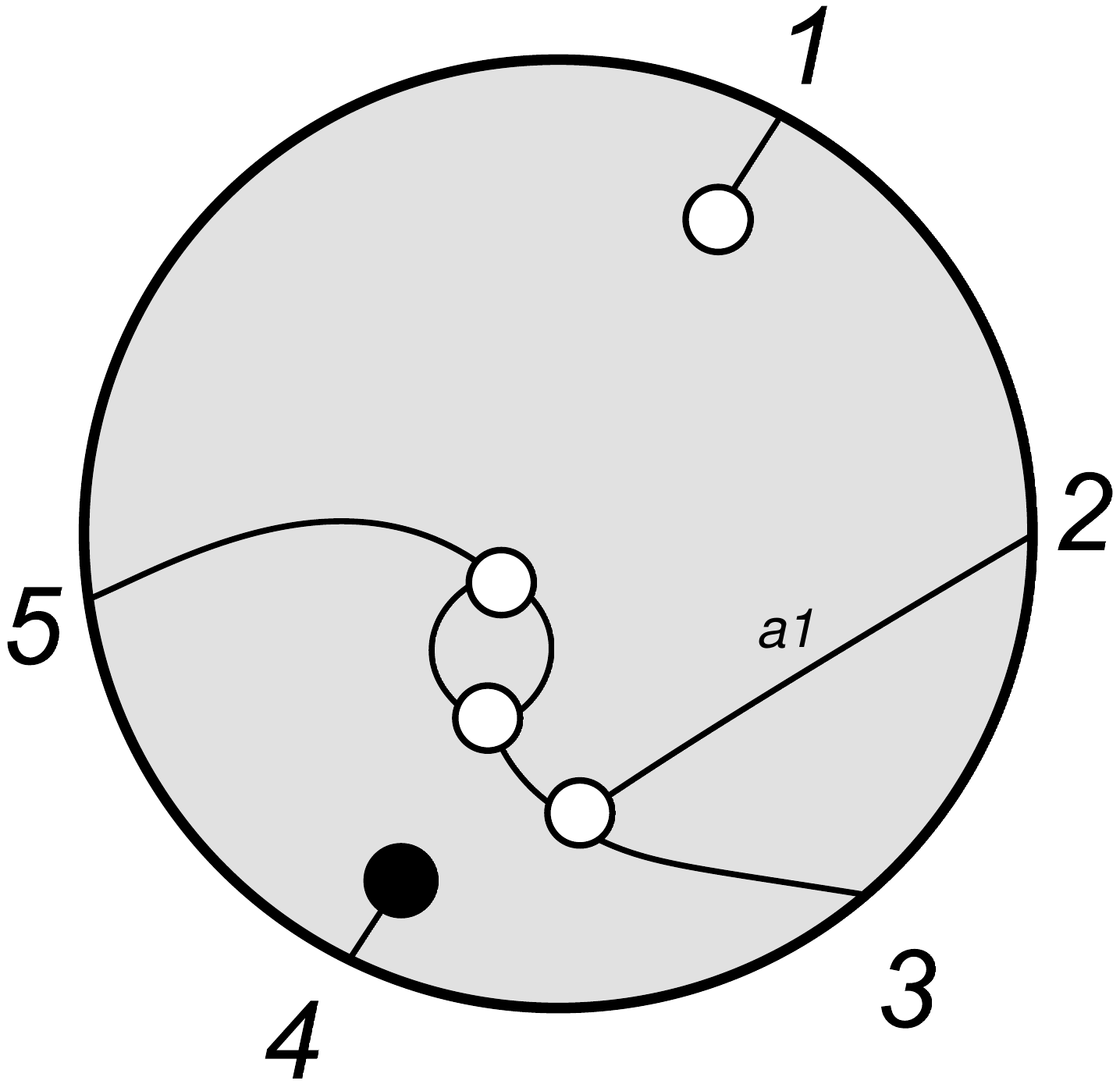}\;\;\;\;\;\includegraphics[width=3.5cm]{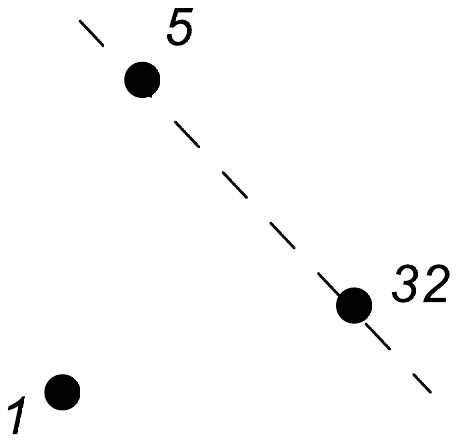}}}
$$

\noindent
(2) Apply the shift $(4 \to 5)$.  This places $P_4$ at the same location as $P_5$.  

$$
P=\begin{pmatrix}0&a_1&1& 0 & 0\\
0&0&0&a_2&1 \\
1& 0&0&0&0\end{pmatrix}
\;\;\;\;\;
\vcenter{\hbox{\includegraphics[width=3.5cm]{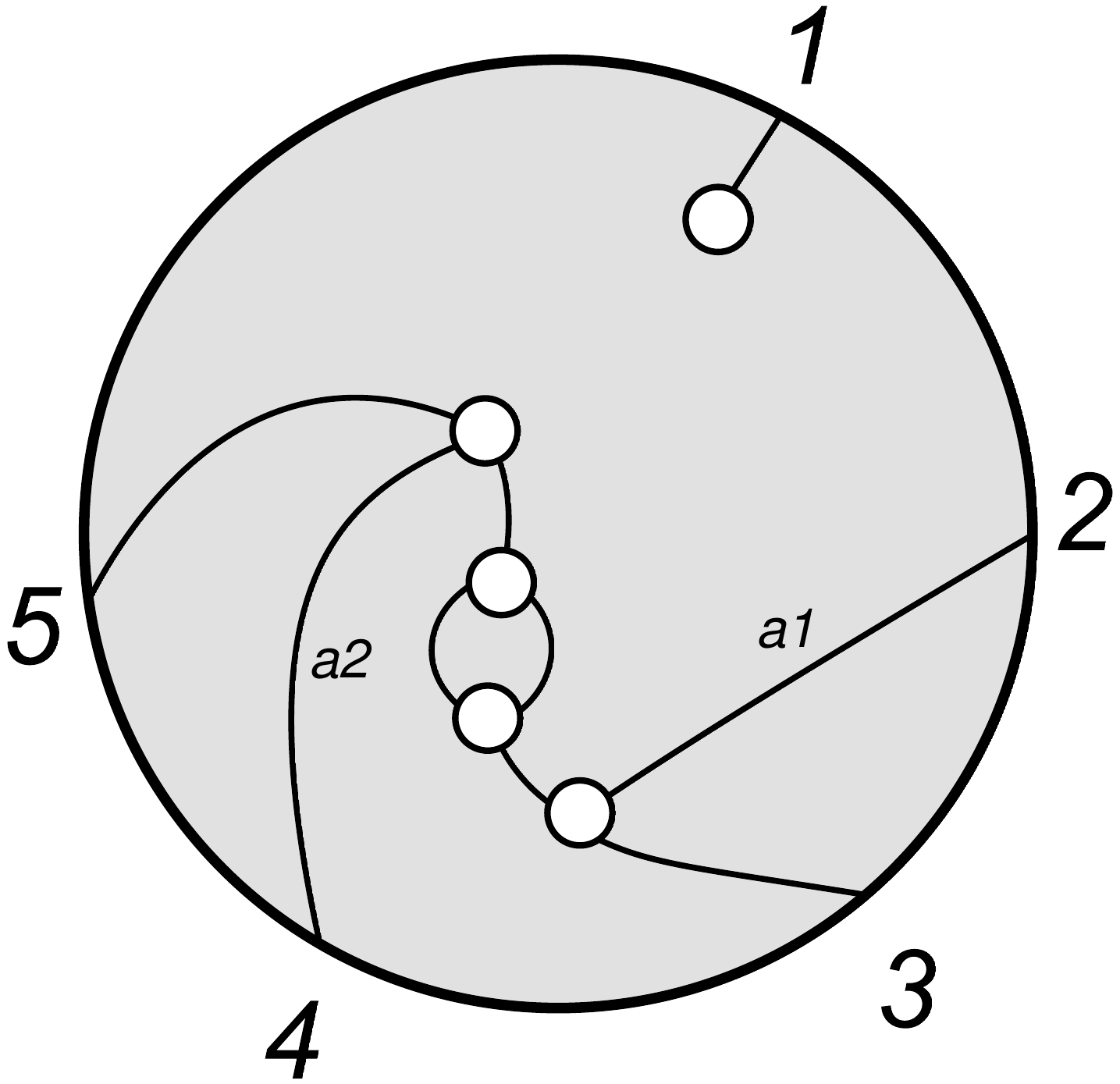}\;\;\;\;\;\includegraphics[width=3.5cm]{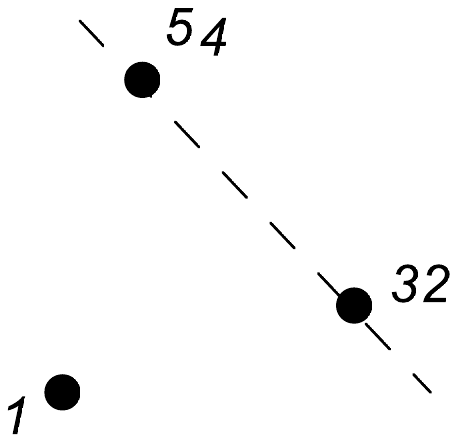}}}
$$

\noindent
(3) Apply the shift $(2 \to 5)$.   This moves $P_2$ towards $P_5$.  To keep the arrangement convex, this is visualized with $P_2$ being pushed away from $P_5$. This is an example of a non-adjacent shift with a negative bridge variable $P_2\rightarrow P_2-a_2 P_5$. So we have the space of degenerate pentagons $(P_1,P_2,P_3,P_4,P_5)$ in $\P^2$, where $P_2$, $P_3$, and $P_4 = P_5$ lie at infinity and $P_1$ lies in $\A^2$.

$$
P=\begin{pmatrix}0&a_1&1& 0 & 0\\
0&-a_3&0&a_2&1 \\
1& 0&0&0&0\end{pmatrix}
\;\;\;\;\;
\vcenter{\hbox{\includegraphics[width=3.5cm]{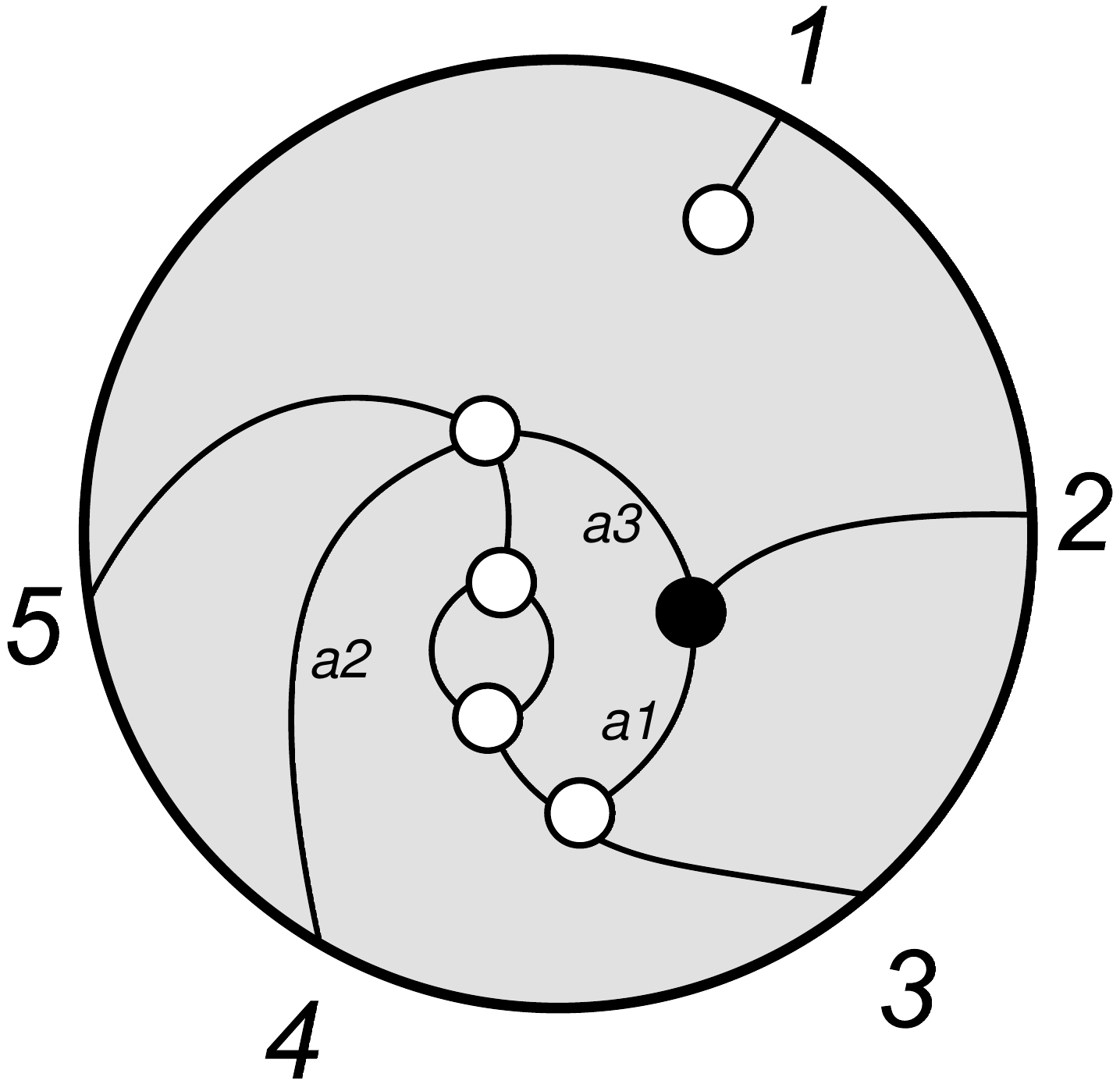}\;\;\;\;\;\includegraphics[width=3.5cm]{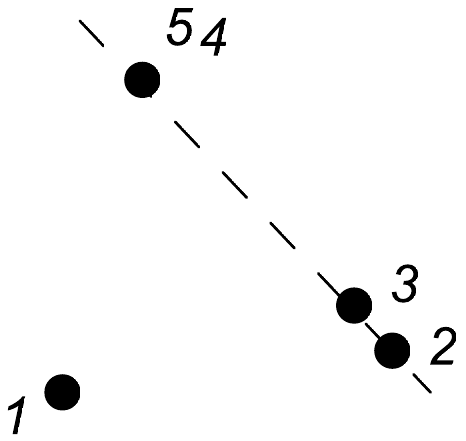}}}
$$

\noindent
(4) Apply the shift $(5 \to 1)$.   Thus $P_5$ is moved off the line at infinity towards $P_1$.  So we have the space of degenerate pentagons $(P_1,P_2,P_3,P_4,P_5)$ in $\P^2$, where $P_2$, $P_3$, and $P_4$ lie at infinity and $P_5$ lies on the line joining $P_1$ and $P_4$.

$$
P=\begin{pmatrix}0&a_1&1& 0 & 0\\
0&-a_3&0&a_2&1 \\
1& 0&0&0&a_4\end{pmatrix}
\;\;\;\;\;
\vcenter{\hbox{\includegraphics[width=3.5cm]{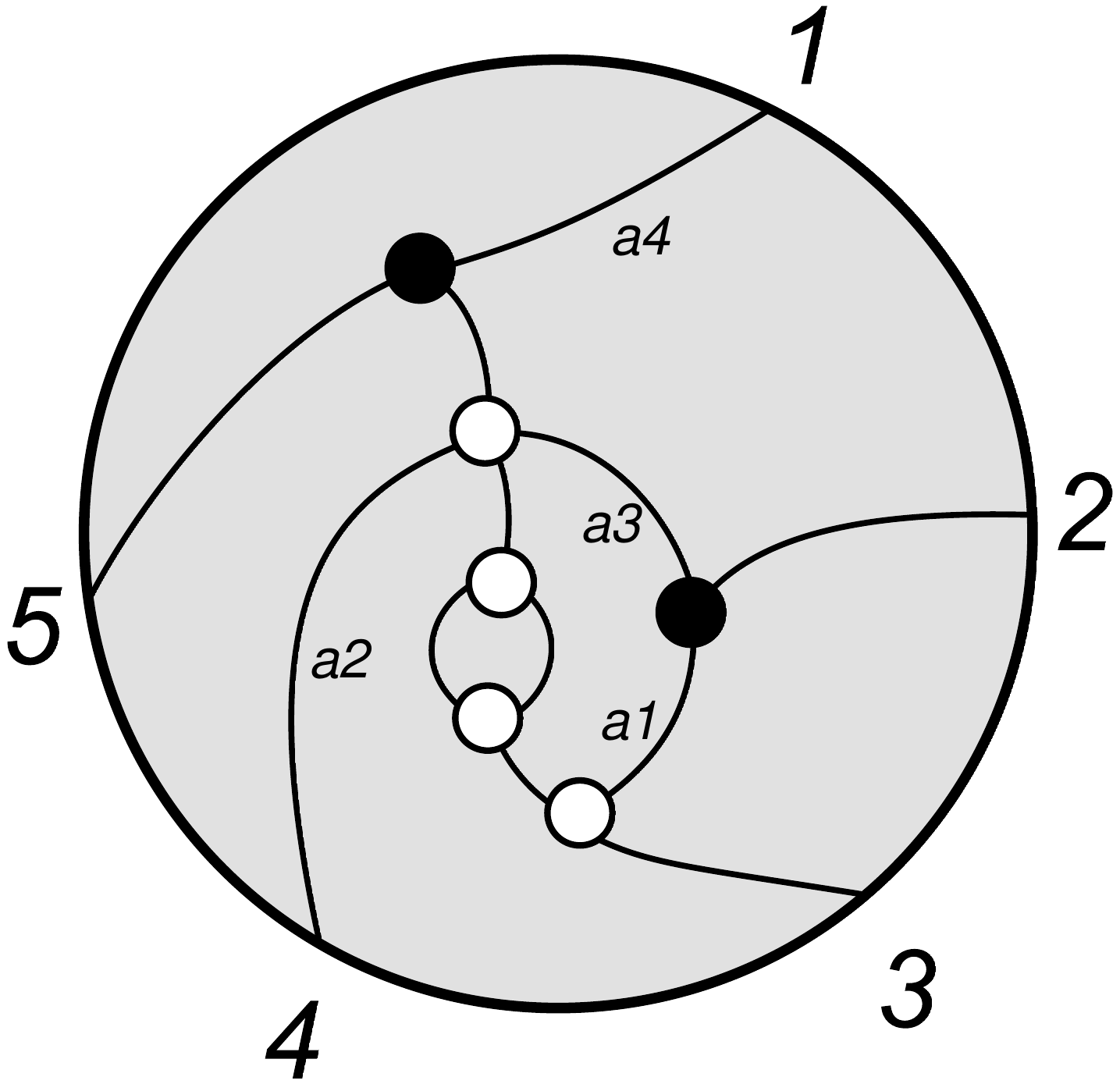}\;\;\;\;\;\includegraphics[width=3.5cm]{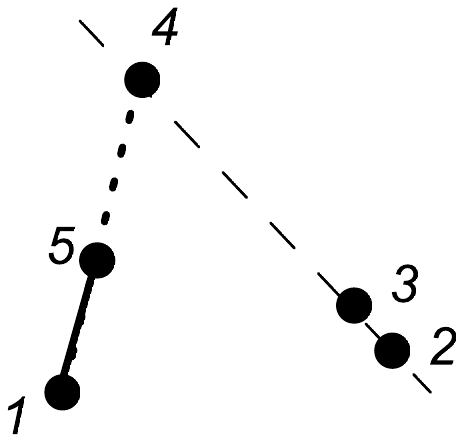}}}
$$

\noindent
(5) Apply the shift $(4 \to 5)$.  Thus $P_4$ is moved towards $P_5$.  So we have the space of degenerate pentagons $(P_1,P_2,P_3,P_4,P_5)$ in $\P^2$, where $P_2$ and $P_3$ lie at infinity and $P_5$ lies on the line joining $P_1$ and $P_4$.

$$
P=\begin{pmatrix}0&a_1&1& 0 & 0\\
0&-a_3&0&a_2+a_5&1 \\
1& 0&0&a_4a_5&a_4\end{pmatrix}
\;\;\;\;\;
\vcenter{\hbox{\includegraphics[width=3.5cm]{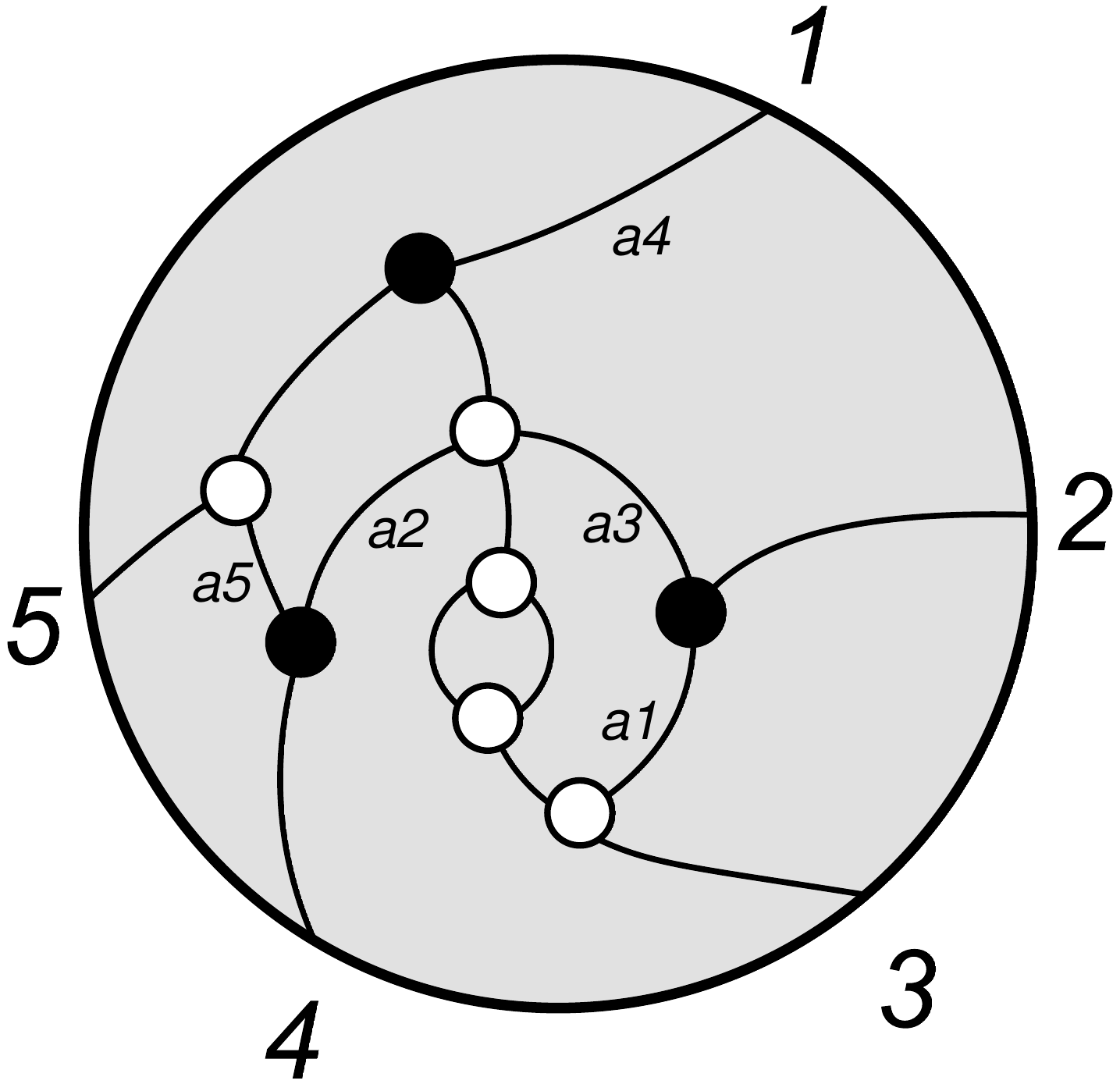}\;\;\;\;\;\includegraphics[width=3.5cm]{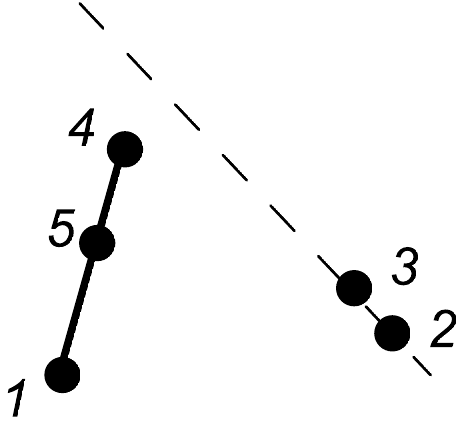}}}
$$

\noindent
(6) Apply the shift $(4 \to 3)$.  Thus $P_4$ is moved towards $P_3$.  So we now have the space of pentagons $(P_1,P_2,P_3,P_4,P_5)$ in $\P^2$, where $P_2$ and $P_3$ lie at infinity.
$$
P=\begin{pmatrix}0&a_1&1&a_6 & 0\\
0&-a_3&0&a_2+a_5&1 \\
1& 0&0&a_4a_5&a_4\end{pmatrix}
\;\;\;\;\;
\vcenter{\hbox{\includegraphics[width=3.5cm]{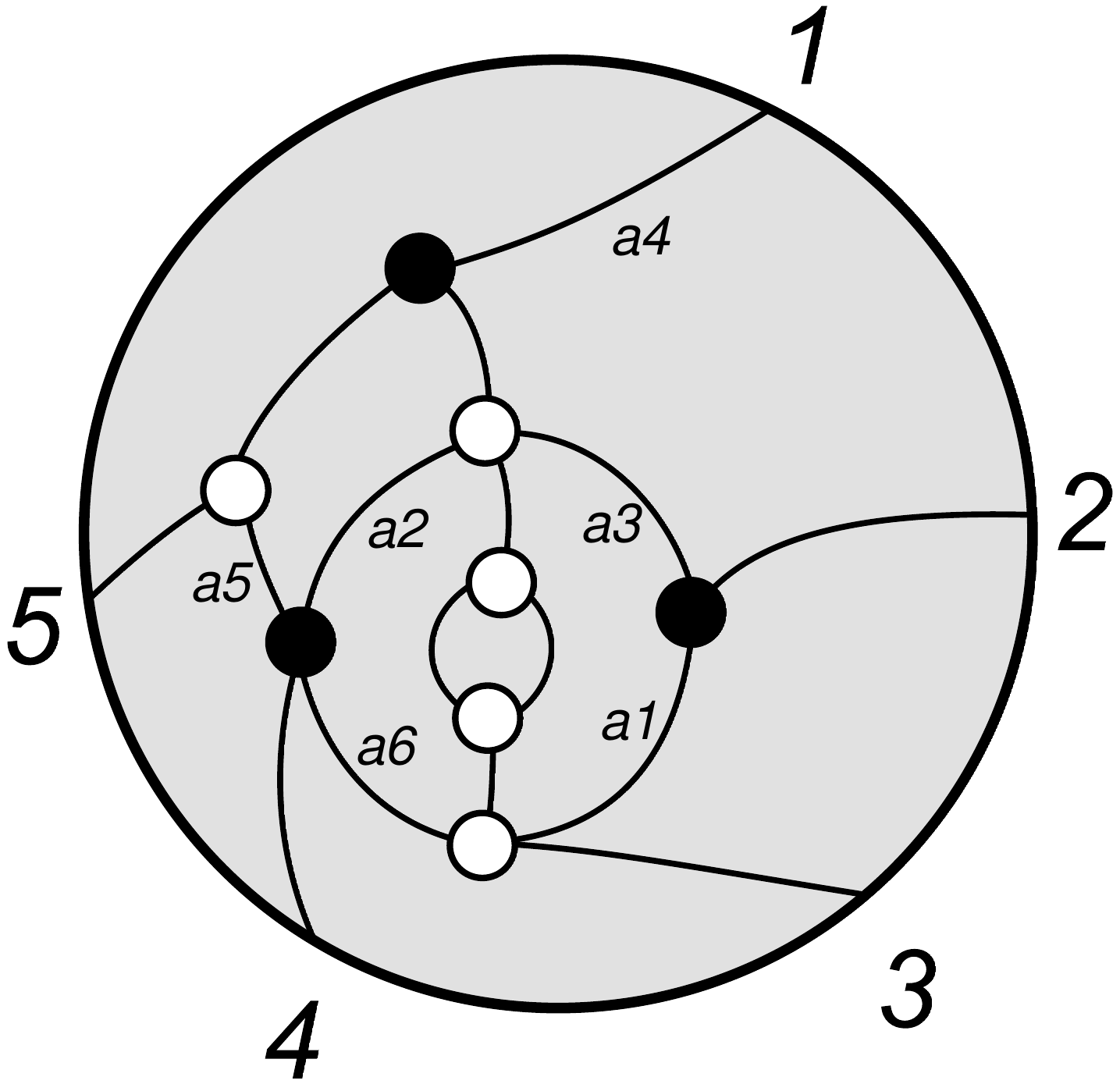}\;\;\;\;\;\includegraphics[width=3.5cm]{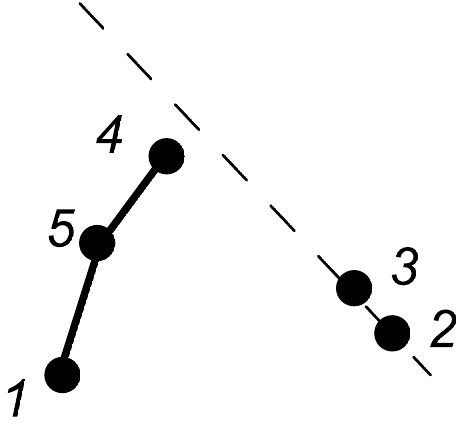}}}
$$

\noindent
(7) Apply the shift $(2 \to 1)$.  Thus $P_2$ is moved towards $P_1$.  So we now have the space of pentagons $(P_1,P_2,P_3,P_4,P_5)$ in $\P^2$, where $P_3$ is on the line at infinity. In the diagram below, you should imagine that $P_3$ is placed infinitely away so that the two dashed lines are parallel, and so that the interior angles at $P_4$ and $P_2$ are less than $\pi$. In particular, given any point $P_*$ on the half-line $(2,3)$, the extended lines $(2,3)$ and $(4,5)$ intersect on the side of $(4,*)$ containing $P_2$ and $P_5$. This observation is crucial for the next step.

$$
P=\begin{pmatrix}0&a_1&1&a_6 & 0\\
0&-a_3&0&a_2+a_5&1 \\
1& a_7&0&a_4a_5&a_4\end{pmatrix}
\;\;\;\;\;
\vcenter{\hbox{\includegraphics[width=3.5cm]{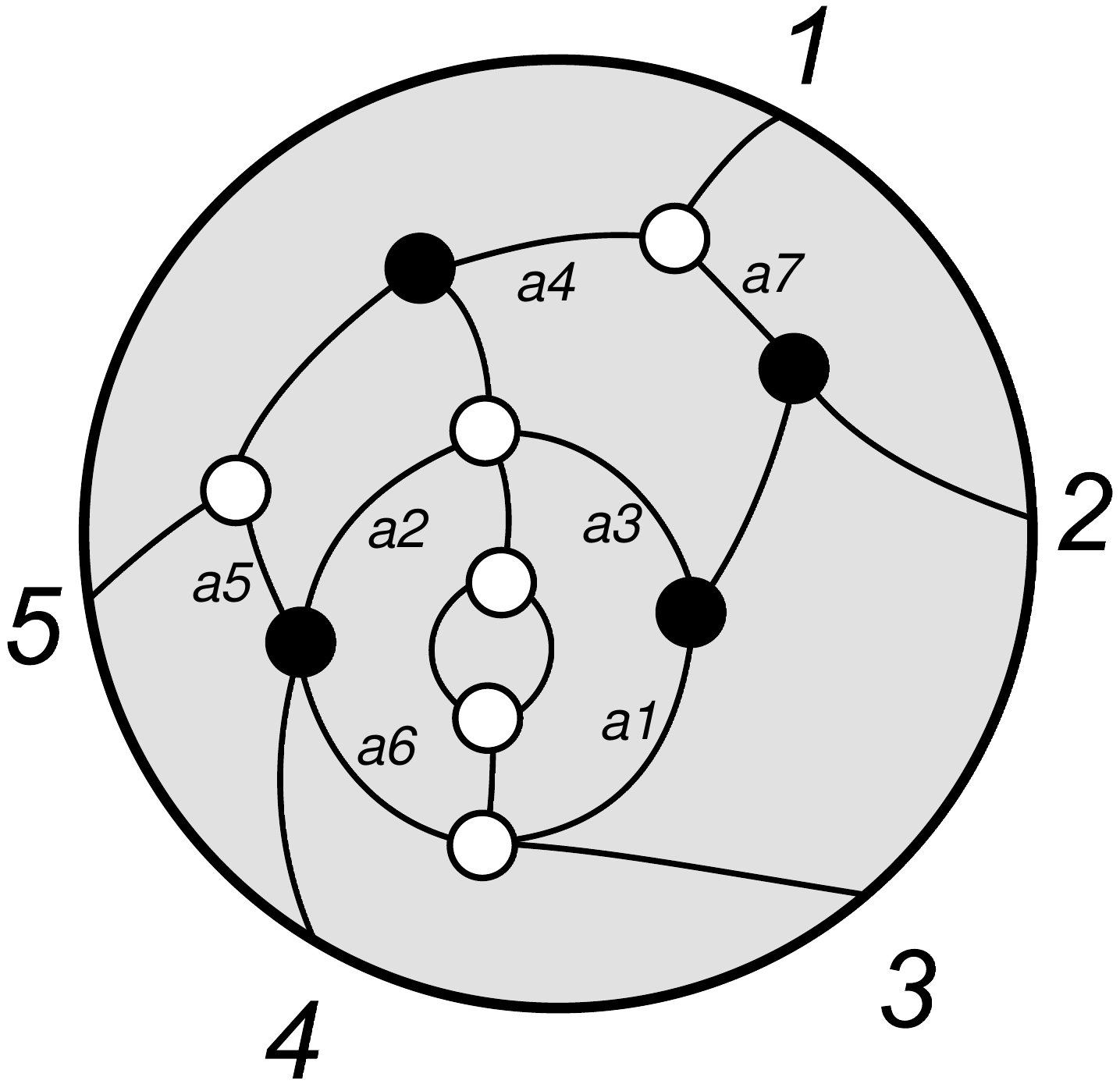}\;\;\;\;\;\includegraphics[width=3.5cm]{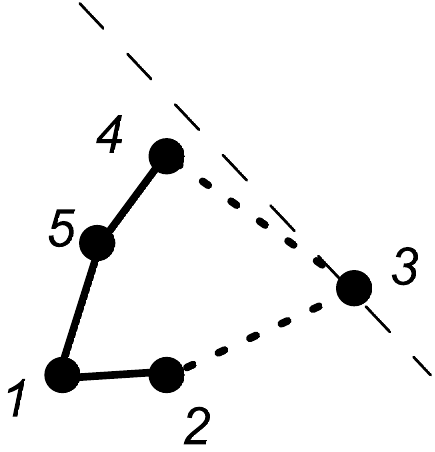}}}
$$

\noindent 
(8) Apply the shift $(3 \to 2)$.  This moves $P_3$ off the line at infinity towards $P_2$ along $(2,3)$.  As noted earlier, the lines $(2,3)$ and $(4,5)$ intersect on the side of $(3,4)$ containing $P_2$ and $P_5$.  In other words, $[4,5;2,3]$ is positive.  So we now have the space of pentagons $(P_1,P_2,P_3,P_4,P_5)$ in $\A^2$, where $(4,5)$ and $(2,3)$ intersect positively.  We have thus obtained the top cell $\Pi_S$ where $S = \{2,3,4\}$.  (The conditions that $(2,3)$ intersects $(3,4)$ positively, and $(3,4)$ intersects $(4,5)$ positively are trivially satisfied.)

$$
P=\begin{pmatrix}0&a_1&1+a_1a_8&a_6 & 0\\
0&-a_3&-a_3a_8&a_2+a_5&1 \\
1& a_7&a_7a_8&a_4a_5&a_4\end{pmatrix}
\;\;\;\;\;
\vcenter{\hbox{\includegraphics[width=3.5cm]{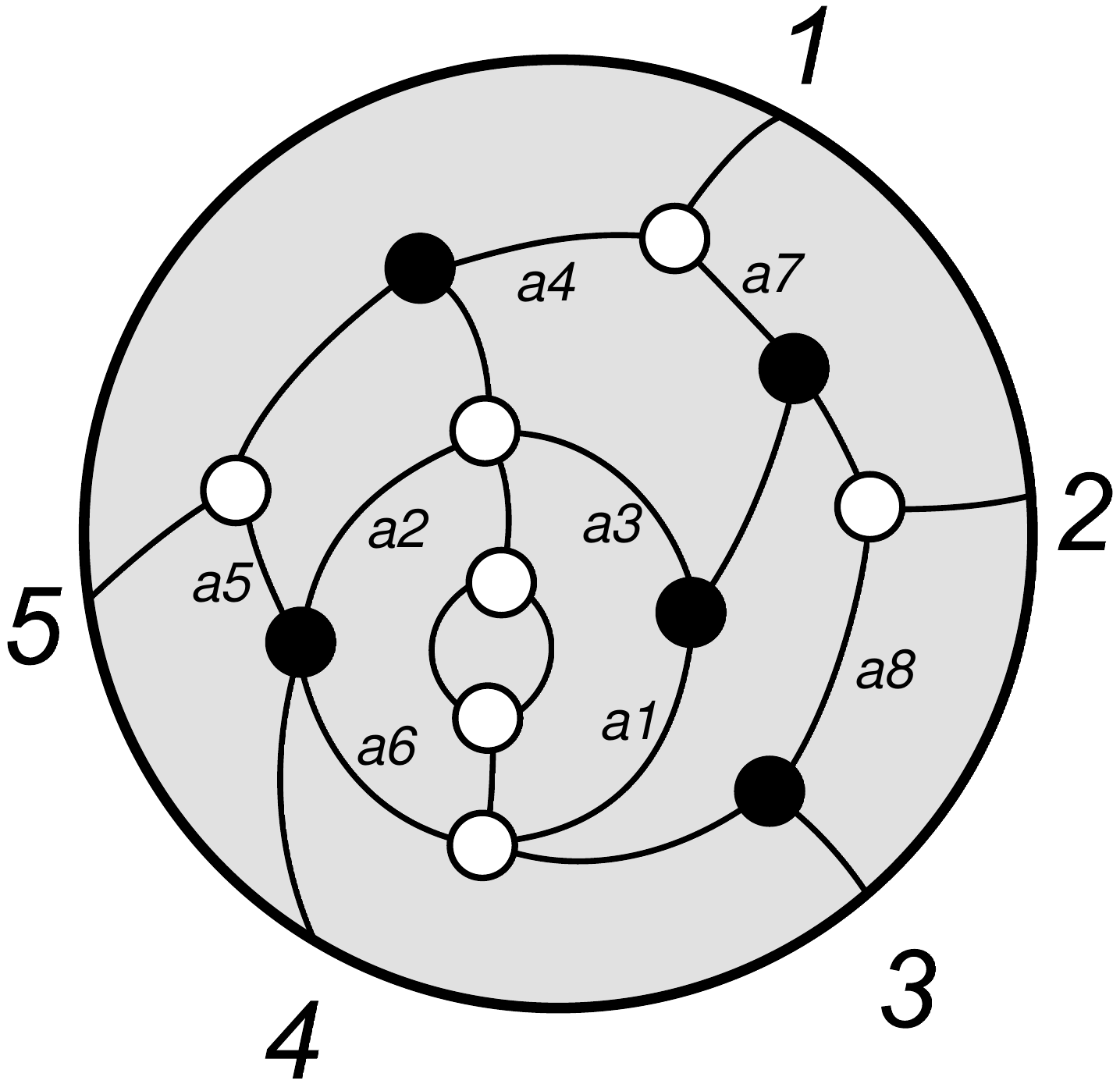}\;\;\;\;\;\includegraphics[width=3.5cm]{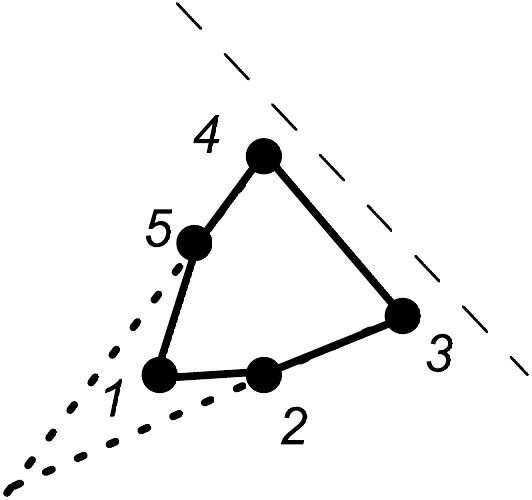}}}
$$

The above diagram is identical to the B-1 diagram in the Appendix with the black lollipop at particle 6 stripped off, since we are only concerned with the 5 point integrand. However, the above matrix $P$ is not obviously the same as the B-1 matrix (with the sixth column removed) in the Appendix.  The two are related as follows.  First change the gauge of the above $P$ by left multiplying by
$$
g = \begin{pmatrix}
 1 & 0 & 0 \\
0 & a_4 &0 \\
0 &0 & 1
\end{pmatrix} \in \GL(1;1).
$$
then perform the monomial coordinate change $a_1 \to c_1 c_6, a_2 \to c_8/c_4, a_3 \to c_1 c_7/c_4, a_4 \to c_4, a_5 \to c_3/c_4, 
 a_6 \to c_5, a_7 \to c_1, a_8 \to c_2/c_1$. Note that this change of variables is a diffeomorphism of $\mathbb{R}_{>0}^8$, which must be the case for any change of gauge between positive variables.

All the $\binom{n}{3}$ top cells $\Pi_S$ can be constructed in this way.  They are $(3n-7)$-dimensional spaces isomorphic to $\R_{>0}^{3n-7}$.  Each top cell $\Pi_S$ has an associated canonical form $\M_{\Pi_S} = \prod_i \frac{d\alpha_i}{\alpha_i}$, where $\alpha_i$ are the coordinates used in the shifts.

\subsection{A 6-point example}\label{ssec:ex2}
The 8-dimensional cell $\Pi_{\text{FL-2}}$ corresponding to the FL-2 term described in the Appendix can be constructed via shifts as follows.\\

\noindent
(0) Start with the 0-dimensional cell $\Pi_{4;1,6}$.  Thus we have a point $P_4 \in \A^2$ and distinct points $P_1,P_6 \in L_\infty$.  There is a unique such configuration up to affine transformations.  The points $P_2,P_3,P_5$ are not on the projective plane.  A representative matrix is 
$$
P=\begin{pmatrix}1&0&0& 0 & 0&0\\
0&0&0&0&0&-1 \\
0& 0&0&1&0&0\end{pmatrix}
\;\;\;\;\;
\vcenter{\hbox{\includegraphics[width=3.5cm]{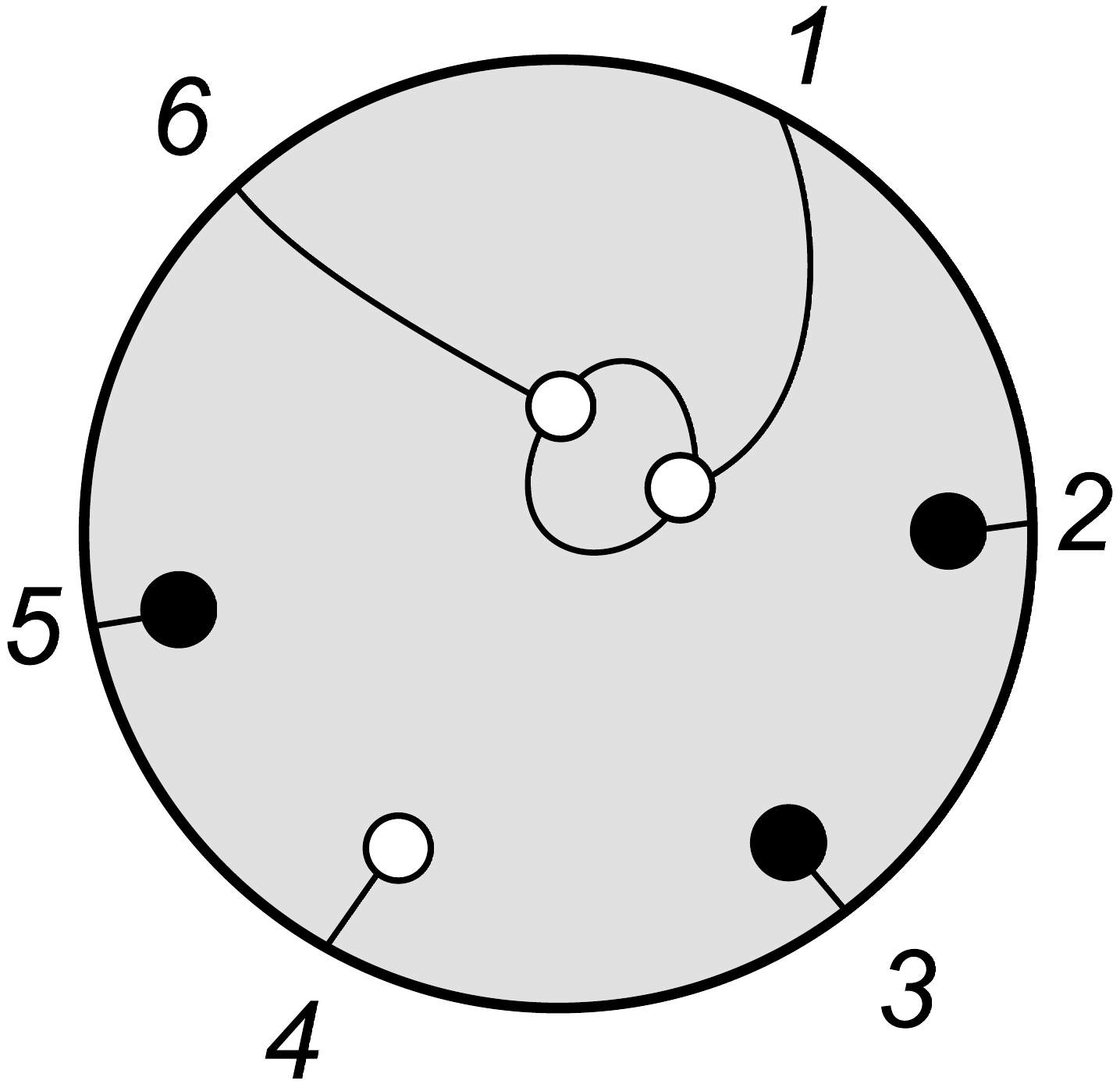}\;\;\;\;\;\includegraphics[width=3.5cm]{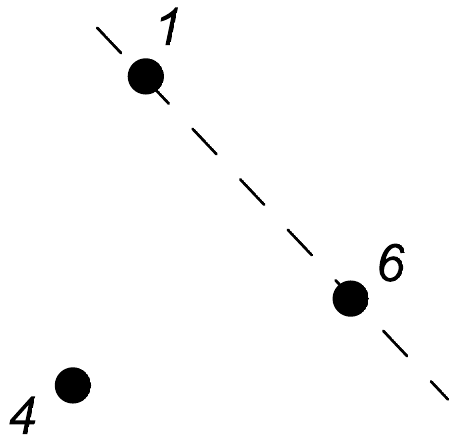}}}
$$

\noindent
(1) Perform the shift $(2 \to 1)$.  This places $P_2$ on top of $P_1$.  Thus we have the space of degenerate quadrilaterals $(P_1,P_2,P_4,P_6)$ where $P_1 = P_2$ and $P_1,P_2,P_6$ lie on $L_\infty$.  A representative matrix for this one-dimensional cell is 
$$
P=\begin{pmatrix}1&a_1&0& 0 & 0&0\\
0&0&0&0&0&-1 \\
0& 0&0&1&0&0\end{pmatrix}
\;\;\;\;\;
\vcenter{\hbox{\includegraphics[width=3.5cm]{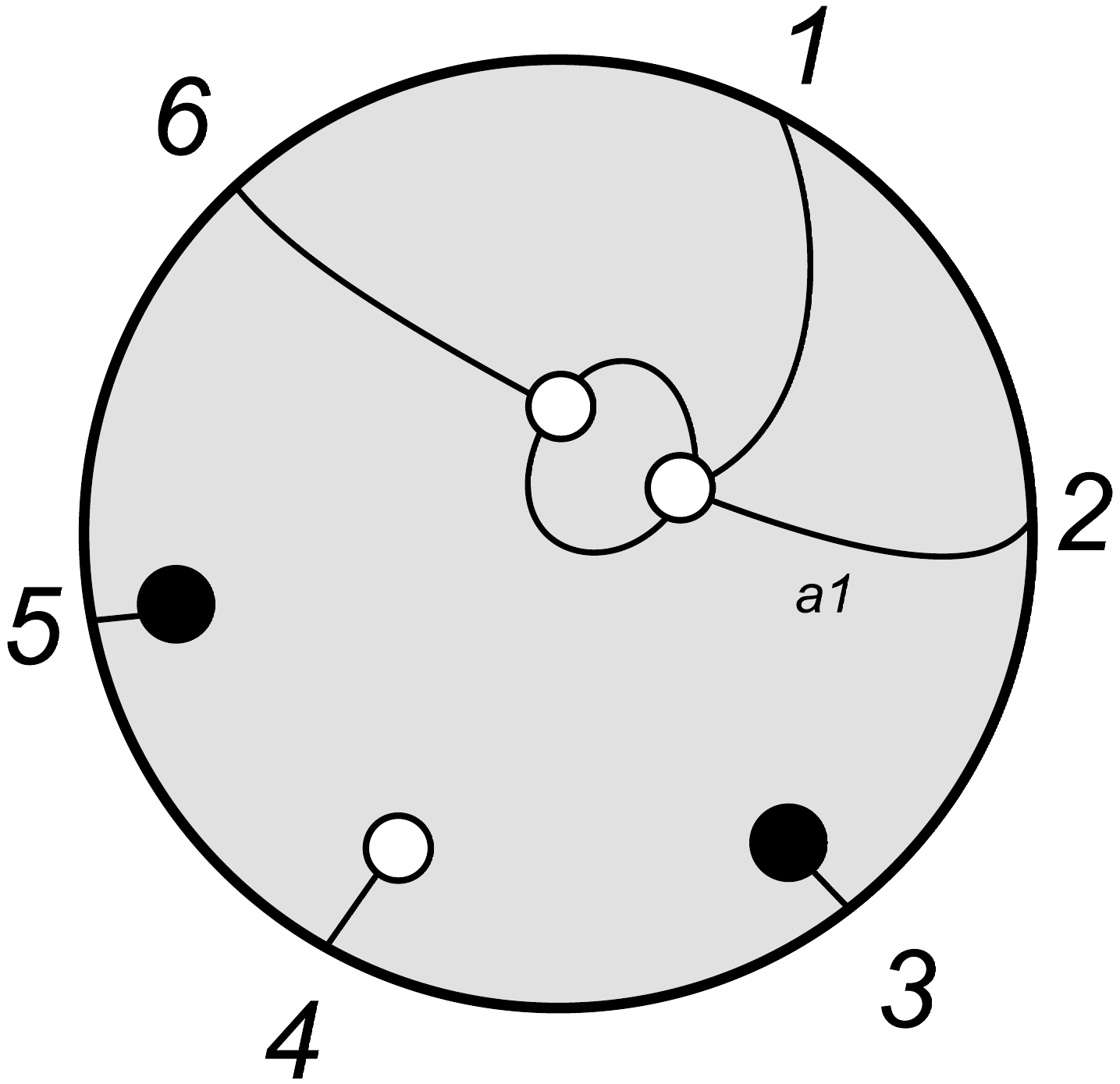}\;\;\;\;\;\includegraphics[width=3.5cm]{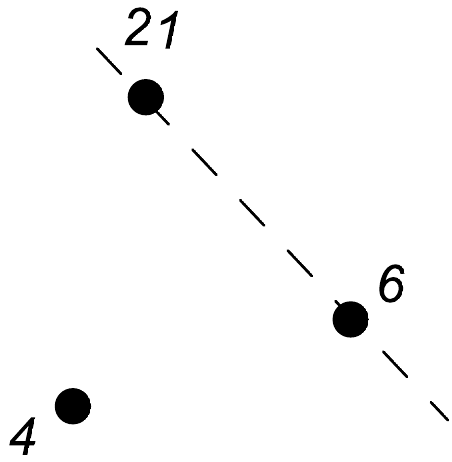}}}
$$

\noindent
(2) Perform the shift $(6 \to 4)$.  This moves $P_6$ towards $P_4$ along the line joining $P_4$ and $P_6$.  Now we have a space of degenerate quadrilaterals $(P_1,P_2,P_4,P_6)$, where two vertices $P_4,P_6$ lie in $\A^2$, and $P_1 = P_2$ lies on $L_\infty$.  A representative matrix for this two-dimensional cell is 
$$
P=\begin{pmatrix}1&a_1&0& 0 & 0&0\\
0&0&0&0&0&-1 \\
0& 0&0&1&0&a_2\end{pmatrix}
\;\;\;\;\;
\vcenter{\hbox{\includegraphics[width=3.5cm]{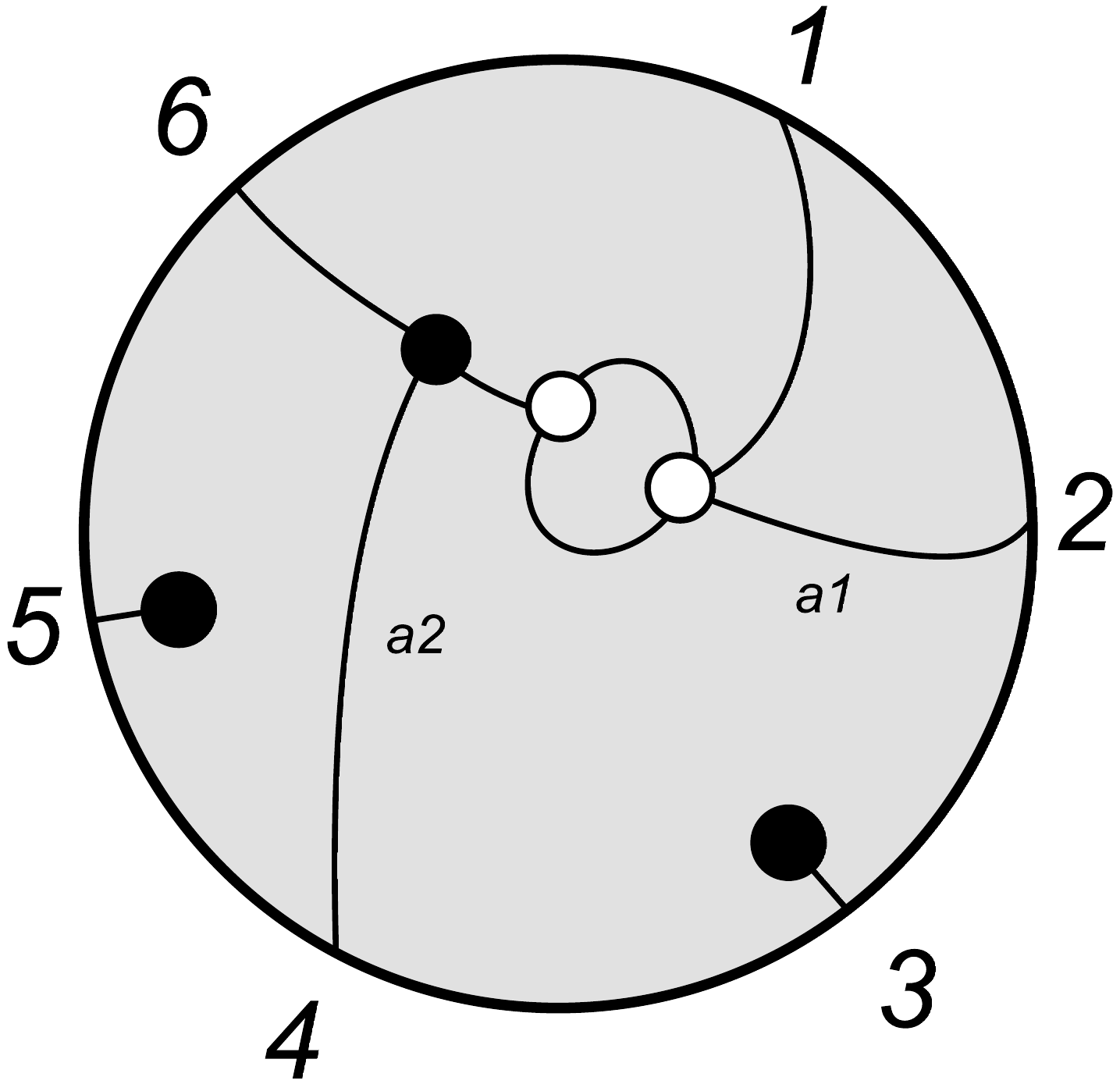}\;\;\;\;\;\includegraphics[width=3.5cm]{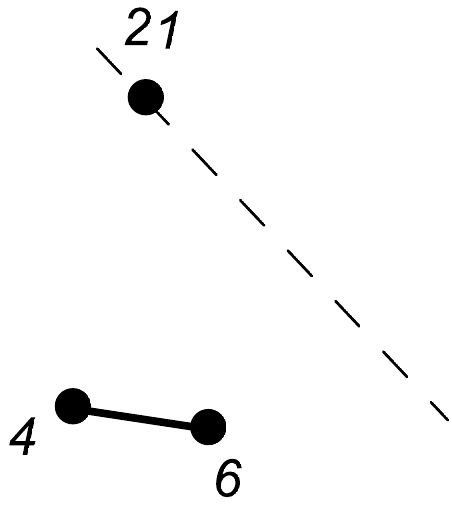}}}
$$

\noindent
(3) Perform the shift $(2 \to 4)$.  This moves $P_2$ towards $P_4$ along the line joining $P_2$ and $P_4$.  So now we have the space of degenerate quadrilaterals $(P_1,P_2,P_4,P_6)$, where $P_2$ lies on the line segment $(1,4)$, and $P_1$ lies on $L_\infty$.  A representative matrix for this three-dimensional cell is 
$$
P=\begin{pmatrix}1&a_1&0& 0 & 0&0\\
0&0&0&0&0&-1 \\
0&a_3&0&1&0&a_2\end{pmatrix}
\;\;\;\;\;
\vcenter{\hbox{\includegraphics[width=3.5cm]{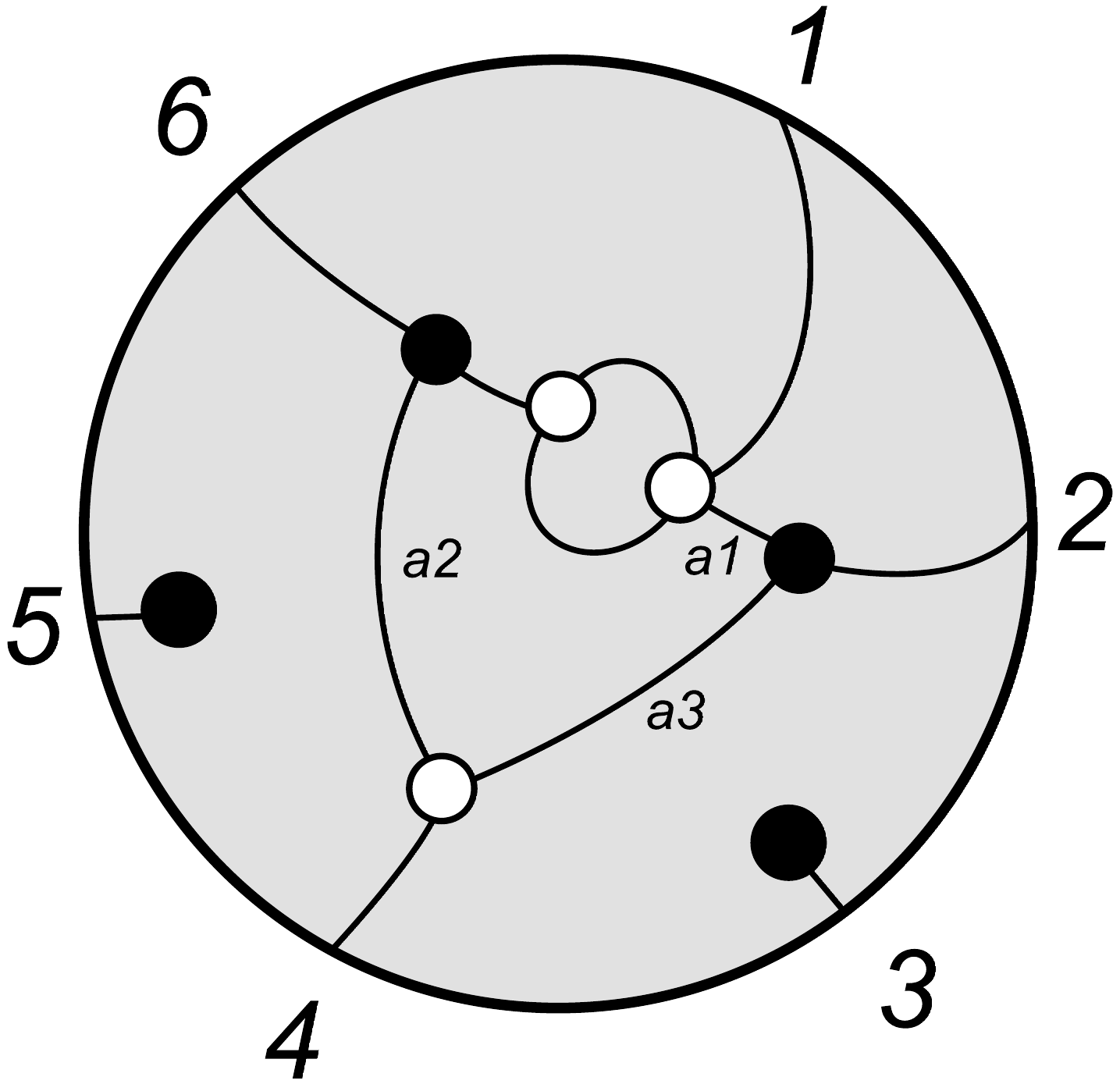}\;\;\;\;\;\includegraphics[width=3.5cm]{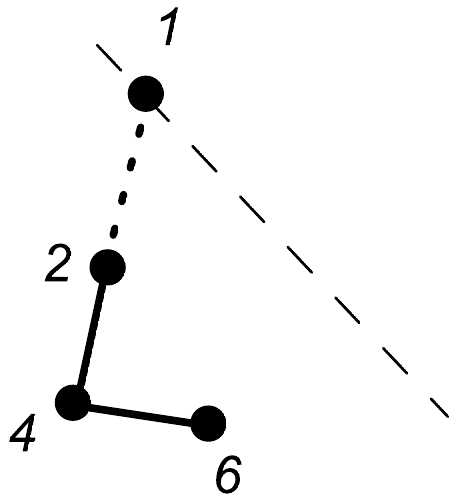}}}
$$

\noindent
(4) Perform the shift $(5 \to 6)$.  This places $P_5$ at the same location as $P_6$.  Now we have the space of degenerate pentagons $(P_1,P_2,P_4,P_5, P_6)$, where $P_2$ lies on the line segment $(1,4)$, and $P_1$ lies on $L_\infty$, and $P_5 = P_6$.  A representative matrix for this four-dimensional cell is 
$$
P=\begin{pmatrix}
1&a_1&0& 0 & 0&0\\
0&0&0&0&-a_4&-1 \\
0&a_3&0&1&a_4a_2&a_2
\end{pmatrix}
\;\;\;\;\;
\vcenter{\hbox{\includegraphics[width=3.5cm]{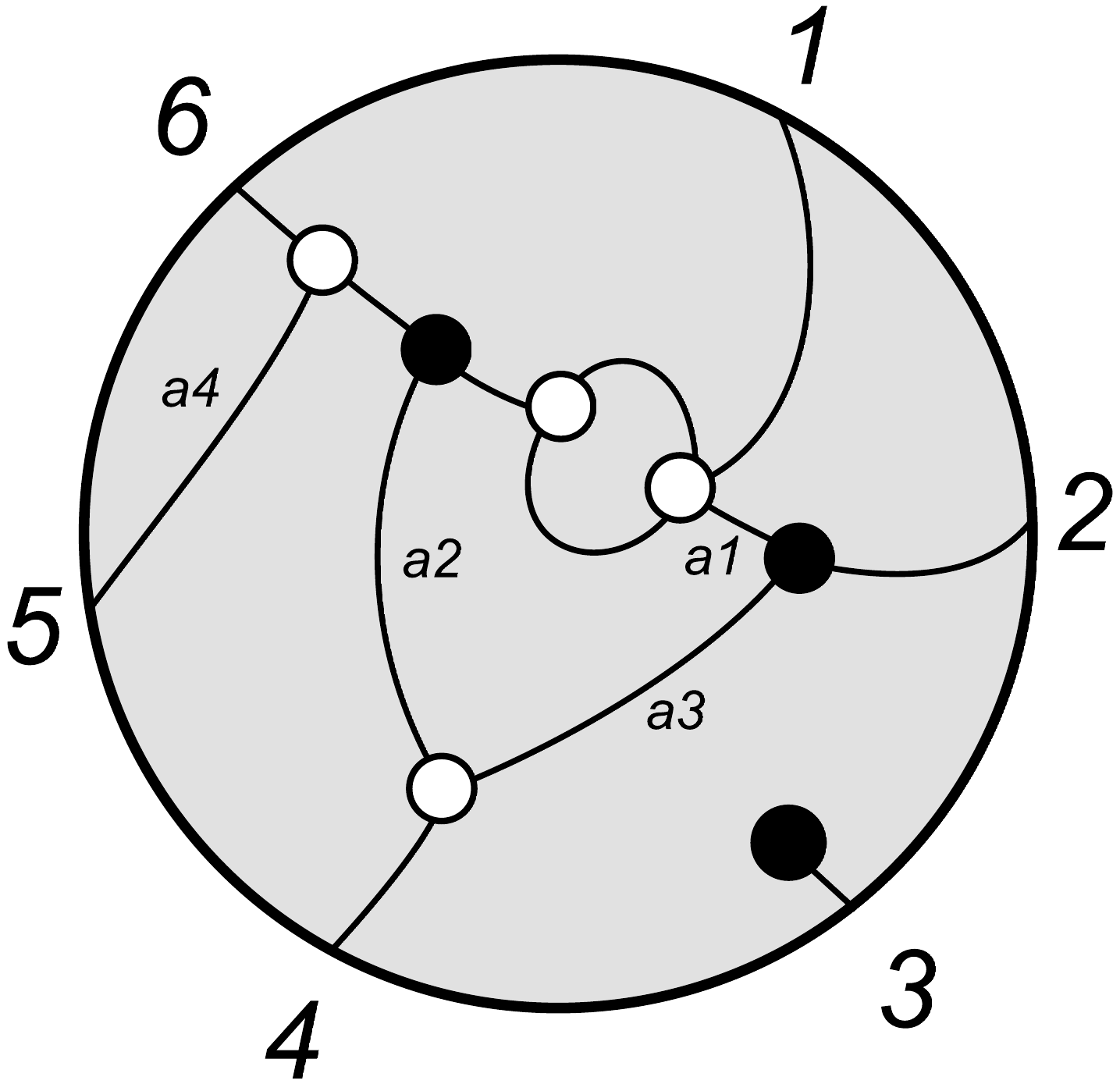}\;\;\;\;\;\includegraphics[width=3.5cm]{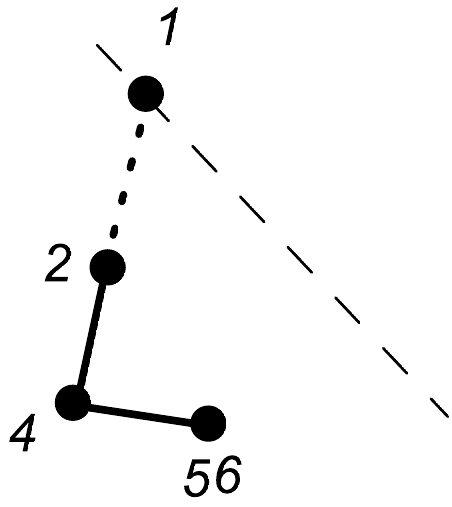}}}
$$

\noindent
(5) Perform the shift $(1 \to 6)$.  This moves $P_1$ towards $P_6$ along the line joining $P_1$ and $P_6$.  So now we have the space of degenerate pentagons $(P_1,P_2,P_4,P_5,P_6)$ where all points lie in $\A^2$, and in addition the edge $(1,6)$ is parallel to the edge $(2,4)$, and $P_5 = P_6$.  A representative matrix for this five-dimensional cell is 
$$
P=\begin{pmatrix}
1&a_1&0& 0 & 0&0\\
-a_5&0&0&0&-a_4&-1 \\
a_5 a_2&a_3&0&1&a_4a_2&a_2
\end{pmatrix}
\;\;\;\;\;
\vcenter{\hbox{\includegraphics[width=3.5cm]{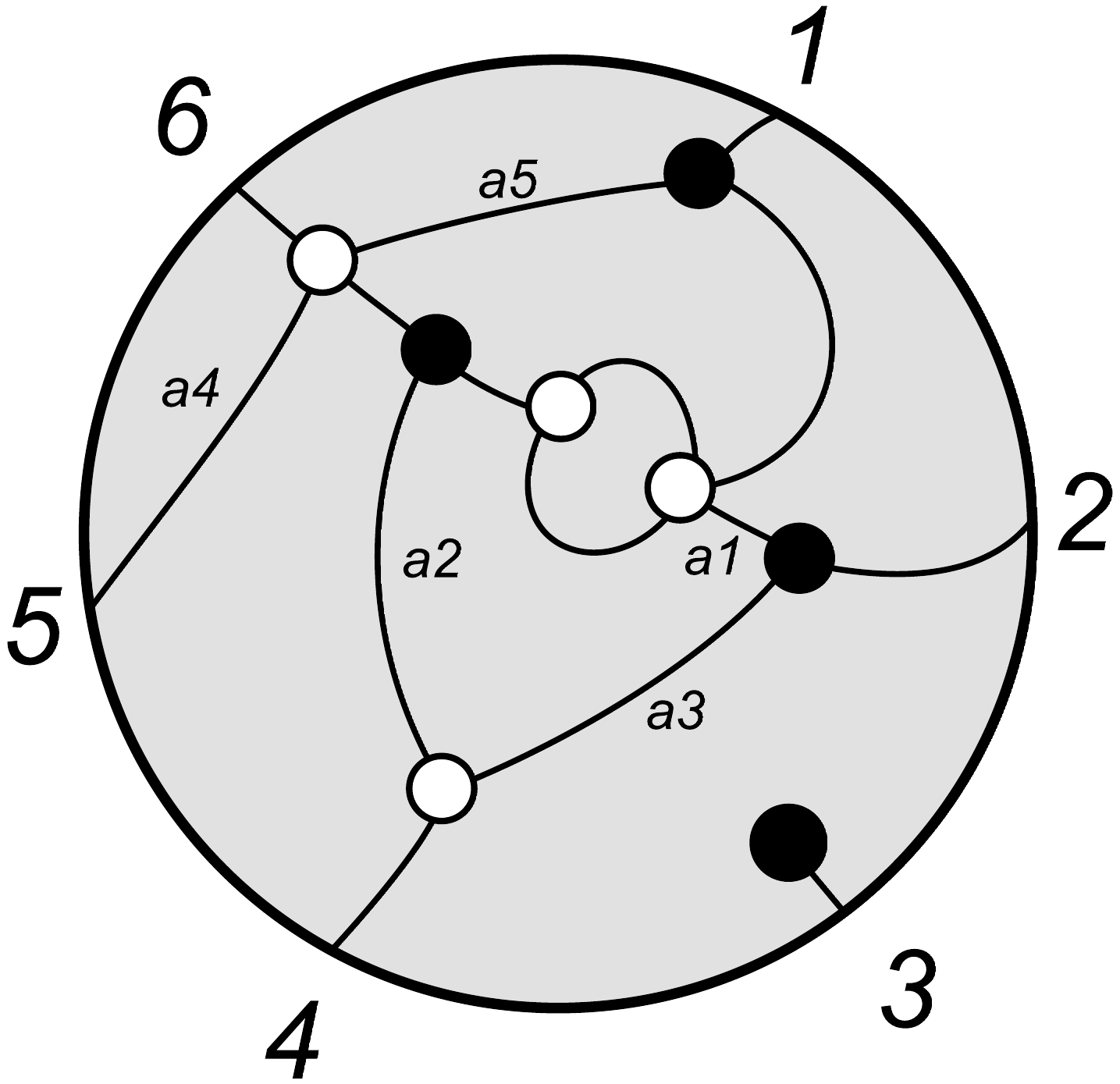}\;\;\;\;\;\includegraphics[width=3.5cm]{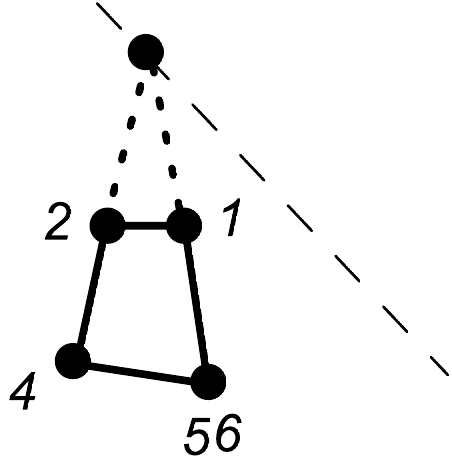}}}
$$

\noindent
(6) Perform the shift $(5 \to 4)$.  This moves $P_5$ towards $P_4$ along the line joining $P_5$ and $P_4$.  So now we have the space of degenerate pentagons $(P_1,P_2,P_4,P_5,P_6)$ where all points lie in $\A^2$, the edge $(1,6)$ is parallel to the edge $(2,4)$, and the point $P_5$ lies on the edge $P_4P_6$.  A representative matrix for this six-dimensional cell is 
$$
P=\begin{pmatrix}
1&a_1&0& 0 & 0&0\\
-a_5&0&0&0&-a_4&-1 \\
a_5a_2&a_3&0&1&a_4a_2+a_6&a_2
\end{pmatrix}
\;\;\;\;\;
\vcenter{\hbox{\includegraphics[width=3.5cm]{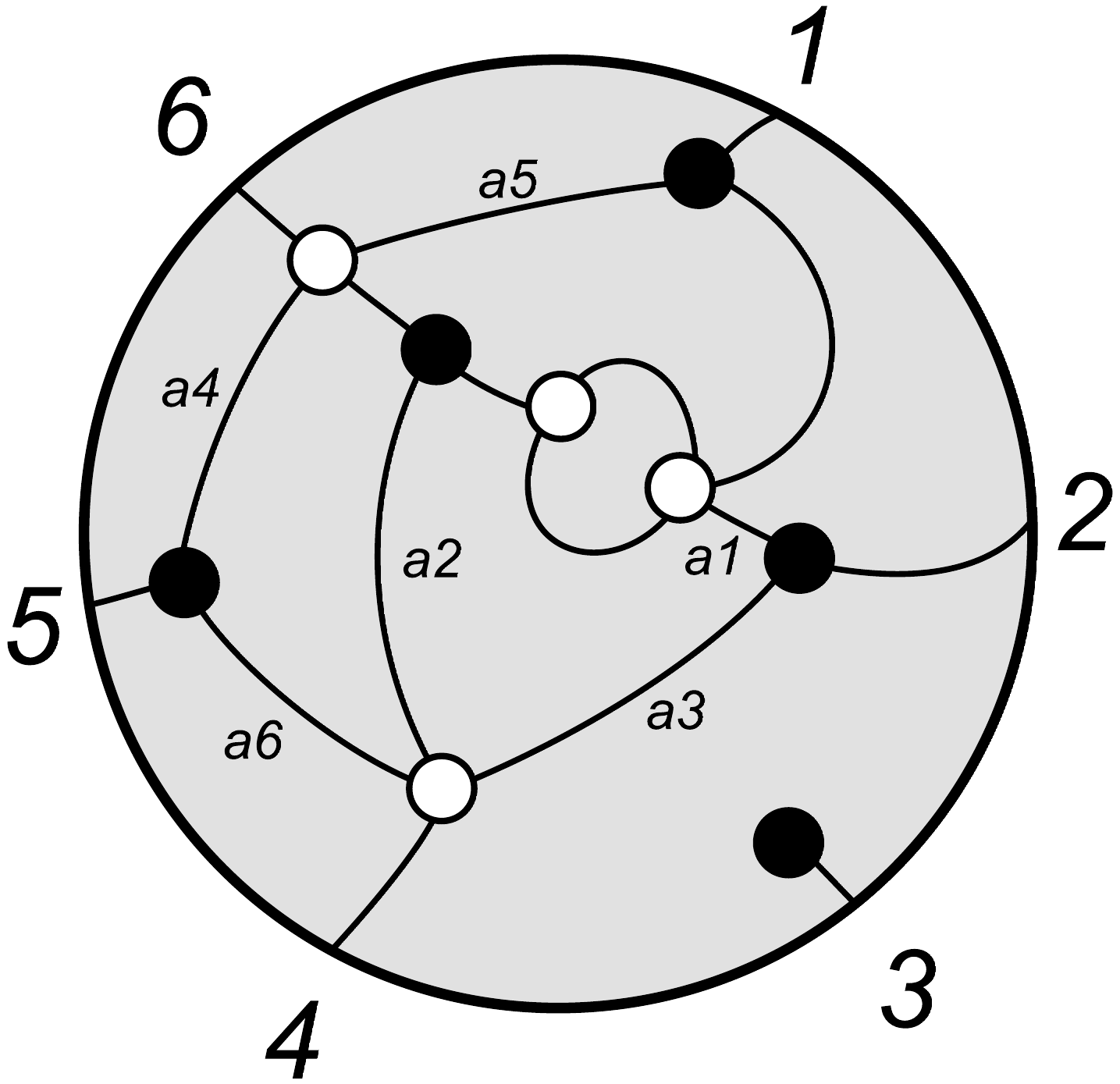}\;\;\;\;\;\includegraphics[width=3.5cm]{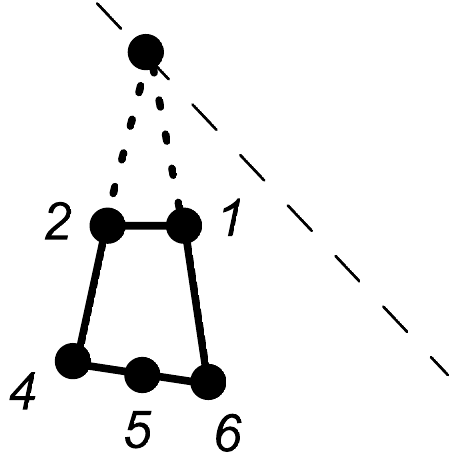}}}
$$

\noindent
(7) Perform the shift $(3 \to 4)$.  This places $P_3$ at the same location as $P_4$.  So now we have the space of degenerate convex hexagons $(P_1,P_2,P_3,P_4,P_5,P_6)$ where all points lie in $\A^2$, the edge $(1,6)$ is parallel to the edge $(2,4)$, the point $P_5$ lies on the edge $P_4P_6$, and $P_3 = P_4$.  A representative matrix for this seven-dimensional cell is 
$$
P=\begin{pmatrix}
1&a_1&0& 0 & 0&0\\
-a_5&0&0&0&-a_4&-1 \\
a_5a_2&a_3&a_7&1&a_4a_2+a_6&a_2
\end{pmatrix}
\;\;\;\;\;
\vcenter{\hbox{\includegraphics[width=3.5cm]{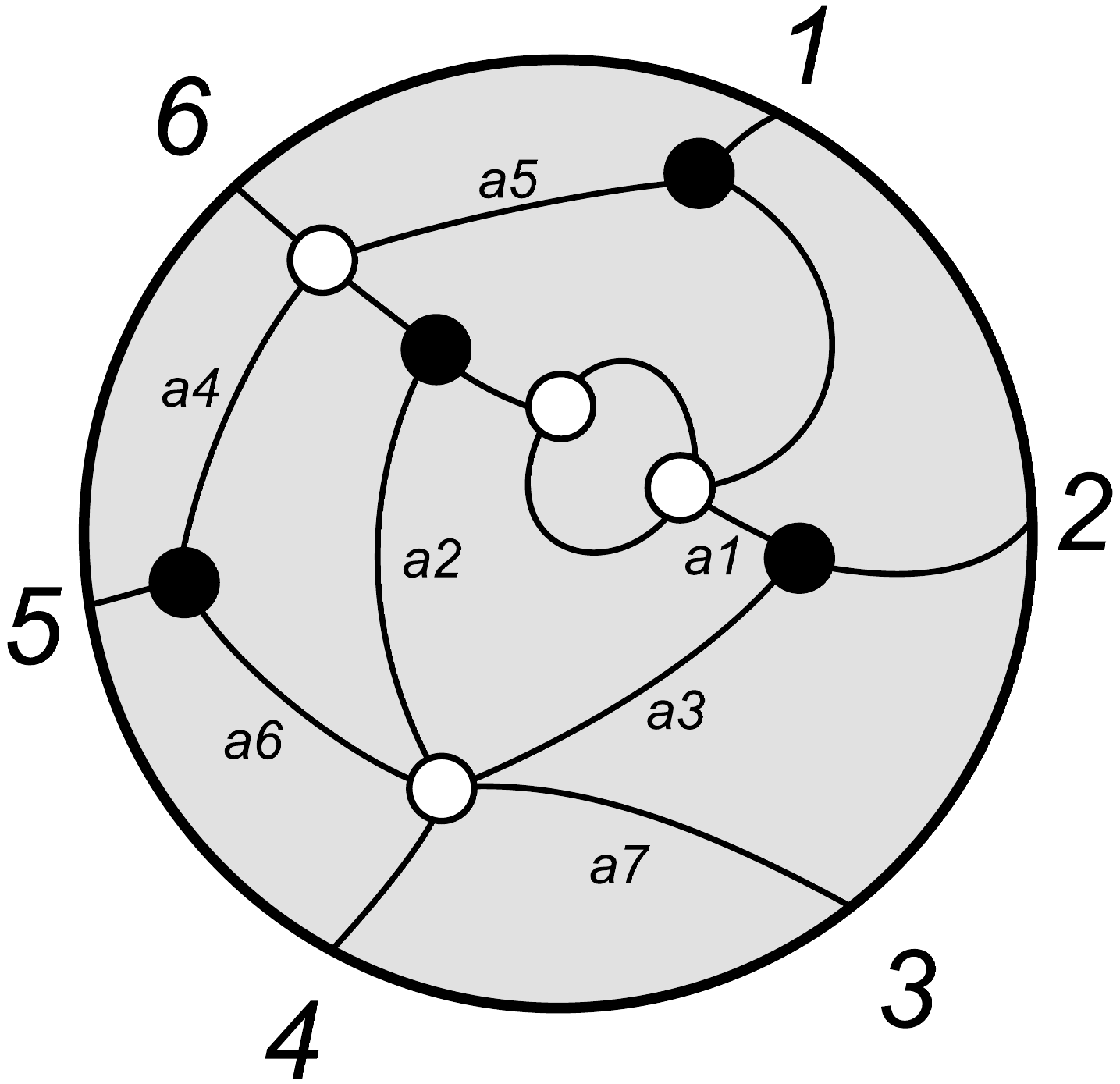}\;\;\;\;\;\includegraphics[width=3.5cm]{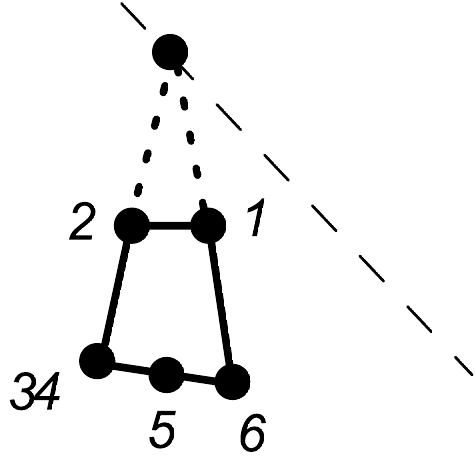}}}
$$

\noindent
(8) Perform the shift $(3 \to 2)$.  This moves $P_3$ towards $P_2$ along the line joining $P_2$ and $P_3$.  So now we have the space of degenerate convex hexagons $(P_1,P_2,P_3,P_4,P_5,P_6)$ where all points lie in $\A^2$, the edge $(1,6)$ is parallel to the edge $(2,4)$, the point $P_5$ lies on the edge $(4,6)$, and the point $P_3$ lies on the edge $(2,4)$.  A representative matrix for this eight-dimensional cell is 
$$
P=\begin{pmatrix}
1&a_1&a_1 a_8& 0 & 0&0\\
-a_5&0&0&0&-a_4&-1 \\
a_5a_2&a_3&a_7+a_3 a_8&1&a_4a_2+a_6&a_2
\end{pmatrix}
\;\;\;\;\;
\vcenter{\hbox{\includegraphics[width=3.5cm]{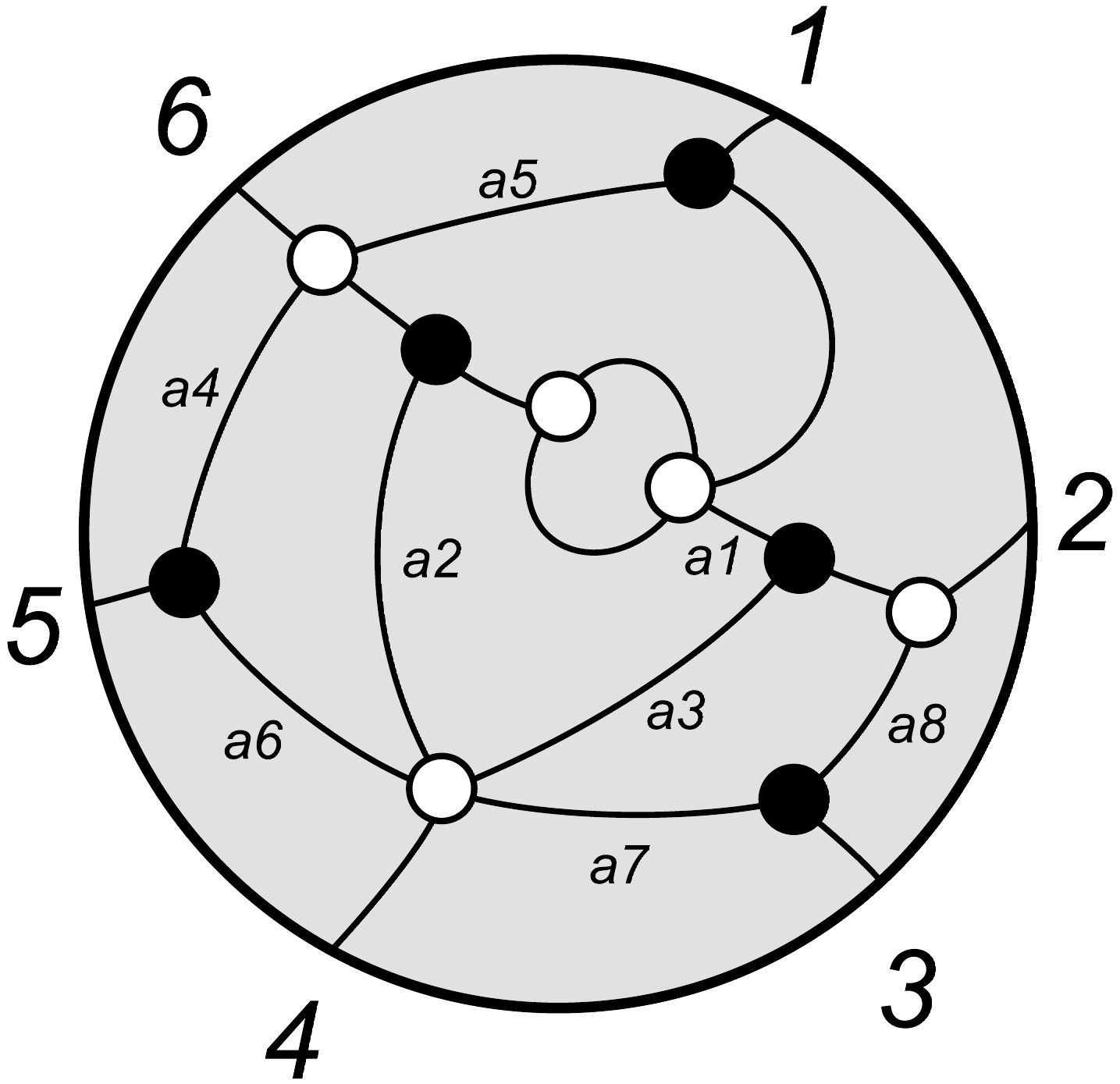}\;\;\;\;\;\includegraphics[width=3.5cm]{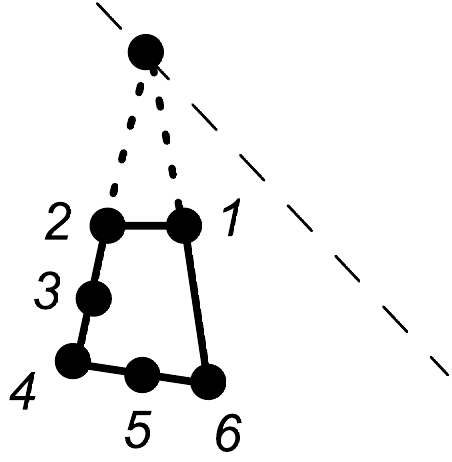}}}
$$
Note that vanishing of the minors $(234)$ and $(456)$ is immediate from the geometric description.  The vanishing of $[(16) \cap (24)]$ is equivalent to the statement that the edge $(1,6)$ is parallel to the edge $(2,4)$. Again, in the polygon diagram above, we should imagine that the intersection $(1,6)\cap (2,4)$ is placed infinitely away so that the dashed lines are parallel.

The above matrix $P$ is not obviously the same as the FL-2 matrix in the Appendix.  The two are related as follows.  First change the gauge of the above $P$ by left multiplying by
$$
g = \begin{pmatrix}
 0 & \frac{1}{a_5} & -\frac{1}{a_2^2 a_5^2} \\
 -1 & 0 & \frac{1}{a_2^2 a_5} \\
 0 & 0 & \frac{1}{a_2 a_5} \\
\end{pmatrix} \in \GL(1;1).
$$
then perform the monomial coordinate change $a_1 \to c_5, a_2\to c_8/c_3,  a_3 \to c_1 c_5/c_3, a_4 \to c_7/c_8, a_5 \to 1/c_8, a_6 \to c_4 c_7/c_3,a_7 \to c_2 c_6/c_3,a_8 \to c_6/c_5
$. Again, this is a diffeomorphism of $\mathbb{R}_{>0}^8$

\subsection{FL-2 as a boundary of a top cell}\label{ssec:ex2b}
Let $\Pi_{\text{FL-2}} \subset G_+(1,n;1)$ be the subset described as the the space of degenerate convex hexagons $(P_1,P_2,P_3,P_4,P_5,P_6)$ where all points lie in $\A^2$, the edge $(1,6)$ is parallel to the edge $(2,4)$, the point $P_5$ lies on the edge $(4,6)$, and the point $P_3$ lies on the edge $(2,4)$. In other words, this is the space of polygons swept out by the 8 positive variables $a_1,...,a_8$ appearing in the last step of our preceding construction.
%
%
%

We claim that $\Pi_{\text{FL-2}}$ is a 3-codimensional boundary of the eleven-dimensional top cell $\Pi_{\{2,4,6\}}$.  This top cell is the space of convex hexagons that can be inscribed into a big triangle, so that the edges $(2,3)$, $(4,5)$ and $(6,1)$ lie on the edges of the big triangle, as illustrated below on the left.   
\be
\vcenter{\hbox{\includegraphics[width=8cm]{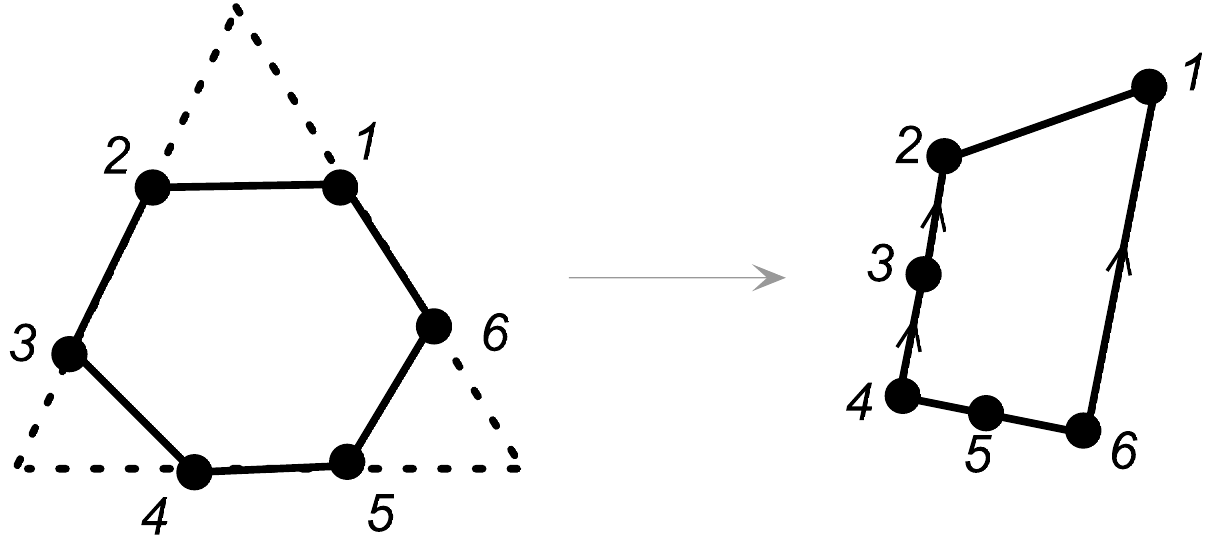}}}
\ee
To obtain $\Pi_{\text{FL-2}}$ as a boundary of $\Pi_{\{2,4,6\}}$, first move $P_4$ away from $P_5$ until it lies on the line $(2,3)$, then move $P_6$ away from $P_1$ until it lies on $(4,5)$.  Thus, $P_2,P_3,P_4$ are now collinear and $P_4,P_5,P_6$ are now collinear.  Finally, move the point $P_1$ {\it away} from $P_2$ until $(1,6)$ is parallel to $(2,3)$.  The three steps are equivalent to taking $(234),(456),[(16)\cap(24)]$ to zero. The resulting space of degenerate hexagons, as shown on the upper right diagram where the directed lines are parallel, is the cell $\Pi_{\text{FL-2}}$.

We should note that $\Pi_{\{2,4,6\}}$ is not the only top cell from which $\Pi_\text{FL-2}$ can be obtained. Another possibility is $\Pi_{\{3,4,6\}}$. 

\subsection{Permutations for cells of $G_+(1,n;1)$}
The cells of $G_+(1,n;1)$ (and more generally, $G_+(k,n;1)$) can be indexed by ``affine permutations'' $f$, generalizing the indexing for the positive Grassmannian.  We briefly describe this indexing.

An affine permutation is a bijection $f : \ZZ \to \ZZ$ satisfying the periodicity condition $f(i+n) = f(i) + n$.  An affine permutation $f$ is completely determined by the window $[f(1),f(2),\ldots,f(n)]$.

The cells of $G_+(1,n;1)$ can be indexed by certain affine permutations $f$ satisfying the additional condition that $f(1) + f(2) + \cdots + f(n) = 1+ 2 + \cdots + n + 4n$.  The indexing is completely determined by the following properties:
\begin{enumerate}
\item
The 0-dimensional cell $\Pi_{a;b,c}$ is indexed by the affine permutation
$$
f(i) = \begin{cases} i+2n & \mbox{if $i = a$,} \\
i+n & \mbox{if $i = b$ or $i = c$,} \\
i & \mbox{otherwise.} 
\end{cases}
$$
For example the 0-dimensional cell $\Pi_{3;1,4}$ is indexed by $f = [6,2,13,9,5]$.
\item
Suppose a cell $\Pi$ is labeled by $f$ satisfying $f(i) > f(i+1)$.  Then the shift $i+1 \to i$ produces a cell $\Pi'$ labeled by $g$, where $g$ is obtained from $f$ by swapping $f(i)$ and $f(i+1)$.
\item
Suppose a cell $\Pi$ is labeled by $f$ satisfying $f^{-1}(i) > f^{-1}(i+1)$.  Then the shift $i \to i+1$ produces a cell $\Pi'$ labeled by $g$, where $g$ is obtained from $f$ by swapping the values $i$ and $i+1$.
\end{enumerate}
This completely determines the permutations of all cells that can be obtained by these shifts, though it is not obvious that this recipe is well-defined.  In some cases, the rules (2) and (3) from above can also be applied to non-adjacent shifts, as we shall illustrate below.

For example, taking $n = 5$, the permutation of the top cell $\Pi_{\{2,3,4\}}$ (see Section \ref{ssec:ex1}) is calculated as follows:
\begin{align*}
&[11,2,8,4,10] \xrightarrow{2 \to 3} [11,3,7,4,10] \xrightarrow{4 \to 5} [11,3,7,5,9] \xrightarrow{2 \to 5} [11,4,7,5,8] \xrightarrow{5 \to 6}\\ & [10,4,7,6,8] \xrightarrow{4 \to 5} [9,5,7,6,8] \xrightarrow{4 \to 3} [9,5,6,7,8] \xrightarrow{2 \to 1} [5,9,6,7,8] \xrightarrow{3 \to 2} [5,6,9,7,8].
\end{align*}


The permutation $[8,4,7,6,11,9]$ of the cell $\Pi_{\text{FL-2}}$ (see Section \ref{ssec:ex2}) is calculated as follows:
\begin{align*}
&[7,2,3,16,5,12] \xrightarrow{2 \to 1} [2,7,3,16,5,12] \xrightarrow{6 \to 4} [2,7,3,12,5,16] \xrightarrow{2 \to 4}\\ & [4,7,3,12,5,14] \xrightarrow{5 \to 6}   [4,7,3,11,6,14] \xrightarrow{7 \to 6}[8,7,3,11,6,10] \xrightarrow{5 \to 4}\\ & [8,7,3,6,11,10] \xrightarrow{3 \to 4} [8,7,4,6,11,9]\xrightarrow{3 \to 2} [8,4,7,6,11,9].
\end{align*}
Here we started with the permutation $[7,2,3,16,5,12]$ indexing the cell $\Pi_{4;1,6}$.   The permutation $[8,4,7,6,11,9]$ can also be read off the momentum twistor diagram of FL-2 in the Appendix.  For example, starting at vertex 1, we reach vertex 2 (equal to 8 modulo 6) by turning left at white vertices and turning right at black vertices.  Similarly, starting at vertex 2, we reach vertex 4 by turning left at white vertices and turning right at black vertices.  This process is an extension of the method by which one obtains a permutation from an on-shell diagram.

 The top cell $\Pi_{\{2,4,6\}}$ is then calculated by further shifting $[8,4,7,6,11,9]$:
$$
[8,4,7,6,11,9] \xrightarrow{1 \to 2} [7,4,8,6,11,9] \xrightarrow{4 \to 5} [7,5,8,6,10,9] \xrightarrow{6 \to 7} [6,5,8,7,10,9].
$$

We will leave the detailed analysis of decorated permutations to future work. Here we only wanted to present the basic idea.


\section{Grassmannian measures at one loop}\label{general}
In this section, we discuss our main idea, the Grassmannian measure at one loop. It is well known that the tree level positive Grassmannian $G_+(k,n)$ has a cyclic measure $$\frac{d^{kn}C/{\text{vol}\;\GL(k)}}{\prod_{i=1}^n(i,i{+}1,i{+}2)}$$ whose physical application for amplitudes was first discussed in momentum space in \cite{ArkaniHamed:2009si}.  Here, ``cyclic'' denotes the product of the cyclically consecutive $k \times k$ minors of the $k\times n$ matrix $C$.  We have the identity 
$$
\frac{d^{kn}C/{\text{vol}\;\GL(k)}}{\prod_{i=1}^n(i,i{+}1,i{+}2)} = \prod_i \frac{da_i}{a_i},
$$
where $a_i$ are positive coordinates on $G_+(k,n)$ (see for example \cite[Proposition 13.3]{Lam:2015uma} and \cite{posGrassmannian}). Thus the cyclic measure can be thought of as the logarithmic form on the positive Grassmannian. Furthermore, it was discovered that BCFW terms and leading singularities of loop-level integrands are residues of this Grassmannian form. This is true for both the momentum space Grassmannian and the momentum-twistor space Grassmannian.

Going along this line of thought, we can ask whether the same can be said about loop level. It turns out the answer is yes, at least at one loop. However, there are new subtleties appearing at loop level. For instance, the measure is not unique. There are multiple measures, one for each top cell $\Pi_S$ of $G_+(k,n;1)$.  The residues of these measures give rise to BCFW terms at one loop. It is likely the case that some of these residues also give ``almost" leading singularities of higher loop integrands, where an ``almost" leading singularity is a $4(L-1)$ residue of a $L$ loop integrand.

We begin with a general discussion of the properties that the measure must satisfy before writing down the specific measure. We will mostly focus on the $k=1$ case where the measure has a beautiful interpretation as the volume of polygon, but will also try to make our discussion as general as possible.
 
\subsection{The general setup}
In this section, we construct a family of integration measures on the one-loop Grassmannian (i.e. the space of $P$ matrices) which we claim generate one-loop BCFW terms.  We shall consider measures of the form
$$
\M_{k,n}(P) = \frac{d^{(k+2)n}P/{\text{vol}\;\GL(k;1)}}{\prod_{i=1}^n\;(i\;i{+}1\;i{+}2\; \cdots \;i{+}k{+}1)}\;\mathcal{M}'_{k,n}(P)
$$
on $G_+(k,n;1)$.  The factor $\M'_{k,n}(P)$ will be called the geometric factor and $\M_{k,n}(P)$ the measure.  We propose that BCFW terms can be generated by taking residues of the following integral
\be
\mathcal{Y}_{k,n}= \int \frac{d^{(k+2)n}P}{\prod_{i=1}^n\;(i\;i{+}1\;i{+}2\; \cdots \;i{+}k{+}1)}\;\M'_{k,n}(P)\;
\delta^{(k+2)(k+4)}(\mathcal{Y}{-}P.Z)
\ee\\
We are using the following notation:
\be
(i_1\;i_2\;\cdots\;i_{k+2})&=&\epsilon^{A_1,\ldots,A_{k+2}}P_{A_1i_1}\cdots P_{A_{k+2}i_{k+2}}\\
\;[i_1\;i_2\; \cdots \;i_k] \;&=&\epsilon^{a_1,\ldots,a_k}C_{a_1i_1} \cdots C_{a_k i_k}
\ee

The repeated indices $A_1,\ldots,A_{k+2}=1,\ldots,k+2$ and $a_1,a_2,\ldots,a_k = 1,2,\ldots,k$ are implicitly summed over their respective range. The indices $i_1,i_2,\ldots,i_{k+2}\in\{1,2,\ldots,n\}$ label the columns of the $C$ and $P$ matrices. Thus $(i_1,\ldots,i_{k+2})$ denotes the determinant of the matrix formed by the columns $i_1,\ldots,i_{k+2}$ of the $P$ matrix, while $[i_1,\ldots,i_k]$ denotes the determinant of the matrix formed by the columns $i_1,\ldots,i_k$ of the $C$ matrix. In the definition of our measure, we have not yet used the square bracket. Nonetheless, both brackets are important because they are covariant under $\GL(k;1)$ transformations. Using the notation of $\eqref{gauge}$, the round bracket scales as $\det(G)$ while the square bracket scales as $\det(J)$.

The round brackets appearing in the denominator of the measure is the familiar cyclic factor. However, the most important part of the measure is the geometric factor $\M'_{k,n}(P)$. For $k=0$, the $C$ matrix disappears and we are only left with the $D$ matrix, which is nothing but the usual Grassmannian. Since we are already familiar with the cyclic measure on the Grassmannian, it is easy to see that $\mathcal{M}'_{0,n}=1$ is the only sensible measure. For $k>0$, however, the measure is much more interesting, and is in fact not unique.

Let us begin by describing some simple constraints the $\M'_{k,n}$ must satisfy. First of all, it must be covariant under $\GL(k;1)$ transformations. Namely, by applying the transformation $\eqref{gauge}$, $\M'_{k,n}$ should scale as $\det(G)^k\det(J)^{k+2}$. The exponents are chosen so that a $\GL(k;1)$ transformation on $\mathcal{Y}$ will make the integrand scale as $\det(G)^{4}\det(J)^{k+2}$, which is the expected result. Furthermore, covariance implies that $\M'_{k,n}$ should only depend on round brackets and square brackets as defined above.

The number of integrations we must do after integrating out the bosonic delta functions is $(k+2)(n-k-4)$.  We interpret these integrals as contour integrals, or as a sum over multi-variable residues. We will show later that the residues of this integral are BCFW terms (or sums of them), provided the correct choice of $\M'$. 

In subsequent parts, we will also make use of the following notation:
\be
\;[i_1\;i_2\;\cdots \;i_s]_{r_1,\ldots,r_t} \;&=&{\epsilon_{r_1,\ldots,r_t}}^{a_1,\ldots,a_s}C_{a_1i_1}\cdots C_{a_s i_s}
\ee

\noindent where $s+t=k$. This denotes, up to a sign, the determinant of the matrix formed by columns $i_1,\ldots,i_s$ of the $C$ matrix with rows $r_1,\ldots,r_t$ removed. The indices are raised and lowered with the all plus metric, so we can raise and lower them arbitrarily. For $t=0$, we get back the square bracket defined earlier.

\subsection{The measure $\mathcal{M}_{k,n}(P)$}

The main content of our proposed formula is the geometric factor $\mathcal{M}'_{k,n}(P)$, which we now discuss.

For $k=0$, as noted earlier, the full measure is simply the cyclic factor times unity.
\be
\mathcal{M}'_{0,n}(P) = 1
\ee
In the case $n=4$, there is no contour, and the measure precisely gives the 4-point MHV 1-loop integrand. For higher $n$, there is a $(2n-8)$-dimensional form, and it is a well known fact that residues of this form give BCFW terms, which are also known as Kermit terms in \cite{ArkaniHamed:2010gg}.

Starting at $k=1$, the geometric factor $\M'$ is no longer trivial. To guess the form of $\M'$, we begin by analyzing the simplest case $n=5$. Since there is no contour, we expect the measure to give the exact integrand. Here we will skip the details and just give the result, which is clean and simple.
\be
\M'_{1,5}=\sum_{i=1}^n \frac{(X_1\;i\;i{+}1)}{[X_1][i][i{+}1]}
\ee

Here $X_1\in \mathbb{P}^2$ is an arbitrary point. Readers familiar with polytopes in projective space will immediately recognize $\M'$ as the area of the pentagon whose vertices are formed by the columns of the $P$ matrix. We have checked both analytically and numerically that this measure precisely gives the 5-point 1-loop NMHV integrand.

We now make a curious and surprising observation. It is well known that this integrand consists of three BCFW terms. Naively, we should expect there to exist a measure whose residues give precisely those three terms. However, that turns out to be the incorrect intuition. What we have learned from this measure is that the $\overline{\text{MHV}}$ case does not require a contour. In fact, as we will see in a moment, this is true for higher $k$ as well, and is probably true to all loops.

So if there are no residues, how could there be three BCFW terms? Let us go back to the geometric factor $\M'_{1,5}$. Obviously, the area of a pentagon can be written in multiple ways, depending on the choice of triangulation. For instance, we can write
\be\label{eq:pentagon}
\M'_{1,5} = \{3,4,(23)\cap(45)\}+\{1,2,(23)\cap(15)\}+\{5,(23)\cap(15),(23)\cap(45)\}
\ee

\noindent where $\{a_1,a_2,a_3\}$ denotes the signed area of the triangle whose vertices are the columns $a_1,a_2,a_3$ of the $P$ matrix, and $(a_1 a_2)\cap(a_3 a_4)$ the intersection of lines $(a_1 a_2)$ and $(a_3 a_4)$. See the Appendix for precise definitions.

It is clear that our geometric factor is the sum of the three terms corresponding to the triangulation of the pentagon. The three triangles are illustrated below. 
\be
\vcenter{\hbox{\includegraphics[width=4.5cm]{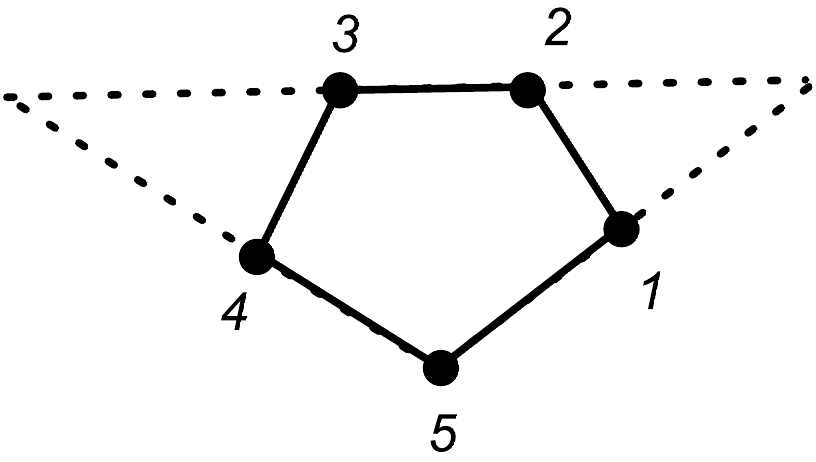}}}
\ee

By now one may readily guess that these three terms are precisely the three BCFW terms, and that is correct.

Even before moving on to more complicated cases, there is a great deal of physical insight that can be learned from this measure. Let us write down the full integral explicitly.

\be\label{local}
\int\;\frac{d^{15}P}{(123)(234)(345)(451)(512)}\sum_{i=1}^5 \frac{(1\;i\;i{+}1)}{[1][i][i{+}1]}
\ee
\noindent where we omitted writing the delta functions, and we have set $X_1\rightarrow 1$.

First, we observe that all the poles of the integrand are manifest in the measure. For example, the $(123)$ factor corresponds to the pole $\left<YAB 45\right>$, which is easy to see on the support of the delta function. In fact, it is generally the case that $(a
\;a{+}1\;a{+}2)$ corresponds to the pole $\left<YAB\;a{+}3\;a{+}4\right>$, where the indices are mod 5 as usual. So all the poles coming from propagators that lie inside the loop are manifest.

Of course there are also propagators that do not lie inside the loop. These are manifest in the square brackets. Namely, the square bracket $[a]$ corresponds to the pole $\left<Y\;a{+}1\;a{+}2\;a{+}3\;a{+}4\right>$.

We see that the pole structure of the amplitudes places very severe constraints on the measure. In fact, one may have even guessed the measure purely from this insight.

We make yet another observation. In the limit where the external data $Z$ is positive and the variables $YAB$ live on the inside of $Z$, we see that the $P$ matrix, which is uniquely constrained by the delta functions, must be positive in the one-loop Grassmannian sense. Geometrically, positivity of the $P$ matrix means that the polygon is convex. This is a rather neat observation, and is in fact connected to the pole structure of the integrand. For instance, the pole $(a\;a{+}1\;a{+}2)\rightarrow 0$ simply corresponds to the vertices $a,a{+}1,a{+}2$ becoming collinear. Moreover, the pole $[a]\rightarrow 0$ corresponds to the vertex $a$ going to infinity. Hence, poles involving the loop correspond to a degenerating polygon while poles not involving the loop correspond to a polygon whose area blows up to infinity.

We make one last observation. There are two classes of triangulations of the pentagon discussed so far, which we will refer to as ``spurious" and ``non-spurious" triangulations respectively.  A spurious triangulation consists of triangles formed by edges (possibly extended in either direction if needed) of the pentagon. These triangles may lie partially outside the pentagon. The BCFW representation shown above is an example of a spurious triangulation. A non-spurious triangulation consists of triangles formed by vertices of the pentagon. These triangles all lie inside the pentagon and tile the pentagon perfectly. An example of a non-spurious triangulation is $\eqref{local}$.

The choice of the word ``spurious" is related to the pole structure of the amplitude. A spurious triangulation gives rise to spurious poles, e.g. $[(12)\cap(34)]$. A non-spurious triangulation, however, does not contain spurious poles. Thus, a triangulation of the form $\eqref{local}$ is term-by-term local. The local form is also term-by-term positive, since each of the triangles in the triangulation has positive area. This is not true, of course, for the BCFW representation. Locality and positivity are probably joined together in a deep way for all integrands, but that is beyond the scope of this article. 

We may now go ahead and propose the measure for all $n$ in the NMHV sector. The most obvious guess is the area of the $n$-gon, as follows.
\be
\mathcal{M}'_{1,n}(P) = \sum_{i=1}^n\;\frac{(X_1\;i\;i{+}1)}{[X_1][i][i+1]}
\ee

Here we have written $\M'_{1,n}$ in local form. We note that this formula has another simple geometric description. When combined with the cyclic measure, it is the unique measure on $P$ space that has logarithmic unit singularities on the boundary of the positive part of the $P$ matrix. In fact, this is a guiding principle for identifying measures in all other sectors, and maybe even to all loops. For now, we can try to answer one simple geometric question for all $k$. What is the top form that has logarithmic unit singularities on the boundary of the positive $P$ space?

For $k=2$, the local form of the measure is given by
\be
\mathcal{M}'_{2,n}(P) = \sum_{i,j=1}^n\;\frac{
\begin{vmatrix}
[i{+}1]_1 & [j{+}1]_1\\
[i{+}1]_2 & [j{+}1]_2
\end{vmatrix}
 \begin{vmatrix}
(X_1\; i\;i{+}1\;i{+}2) & (X_1\; j\;j{+}1\;j{+}2)\\
(X_2\; i\;i{+}1\;i{+}2) & (X_2\; j\;j{+}1\;j{+}2)
\end{vmatrix}
}{
[X_1\;X_2][i\;i{+}1][i{+}1\;i{+}2][j\;j{+}1][j{+}1\;j{+}2]
}
\ee
where $X_1,X_2$ are arbitrary 4-vectors, and the vertical bars denote taking the determinant.

We do not yet have a rigorous proof that this is the correct answer, but our answer is supported by the fact that the measure gives the correct 1-loop integrand for the $n=6$ NNMHV case where there is no contour.

The generalization to higher $k$ is now obvious:
\be
\mathcal{M}'_{k,n}(P) = \sum_{i_1,\ldots,i_k=1}^n\frac{\det_{st}[i_t{+}1,\ldots,i_t{+}k{-}1]_s\;\det_{st}(X_{s}\;i_t\;\cdots\;i_t{+}k)}{[X_1,\ldots,X_k]\prod_{p=1}^k\;[i_p\;i_p{+}1\;\cdots \;i_p{+}k{-}1][i_p{+}1\;i_p{+}2\;\cdots \;i_p{+}k]}
\ee
where $\det_{st}$ denotes the determinant of a matrix whose row and column indices are $s,t$, respectively. The constants $X_1,\ldots,X_k$ are arbitrary $(k+2)$-vectors. We have tested this measure in the $\overline{\text{MHV}}$ case up to $k=4$.

\subsection{Top cells and BCFW terms}\label{ssec:triangle}

We now try to address the problem of obtaining BCFW terms as residues of some measure. Naively, one may be tempted to think that BCFW terms are simply residues of a single measure, which for $k=1$ is given by the area of the $n$-gon. However, this is obviously incorrect. Even for $n=5$, the BCFW terms are given by areas of triangles rather than the full area of the pentagon. This invites us to define a larger class of measures which we will refer to as top cell measures. 

For each top cell $\Pi_S \subset G_+(1,n;1)$, we defined the top cell measure $\mathcal{M}_{\Pi_S}$ as a dlog-form $\prod_i \frac{d\alpha_i}{\alpha_i}$ in Section \ref{ssec:ex1}.  It is not hard to give an explicit description of this top cell measure. Recall that for a top cell $\Pi = \Pi_{\{a,b,c\}}$, there exists a triangle $\Delta$ whose vertices $X_1,X_2,X_3$ are given by \eqref{eq:triangle}, such that the polygon is inscribed inside the triangle. Clearly, we can approach any boundary of $\Pi$ by going to the limit where one of the vertices of $\Delta$ approaches infinity. Thus, it is natural to expect that the measure $\mathcal{M}_\Pi$ should have simple poles at $[X_i]\rightarrow 0$ for $i=1,2,3$. It is thus natural to guess that the geometric factor of the top cell measure $\M_{\Pi}$ is given by the area of $\Delta$:
\be \label{eq:topcell}
\mathcal{M}'_\Pi = \{X_1,X_2,X_3\}
\ee
(Recall that the full top cell measure is given by $\M_\Pi = \frac{d^{(3n}P/{\text{vol}\;\GL(1;1)}}{\prod_{i=1}^n\;(i\;i{+}1\;i{+}2)}\M'_\Pi$.)  We shall prove \eqref{eq:topcell} in Section \ref{sec:topcellmeasure}.

For example, in the 5 point case where there is no residue, the measure for the full integrand is given by the sum of three top cell measures.
\be
\mathcal{M}_{1,5} = \mathcal{M}_{\Pi_{\{2,3,4\}}}+\mathcal{M}_{\Pi_{\{1,2,5\}}}+\mathcal{M}_{\Pi_{\{2,4,5\}}}
\ee
The three top cell measures match term by term with the triangle areas in \eqref{eq:pentagon}. Thus, each BCFW term is given by a top cell measure, at least for $n=5$.

We can now make the conjecture that each BCFW term for $k=1$ and any $n$ is given as some multi-dimensional residue of a top cell measure. Both the top cell and residue can be identified using a series of simple steps that we now illustrate. 

We now make an important observation. Consider again the FL-2 term for the 6 point NMHV 1-loop integrand. Recall from Section \ref{ssec:ex2b} that the top cell $\Pi_\text{FL-2}$ of this term is given by a codimension-3 boundary of $\Pi_{2,4,6}$ by taking $(234),(456),[(16)\cap(24)]\rightarrow 0$. It follows that the logarithmic form on $\mathcal{M}_{\text{FL-2}}$ on $\Pi_{\text{FL-2}}$ must be given as a 3-residue of the logarithmic form $\mathcal{M}_{\Pi_{2,4,6}}$ by taking the same three quantities to zero.

In other words, the FL-2 term is given by
\be
\text{FL-2}_\text{Y space} = \int\underset{\substack{(234)\rightarrow 0\\(456)\rightarrow 0\\\;[(16)\cap(24)]\rightarrow0}}{\text{Res}}\;\frac{d^{18}P}{\prod_{i=1}^6(a,a{+}1,a{+}2)}\;\mathcal{M}'_{\Pi_{\{2,4,6\}}}\delta^{15}(\mathcal{Y}-P.Z)
\ee

In practice, the residue can be computed by using the following change of variables.
\be
\;P_3 &=& \alpha_1 P_2 + \alpha_2 P_4 + z_1 P_1\\
\;P_5 &=& \alpha_3 P_4 + \alpha_4 P_6 + z_2 P_1 \\
\;[(16)\cap(24)] &=& z_3
\ee
Then take residues where $z_1,z_2,z_3\rightarrow 0$, which gives the following.
\be
\text{FL-2}_\text{Y space} =\int\frac{d^4\alpha d^3P_1 d^3P_2 d^3P_4 d^3 P_6}{\alpha_1\alpha_2\alpha_3\alpha_4(124)(246)(612)[4][6]}\;\delta\left([1]-[6]\frac{(124)}{(246)}\right)\delta^{15}(\mathcal{Y}-P.Z)
\ee
The remaining part of the integral is straightforward since there are precisely enough delta functions to localize the integrals.

We should note that the top cell from which the FL-2 cell is obtained is not unique. For each term in the Appendix, we indicate only one possible top cell. In fact, for some BCFW terms, it is possible to obtain the cell for the term as a codimension-3 boundary of the positive $P$ space. In that case, it is canonical to use the full polygon area as the geometric factor, as we have done in the Appendix for appropriate terms.

It appears to be a fact for general $n$ that each BCFW term can be obtained by taking $3(n-5)$ residues of a top cell of the $n$-gon. We provide complete evidence for this for $n=6$ in the Appendix where there are 16 BCFW terms. For each term, we specify the top cell measure and the 3 residues we must take to obtain the term.

\subsection{Top cells as a triangulation of $G_+(1,n;1)$}
An arbitrary point of $G_+(1,n;1)$ (considered up to the $T_+$ torus action) is simply an arbitrary convex $n$-gon in $\A^2$.  Here we insist that all $1 \times 1$ minors of $C$ and all $3 \times 3$ minors of $P$ are strictly positive, so that degenerate $n$-gons do not appear.

We can see directly that $G_+(1,n;1)$ can be triangulated as a union of the top cells $\Pi_S$.  Let $C_n$ be a convex $n$-gon (different to the $n$-gon $(P_1,P_2,\ldots,P_n)$).  Suppose we are given a triangulation $\G=\{S_1,S_2,\ldots,S_{n-2}\}$ of $C_n$, where each $S_i$ is a triangle of $C_n$ and thus a three-element subset of $\{1,2,\ldots,n\}$.  Then we claim that $\T = \{\Pi_{S_1},\ldots,\Pi_{S_{n-2}}\}$ is a collection of top cells of $G_+(1,n;1)$ that triangulate $G_+(1,n;1)$.

For example, if $n = 5$, we could have $\G = \{125,234,245\}$.  The top cells $\T = \{\Pi_{\{1,2,5\}},\Pi_{\{2,3,4\}},\Pi_{\{2,4,5\}}\}$ triangulate $G_+(1,n;1)$.  

\begin{itemize}
\item
The cell $\Pi_{\{1,2,5\}}$ is given by the condition that $[2,3;5,1]$ is positive.

\item
The cell $\Pi_{\{2,3,4\}}$ is given by the condition that $[4,5;2,3]$ is positive.
\item
The cell $\Pi_{\{2,4,5\}}$ is given by the conditions that $[2,3;4,5]$ and $[5,1;2,3]$ are positive.
\end{itemize}

Given a nondegenerate convex pentagon, $[2,3;5,1]$ is either positive or negative.  Similarly, $[4,5;2,3]$ is either positive or negative.  It follows that every nondegenerate convex pentagon belongs to exactly one of $\Pi_{\{1,2,5\}},\Pi_{\{2,3,4\}},\Pi_{\{2,4,5\}}$, so $\T$ is a triangulation of $G_+(1,n;1)$, as claimed.

%
%

Let us now show this for arbitrary $n$.  Suppose that $\{S_1,S_2,\ldots,S_{n-2}\}$ is a triangulation of the $C_n$.  By cyclic rotation, we can assume that one of the triangles, say $S_1$, is $\{1,2,3\}$.  The corresponding top cell $\Pi_{\{1,2,3\}}$ is specified by insisting that $[3,4;1,2]$ is positive.  Let $\Pi = \Pi_{S_i}$ ($i \neq 1$) be any other triangle in the triangulation.  We claim that $[3,4;1,2]$ is negative on $\Pi_{S_i}$ which implies that $\Pi_{S_1}$ and $\Pi_{S_i}$ do not intersect.    Let $S_i = \{a,b,c\}$ where $(a,b)$ is the edge of $C_n$ closest to the edge $(1,3)$.  Then the positivity of $[a,a+1;b,b+1]$ is positive on $S_i$, and since $\{1,2,3\}$ lie cyclically between $a$ and $b$, we have that $[3,4;1,2]$ is negative on $S_i$.  Repeating this argument we see that all the top cells $\Pi_{S_i}$ and $\Pi_{S_j}$ for $i \neq j$ are disjoint.  Very similar logic shows that in fact every nondegenerate convex polygon lies in one of the $\Pi_{S_i}$, so that $\T = \{\Pi_{S_1},\ldots,\Pi_{S_{n-2}}\}$ is a collection of top cells of $G_+(1,n;1)$ that triangulate $G_+(1,n;1)$.

Thus, each triangulation $\T$ of $G_+(1,n;1)$ uses $n-2$ top cells, and the number of such triangulations is the Catalan number (the number of triangulations of a $n$-gon). 

As a consistency check, the fact that $\{S_1,...,S_{n-2}\}$ triangulates $C_n$ implies that the geometric factors $\mathcal{M}'_{\Pi_{S_i}}$ add up to the area of the polygon $P$. However, the Grassmannian measure corresponding to the polygon area is precisely the logarithmic form on the positive $P$ space. It follows that the regions $\Pi_{S_i}$ must triangulate the positive $P$ space. This argument does not show that the regions are mutually disjoint (except for overlapping boundaries), but is nonetheless a non-trivial check.


\subsection{Geometric factor of the top cell measure}\label{sec:topcellmeasure}
We prove \eqref{eq:topcell} that
\be \label{eq:prove}
\M_{\Pi} = \frac{d^{3n}P/{\text{vol}\;\GL(1;1)}}{\prod_{i=1}^n(i,i+1,i+2)} \{X_1,X_2,X_3\}
\ee
where $\M_\Pi$ is the dlog-form defined in Section \ref{ssec:ex1} and $\{X_1,X_2,X_3\}$ is the area of the triangle $\Delta$ defined in Section \ref{ssec:triangle}.  In the following, we will not worry about global signs.

The proof is by induction on $n$, where the base case is $n = 3$.  In this case, there is only one top cell $\Pi=\Pi_{\{1,2,3\}}$, and we can gauge fix $P$ to be
$$
P = \begin{pmatrix}1 & 0 & 0 \\ 0 & 1 & 0 \\ \beta \alpha& \alpha& 1\end{pmatrix}
$$
where $(\alpha,\beta)$ are the positive coordinates.  Thus the LHS of \eqref{eq:prove} is equal to $(1/\alpha\beta)d\alpha d\beta$.  The geometric factor $\{X_1,X_2,X_3\}$ is equal to $1/\alpha^2\beta$ and $d^{(k+2)n}P/{\text{vol}\;\GL(1;1)} = d\alpha d(\beta\alpha)= \alpha d\alpha d\beta$.  Since all the cyclic minors are equal to 1, \eqref{eq:prove} holds in this case.

Now suppose \eqref{eq:prove} holds for some value of $n$.  After cyclically relabelling the $n+1$ vertices, any top cell $\Pi'$ for $G_+(1,n+1;1)$ can be obtained from some top cell $\Pi$ for $G_+(1,n;1)$ in the following way:
\begin{enumerate}
\item
Start with the space of $n$-gons $(P_1,P_2,\ldots,P_n)$ for $\Pi$.
\item
Perform the shift $(n \to n+1)$, placing $P_{n+1}$ on top of $P_n$.
\item 
Perform the shift $(n+1 \to 1)$, moving $P_{n+1}$ towards $P_1$.
\item
Perform the shift $(n \to n-1)$ moving $P_n$ towards $P_{n-1}$.
\end{enumerate}
Note that each of the shifts (2),(3),(4) increases the dimension of the cell by one  (so $\dim \Pi' = \dim \Pi +3$), and introduces a new positive coordinate denoted $\alpha,\beta,\gamma$ respectively.  If $P$ is the original matrix for $\Pi$ with columns $P_1,P_2,\ldots,P_n$, then the new matrix is
$$
P'
 = (P_1, P_2,\ldots,P_{n-1}, P_n+\gamma P_{n-1},\alpha P_n + \beta P_1).
 $$
It is not difficult to check that
$$
d^{3(n+1)}P'/{\text{vol}\;\GL(1;1)} = \pm d^{3n}P\alpha (n-1,n,1) d\alpha d\beta d\gamma/{\text{vol}\;\GL(1;1)}
$$
where $(n-1,n,1)$ denotes a minor of $P$.  But we also have
\begin{align*}
(n-2,n-1,n)' &= (n-2,n-1,n)\\
(n-1,n,n+1)' &= \beta (n-1,n,1) \\
(n,n+1,1)' & = \alpha \gamma (n-1,n,1) \\
(n+1,1,2)' &= \alpha (n,1,2).
\end{align*}
So
$$
\frac{d^{3n}P/{\text{vol}\;\GL(1;1)}}{\prod_{i=1}^{n+1}(i,i+1,i+2)'} = \frac{\alpha}{\alpha^2\beta\gamma}\frac{d^{3n}P/{\text{vol}\;\GL(1;1)}}{\prod_{i=1}^{n}(i,i+1,i+2)} d\alpha d\beta d\gamma =\frac{1}{\{X_1,X_2,X_3\}}M_{\Pi'} 
$$
since $\M_{\Pi'} = \M_{\Pi} \frac{d\alpha d\beta d\gamma}{\alpha\beta\gamma}$.  Now as triangles in $\P^2$, the triangle $\Delta$ for $\Pi$ and $\Delta'$ for $\Pi'$ are identical.  So $\{X_1,X_2,X_3\} = \{X'_1,X'_2,X'_3\}$.  Thus we have $\frac{1}{\{X_1,X_2,X_3\}}M_{\Pi'} = \frac{1}{\{X'_1,X'_2,X'_3\}}M_{\Pi'}$, completing the proof of \eqref{eq:prove}. 


\section{Conclusion and outlook}

In this paper we studied systematically the rich mathematical structures associated with scattering amplitudes of planar $\mathcal{N}=4$ SYM at one-loop level. At tree level, it is well known that BCFW terms of tree amplitudes, or equivalently cells of the tree-level amplituhedron, correspond to residues of a Grassmannian contour integral~\cite{ArkaniHamed:2009si}. We generalized this correspondence by showing that cells of the one-loop amplituhedron are naturally associated with residues of a new class of Grassmannian measures at one-loop, called top-cell measures. The measures contain a naive part built from cyclic minors of the $P$ matrix, and an additional part which can be viewed as a natural generalization of the area of polygons in $\mathbb{P}^2$ in the $k=1$ case. 

Moreover, we studied in detail the correspondence between residues of the top cell measures, cells of one-loop amplituhedron and the BCFW terms. For studying this, we use the momentum-twistor diagrams proposed in~\cite{Bai:2014cna}, which correspond directly to such BCFW terms or cells, and this is the one-loop generalization of the tree-level correspondence between on-shell diagrams and Grassmannian cells.

In particular, the story becomes especially simple for the $k=1$ case, where a geometric interpretation of the one-loop Grassmannian is manifest. Each cell of the $n$-point, $k=1$ amplituhedron corresponds to a configuration of a possibly degenerate $n$-sided polygon in $P^2$, and this geometric picture turns out to be very useful for studying cells/diagrams of the $k=1$ case. Similar to the tree-level case, one can construct any $d$-dimensional cell (or momentum-twistor diagram) by applying $d$ transpositions (or BCFW bridges) to a zero-dimensional cell (or the lollipop diagram). We have shown this construction in detail via various illustrative examples, and geometrically it is equivalent to constructing the corresponding polygon configuration with $d$ moves. We have also discussed, in parallel to the tree-level case, the decorated permutations associated with one-loop cells/diagrams.

Our results open up many avenues for future studies. One of the most pressing issues is to completely understand the case with $k>1$, where the amplituhedron and our one-loop Grassmannian formula must hold.  Although momentum-twistor diagrams and BCFW terms are known for higher $k$, it remains a very interesting open question to see their origin as residues of top cells (still undefined for $k>1$) of the one-loop Grassmannian, and in particular if there exist simple geometric pictures. Another important future direction is to go to higher loops, where we have many questions to ask. For example, can we generalize to higher loops the construction of amplituhedron cells (or momentum-twistor diagrams) by applying transpositions/BCFW bridges to 0-dimensional cells, and is there a combinatorial structure associated with it? Is there again some universal measure for higher-loop Grassmannians, with logarithmic singularities and the top cells given by the canonical decomposition? It would be very exciting to study these problems already for the MHV ($k=0$) case at higher loops.

\section*{Acknowledgments}

We would like to thank Nima Arkani-Hamed for useful discussions. Y.B. was supported by the Department of Physics, Princeton University. S.H. was supported by the Institute of Theoretical Physics, Chinese Academy of Science, China. T.L. was partially supported by NSF grants DMS-1160726, DMS-1464693, and a Simons Fellowship $(\#341949)$. 

\appendix
\section{Appendix}

\be
D_i &=& \text{$i^\text{th}$ column of the $D$ matrix}\nonumber\\
C_i &=& \text{$i^\text{th}$ column of the $C$ matrix}\nonumber\\
P_i &=& \text{$i^\text{th}$ column of the $P$ matrix}\nonumber\\
\{i,j,k\} &=& \frac{(ijk)}{[i][j][k]} \nonumber\\
\{i_1,...,i_m\} &=& \text{area of the polygon with vertices $P_{i_1},...,P_{i_m}$}\nonumber\\
(ij)\cap(kl) &=& P_j(ikl) - P_i (jkl) \nonumber\\
\text{Ordinary SUSY space:  } (ab)\cap(cde) &=& \mathcal{Z}_a \lb bcde\rb- \mathcal{Z}_b \lb acde\rb\nonumber\\
\text{$Y$-space:  } (ab)\cap(cde) &=& Z_a \lb Y bcde\rb - Z_b \lb Y acde\rb - Y \lb abcde\rb\nonumber\\
\text{Ordinary bosonic space: } (abc)\cap(def) &=& Z_a Z_b\lb cdef\rb-Z_c Z_b\lb adef\rb-Z_a Z_c\lb bdef\rb\nonumber\\
\text{$Y$-space: } (abc)\cap(def) &=& Z_a Z_b\lb cdef\rb-Z_c Z_b\lb adef\rb-Z_a Z_c\lb bdef\rb\nonumber
\ee

In order to compute the 6 point 1-loop NMHV integrand, we apply the BCFW shift $\hat{\mathcal{Z}}_6 = \mathcal{Z}_6+w\mathcal{Z}_5$. The $w\rightarrow \infty$ pole gives us the 5 point 1-loop NMHV integrand, which we will refer to as a boundary term B. There are in fact three BCFW terms contributing to B, which we will refer to as B-i for $i=1,2,3$. There are five terms coming from the factorization channel, which we refer to as FAC-i for $i=1,...,5$. Also, there are eight forward limit terms referred to as FL-i for $i=1,...,8$.

In this section, we list the BCFW terms and their associated momentum-twistor diagrams. For each term, we display the corresponding $C$ and $D$ matrices in the following form.
\be
P=\begin{pmatrix}
D \\
C
\end{pmatrix}
\ee
We note that the $C$ and $D$ matrices are derived by performing boundary measurements on the diagram. In doing so, we have implicitly adjusted the signs of some of the bridge variables so that the matrix obeys the necessary positivity conditions.

Each term, when written in $Y$-space and multiplied by the 8-form
\be 
\frac{1}{4! 2! 2!}\lb Y \;dY \;dY \;dY \;dY\rb\lb Y AB\; dA\;dA\rb\lb Y A B \;dB\;dB\rb
\ee
is equivalent to the form 
\be
d\log c_1\; d\log c_2\;...\;d\log c_8
\ee
where $c_1,...,c_8$ are the bridge variables appearing in the diagram corresponding to the term. This makes manifest the important and non-trivial property that each BCFW term is simply a single wedge product of $d\log$'s. An important observation is that the diagrammatic approach makes this obvious, while the usual Lagrangian approach obscures it. The bridge variables are ordered so that the logarithmic form is positively oriented when the variables are positive.

Furthermore, for each term, we write down the sequence of residues we must take in order to obtain the term. We include one of possibly several geometric factors. In these computations, we parametrize part of our $P$ matrix using variables $z_s, (s=1,2,3)$, and we take the residues $z_s\rightarrow 0$ in order of increasing $s$. While doing these computations, we may occasionally introduce auxiliary variables $\alpha_1,\alpha_2,...$ which are additional variables on the $P$ matrix. We do not take residues in these variables; they are instead localized by the delta function. Finally, a symbol such as $(a\bullet\bullet)$ denotes all the $3\times 3$ minors of the $P$ matrix which involve the index $a$. In general, $\bullet$ represents an arbitrary particle index.\\

\newpage

\be
\vcenter{\hbox{\includegraphics[width=6cm]{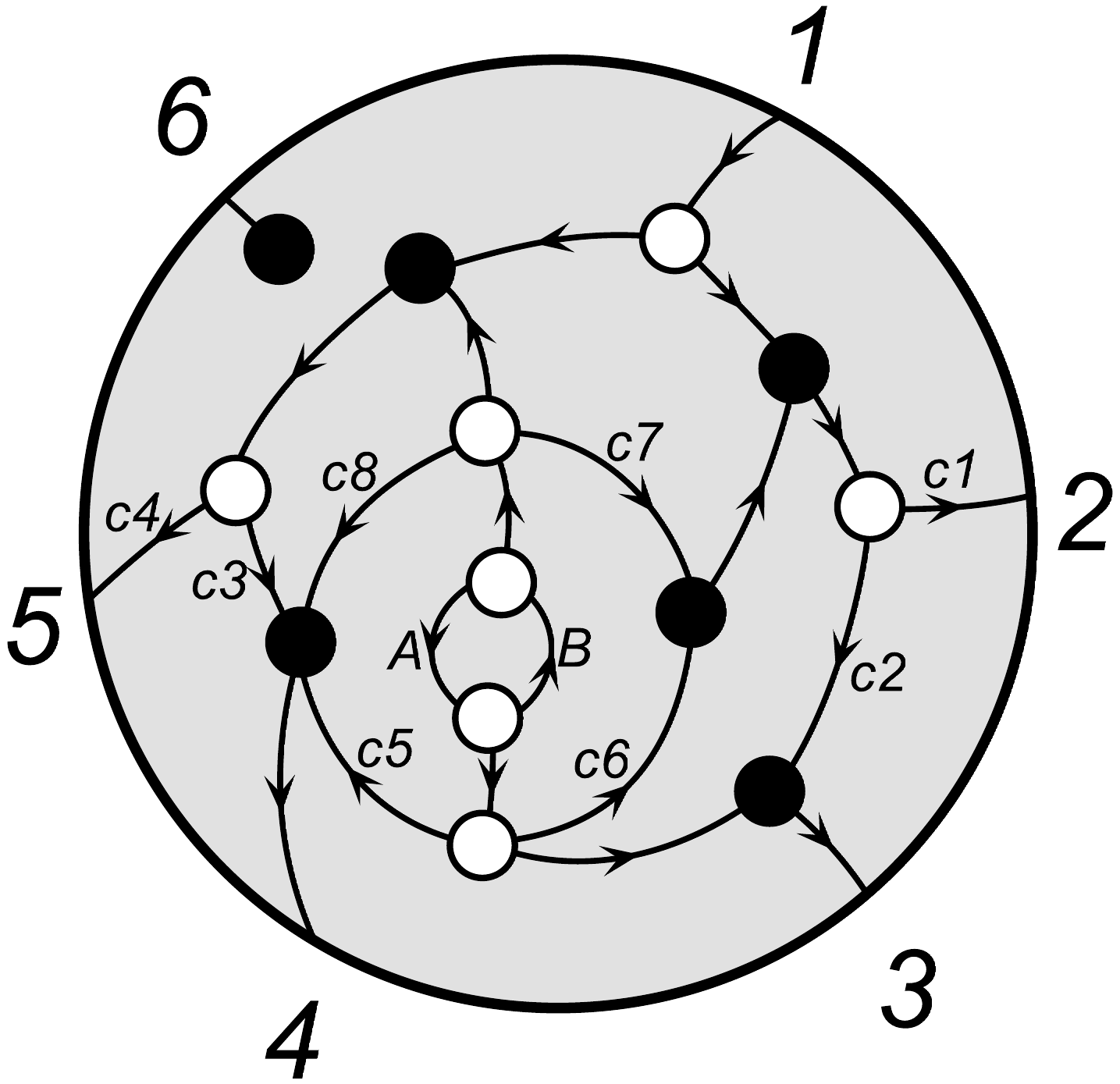}}}\;\;\;\;\;
\begin{tabular}{ll}
$\begin{pmatrix}
0 & c_1 c_6 & 1+c_2 c_6 & c_5 & 0 & 0\\
0 & -c_1c_7 & -c_2 c_7 & c_3+c_8 & c_4 & 0\\
1 & c_1 & c_2 & c_3 & c_4 & 0
\end{pmatrix}$\\ \\
$\text{Vanishing minors: } (6\bullet\bullet),[6]$\\\\
$\text{Residues: }\begin{cases}
P_6 = z_1 P_5+z_2 P_1 + z_3 P_2 & z_1,z_2,z_3\rightarrow 0 
\end{cases}$\\\\
$\mathcal{M}'_{\text{B-1}} = \{4,(23)\cap(45),3\} = \mathcal{M}'_{\Pi_{\{2,3,4\}}}$
\end{tabular}\nonumber
\ee
\be
\text{B-1}&=&\frac{\delta^{0|4}(\eta_1 \left<2345\right>+\text{cyclic})}{
\left<1245\right>\left<1235\right>\left<AB23\right>\left<AB34\right>\left<AB45\right>\left<AB1(45)\cap(123)\right>}\nonumber
\\
\text{B-1}_\text{Y space}&=&\frac{\lb 12345\rb^4}{
\left<Y1245\right>\left<Y1235\right>\left<YAB23\right>\left<YAB34\right>\left<YAB45\right>\left<YAB1(45)\cap(123)\right>}\nonumber
\ee

\be
\vcenter{\hbox{\includegraphics[width=6cm]{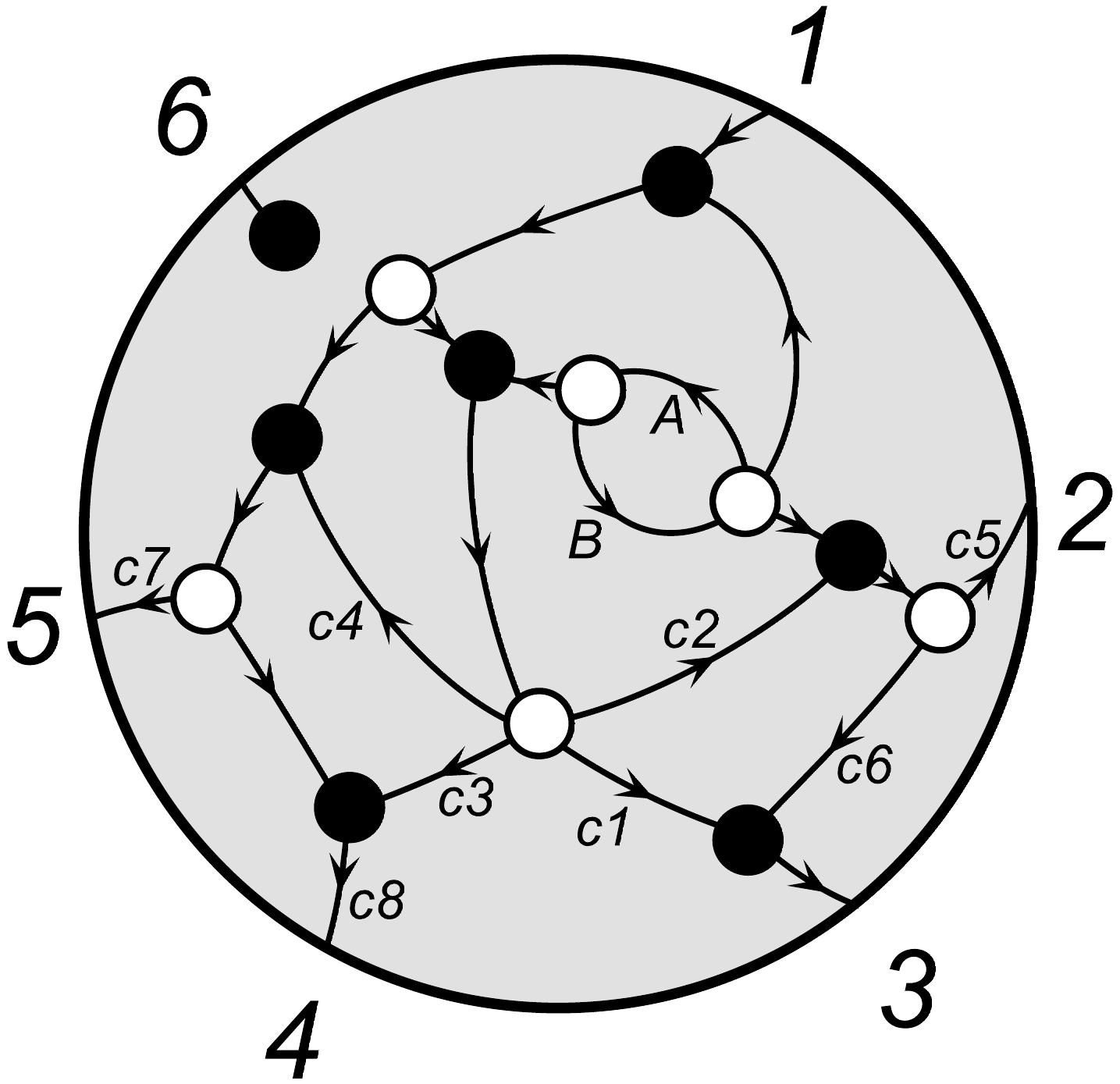}}}\;\;\;\;\;
\begin{tabular}{ll}
$\begin{pmatrix}
0 & c_2c_5 & c_1{+}c_2c_6 & c_8(c_3{+}c_4) & c_4 c_7 & 0 \\
0 & c_5({-}1{+}c_2) & c_1+c_6(-1{+}c_2) & c_8(1{+}c_3{+}c_4) & c_7(1{+}c_4) & 0\\
1 & c_2 c_5 & c_1{+}c_2 c_6 & c_8(1{+}c_3{+}c_4) & c_7(1{+}c_4) & 0
\end{pmatrix}$\\ \\
$\text{Vanishing minors: } (6\bullet\bullet),[6]$\\\\
$\text{Residues: }\begin{cases}
P_6 = z_1 P_5+z_2 P_1 + z_3 P_2 & z_1,z_2,z_3\rightarrow 0 
\end{cases}$\\\\
$\mathcal{M}'_{\text{B-2}} = \{2,(23)\cap(15),1\} = \mathcal{M}'_{\Pi_{\{1,2,5\}}}$
\end{tabular}\nonumber
\ee
\be
\text{B-2}&=&\frac{\delta^{0|4}(\eta_1 \left<2345\right>+\text{cyclic})}{\left<1345\right>\left<2345\right>\left<AB12\right>\left<AB23\right>\left<AB15\right>\left<AB(15)\cap(234)4\right>}\nonumber
\\
\text{B-2}_\text{Y space}&=&\frac{\lb 12345\rb^4}{\left<Y1345\right>\left<Y2345\right>\left<YAB12\right>\left<YAB23\right>\left<YAB15\right>\left<YAB(15)\cap(234)4\right>}\nonumber
\ee

\be
\vcenter{\hbox{\includegraphics[width=6cm]{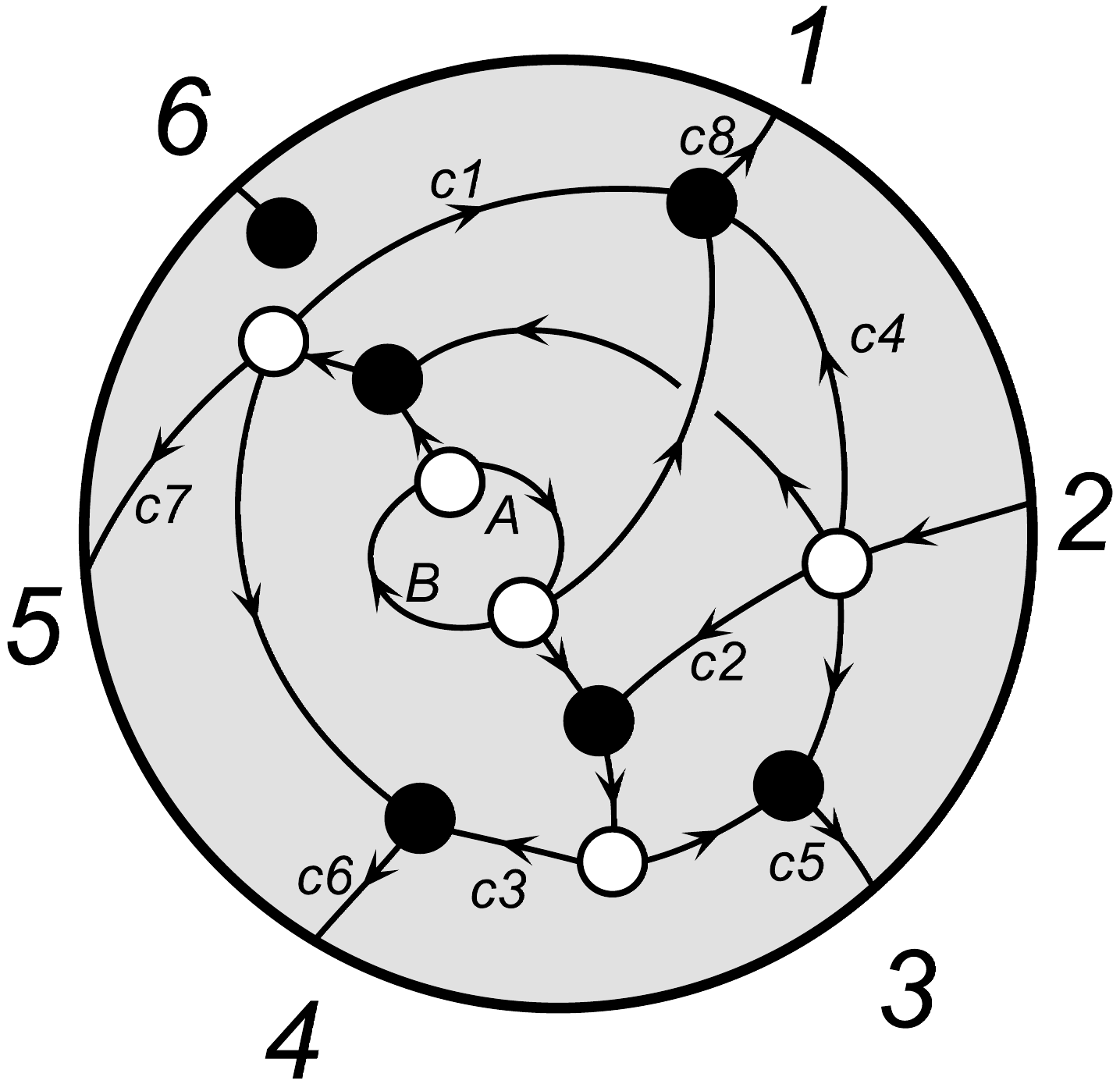}}}\;\;\;\;\;
\begin{tabular}{ll}
$\begin{pmatrix}
-c_8 & 0 & c_5 & c_3c_6 & 0 & 0\\
c_1 c_8 & 0 & 0 & c_6 & c_7 & 0\\
c_8(c_1+c_4) & 1 & c_5(1+c_2) & c_6(1+c_2c_3) & c_7 & 0
\end{pmatrix}$\\ \\
$\text{Vanishing minors: } (6\bullet\bullet),[6]$\\\\
$\text{Residues: }\begin{cases}
P_6 = z_1 P_5+z_2 P_1 + z_3 P_2 & z_1,z_2,z_3\rightarrow 0 
\end{cases}$\\\\
$\mathcal{M}'_{\text{B-3}} = \{5,(23)\cap(15),(23)\cap(45)\} = \mathcal{M}'_{\Pi_{\{2,4,5\}}}$
\end{tabular}\nonumber
\ee
\be
\text{B-3}&=&\frac{\delta^{0|4}(\eta_1 \left<2345\right>+\text{cyclic})\left<AB14\right>^2}{\left<1234\right>\left<AB12\right>\left<AB34\right>\left<AB45\right>\left<AB15\right>\left<AB(45)\cap(123)1\right>\left<AB4(15)\cap(234)\right>}\nonumber
\\
\text{B-3}_\text{Y space}&=&\frac{\lb 12345\rb^4\left<YAB14\right>^2}{\parbox{5in}{$\left<Y1234\right>\left<YAB12\right>\left<YAB34\right>\left<YAB45\right>\left<YAB15\right>\left<YAB(45)\cap(123)1\right>$ \hspace*{7cm} $\left<YAB4(15)\cap(234)\right>$}}\nonumber
\ee

\be
\vcenter{\hbox{\includegraphics[width=6cm]{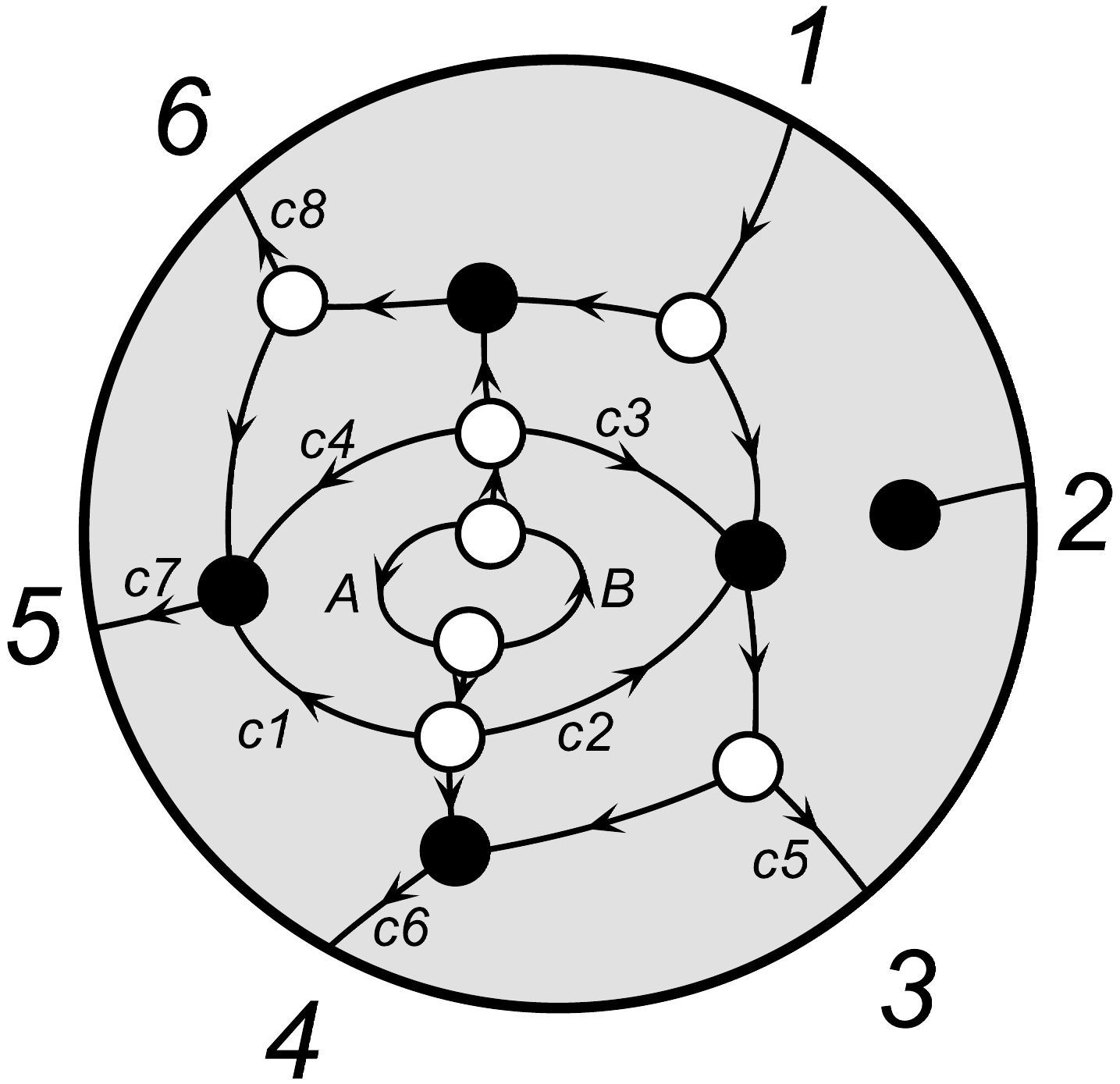}}}\;\;\;\;\;
\centering
\begin{tabular}{ll}
$\begin{pmatrix}
0 & 0 & c_2 c_5 & c_6(c_2+1) & c_1 c_7 & 0\\
0 & 0 & -c_3 c_5 & -c_3 c_6 & c_7(1+c_4) & c_8 \\
1 & 0 & c_5 & c_6 & c_7 & c_8
\end{pmatrix}$\\ \\
$\text{Vanishing minors: } (2\bullet\bullet), [2]$\\\\
$\text{Residues }\begin{cases}
P_2 = z_1 P_1 + z_2 P_3 + z_3 P_4 & z_1,z_2,z_3\rightarrow 0
\end{cases}$\\\\
$\mathcal{M}'_{\text{FAC-1}} = \{4,5,(34)\cap(56)\} = \mathcal{M}'_{\Pi_{\{3,4,5\}}}$\\\\
\end{tabular}\nonumber
\ee
\be
\text{FAC-1}&=&\frac{\delta^{0|4}(\eta_1\lb 3456\rb+\text{cyclic})}{\lb 1356\rb\lb 1346\rb\lb AB34\rb\lb AB45\rb\lb AB56\rb\lb AB(34)\cap(156)1\rb}\nonumber\\
\text{FAC-1}_\text{Y space}&=&	\frac{\lb 13456 \rb^4}{\lb Y1356\rb\lb Y1346\rb\lb YAB34\rb\lb YAB45\rb\lb YAB56\rb\lb YAB(34)\cap(156)1\rb}\nonumber
\ee
\\

\be
\vcenter{\hbox{\includegraphics[width=6cm]{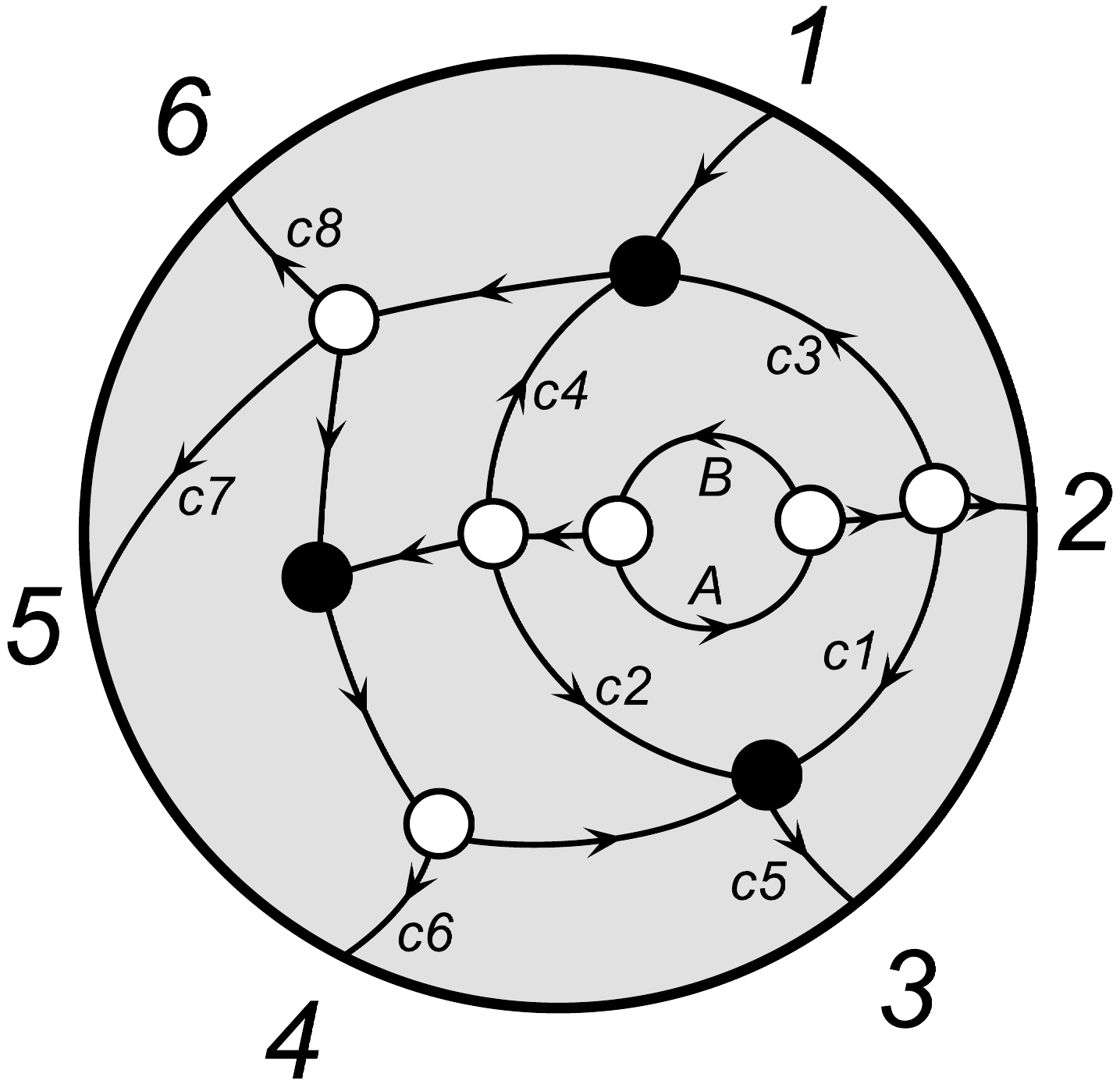}}}\;\;\;\;\;
\begin{tabular}{ll}
$\begin{pmatrix}
0 & 1 & c_5(c_1-c_3) & -c_3 c_6 & -c_3 c_7 & -c_3 c_8\\
0 & 0 & c_5(1+c_2+c_4) & c_6(1+c_4) & c_4 c_7 & c_4 c_8\\
1 & 0 & c_5 & c_6 & c_7 & c_8
\end{pmatrix}$\\ \\
$\text{Vanishing minors: }(56\bullet),[2]$\\\\
$\text{Residues: }\begin{cases}
P_6 = \alpha_1 P_5 + z_1 P_1+z_2 P_4 & z_1,z_2\rightarrow 0\\
C_2 = z_3 & z_3\rightarrow 0
\end{cases}$\\\\
$\mathcal{M}'_{\text{FAC-2}} = \{1,2,3,4,5,6\}$
\end{tabular}\nonumber
\ee
\be
\text{FAC-2}&=&\frac{\delta^{0|4}(\eta_1\lb 3456\rb+\text{cyclic})\lb 1234\rb^2}{\lb 1345\rb \lb 3456\rb \lb 1456\rb \lb 1346\rb \lb AB12\rb \lb AB23\rb \lb AB34\rb \lb AB1(34)\cap(156)\rb}\nonumber\\
\text{FAC-2}_\text{Y space}&=&	\frac{\lb 13456 \rb^4\lb Y1234\rb^2}{\parbox{4.5in}{$\lb Y1345\rb \lb Y3456\rb \lb Y1456\rb \lb Y1346\rb \lb YAB12\rb \lb YAB23\rb \lb YAB34\rb$ \hspace*{7cm} $\lb YAB1(34)\cap(156)\rb$}}\nonumber
\ee
\\

\be
\vcenter{\hbox{\includegraphics[width=6cm]{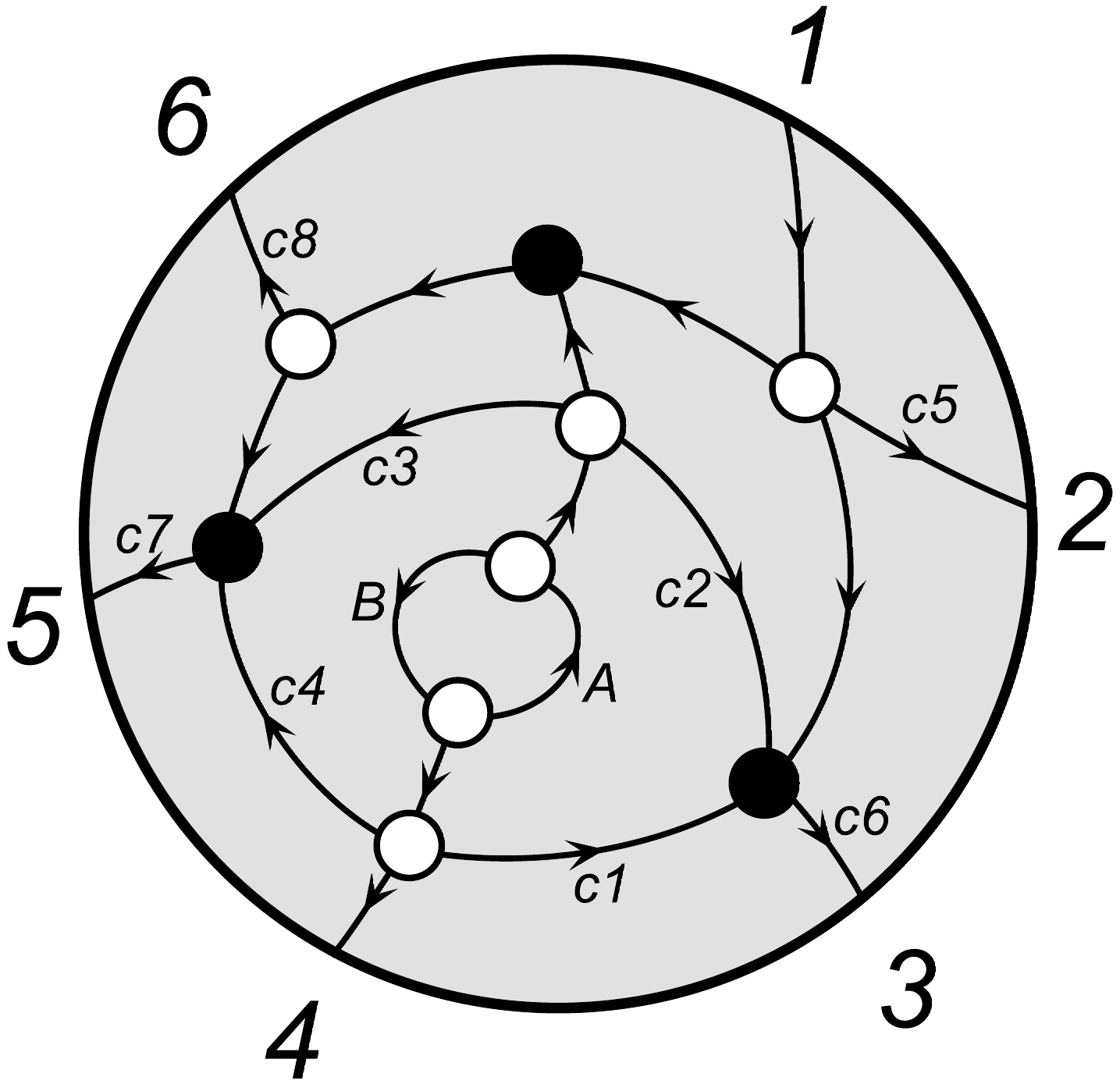}}}\;\;\;\;\;
\begin{tabular}{ll}
$\begin{pmatrix}
0 & 0 & c_1 c_6 & 0 & -c_7(1+c_3) & -c_8 \\
0 & 0 & c_2 c_6 & 1 & c_4 c_7 & 0 \\
1 & c_5 & c_6 & 0 & c_7 & c_8
\end{pmatrix}$\\ \\
$\text{Vanishing minors: } (12\bullet),[4] $\\\\
$\text{Residues: }\begin{cases}
P_2 = \alpha_1 P_1 + z_1 P_3 + z_2 P_6 & z_1,z_2\rightarrow 0 \\
C_4 = z_3 & z_3\rightarrow 0
\end{cases}$\\\\
$\mathcal{M}'_{\text{FAC-3}} = \{1,2,3,4,5,6\}$
\end{tabular}\nonumber
\ee
\be
\text{FAC-3}&=& \frac{\delta^{0|4}(\eta_1\lb 2356\rb+\text{cyclic})\lb 3456\rb^2}
{\lb 2356\rb\lb 1356\rb\lb 1256\rb\lb 1236\rb\lb AB34\rb\lb AB45\rb\lb AB56\rb\lb AB3(56)\cap(123)\rb}\nonumber\\
\text{FAC-3}_\text{Y space}&=& \frac{\lb 12356\rb^4\lb Y3456\rb^2}
{\parbox{4.5in}{$\lb Y2356\rb\lb Y1356\rb\lb Y1256\rb\lb Y1236\rb\lb YAB34\rb\lb YAB45\rb\lb YAB56\rb$ \hspace*{8cm} $\lb YAB3(56)\cap(123)\rb$}}\nonumber
\ee
\\

\be
\vcenter{\hbox{\includegraphics[width=6cm]{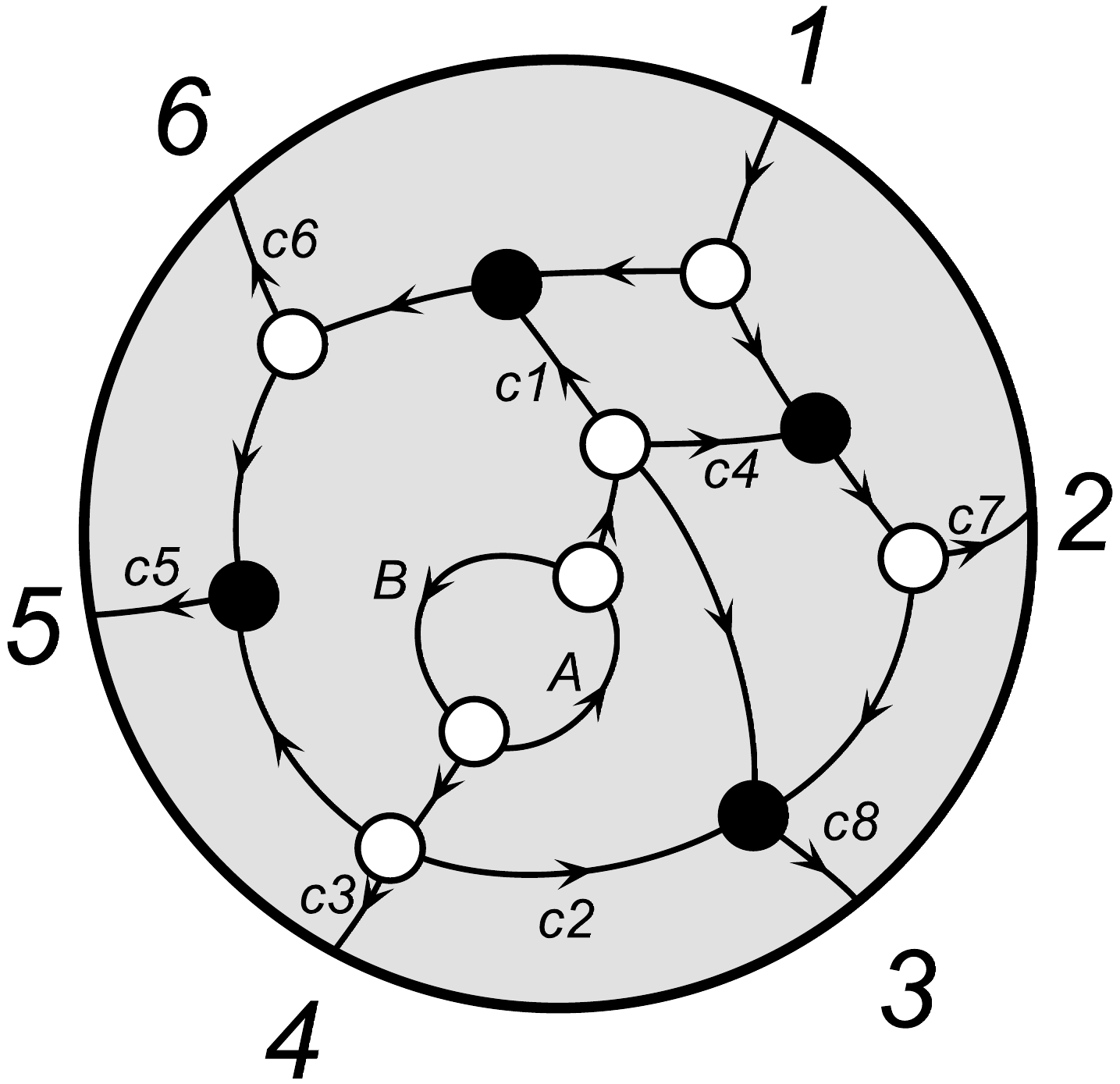}}}\;\;\;\;\;
\begin{tabular}{ll}
$\begin{pmatrix}
0 & c_4 c_7 & c_8(1+c_4) & 0 & -c_1 c_5 & -c_1 c_6 \\
0 & 0 & c_2 c_8 & c_3 & c_5 & 0\\
1 & c_7 & c_8 & 0 & c_5 & c_6
\end{pmatrix}$\\ \\
$\text{Vanishing minors: } (126),(456),[4] $\\\\
$\text{Residues: }\begin{cases}
P_1 = \alpha_1 P_2+\alpha_2 P_6+z_1 P_3 & z_1\rightarrow 0\\
P_5 = \alpha_3 P_4 + \alpha_4 P_6 + z_2 P_3 & z_2 \rightarrow 0\\
C_4 = z_3 & z_3\rightarrow 0
\end{cases}$\\\\
$\mathcal{M}'_{\text{FAC-4}} = \{1,2,3,4,5,6\}$
\end{tabular}\nonumber
\ee
\be
\text{FAC-4}&=& \frac{\delta^{0|4}(\eta_1\lb 2356\rb+\text{cyclic})\lb AB3(45)\cap(123)\rb^2\lb 2356\rb}
{\parbox{5in}{$\lb 1235\rb \lb 1256\rb \lb 1236\rb \lb AB23\rb \lb AB34\rb \lb AB35\rb \lb AB45\rb \lb AB3(56)\cap(123)\rb$\\
 \hspace*{5cm}$ \lb AB(56)\cap(123) (23)\cap(156)\rb $}}\nonumber\\
\text{FAC-4}_\text{Y space}&=& \frac{\lb 12356\rb^4\lb YAB3(45)\cap(123)\rb^2\lb Y2356\rb}{\parbox{5in}{$\lb Y1235\rb \lb Y1256\rb \lb Y1236\rb \lb YAB23\rb \lb YAB34\rb \lb YAB35\rb \lb YAB45\rb $ \\
 \hspace*{3cm}$\lb YAB3(56)\cap(123)\rb\lb YAB(56)\cap(123) (23)\cap(156)\rb$}}\nonumber
\ee\\

\be
\vcenter{\hbox{\includegraphics[width=6cm]{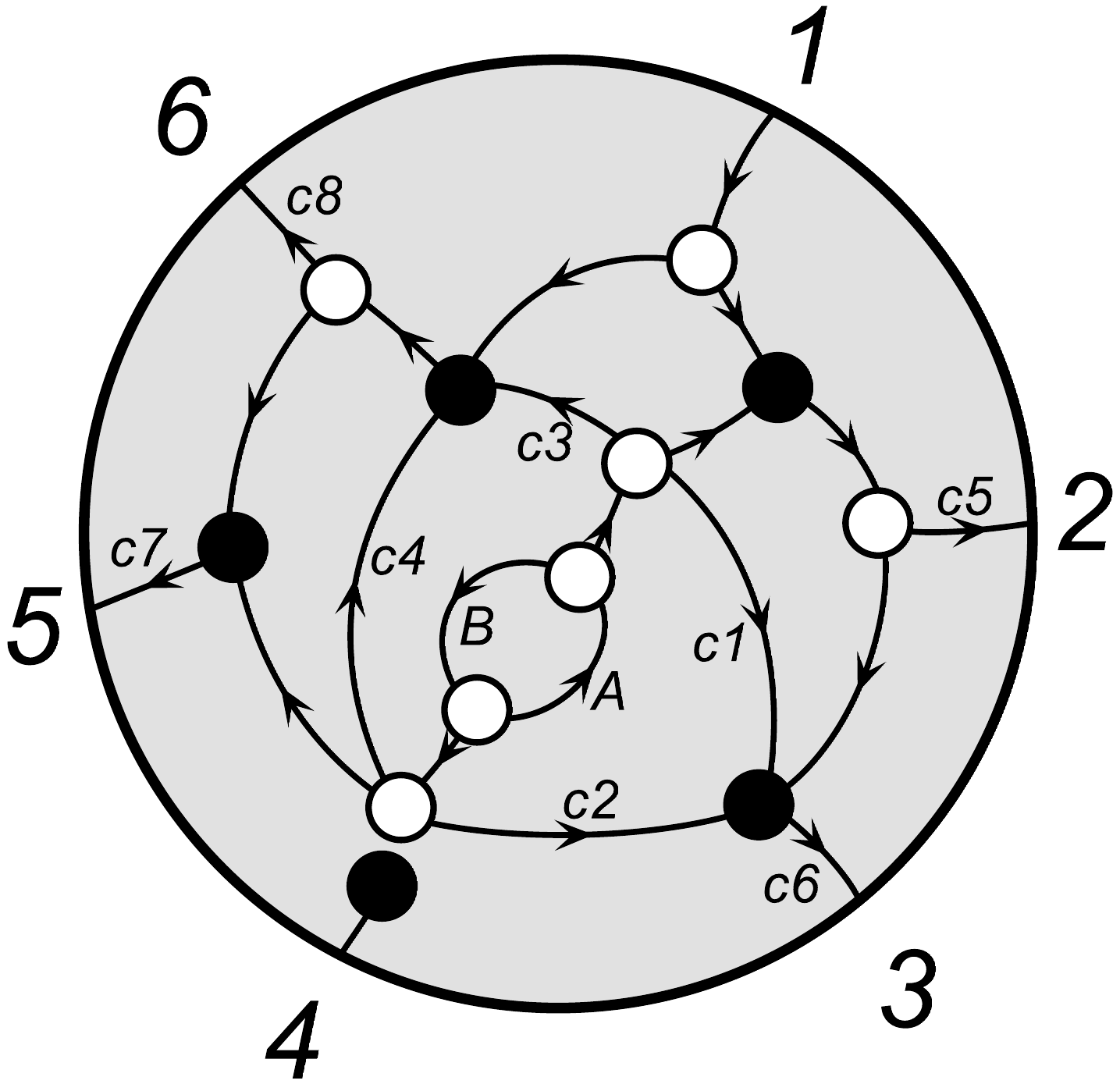}}}\;\;\;\;\;
\begin{tabular}{ll}
$\begin{pmatrix}
0 & c_5 & c_6(1+c_1) & 0 & -c_3 c_7 & -c_3 c_8 \\
0 & 0 & c_2 c_6 & 0 & c_7(1+c_4) & c_4 c_8 \\
1 & c_5 & c_6 & 0 & c_7 & c_8
\end{pmatrix}$\\ \\
$\text{Vanishing minors: } (4\bullet\bullet), [4]$\\\\
$\text{Residues: }\begin{cases}
P_4 = z_1 P_3 +z_2 P_5 + z_3 P_6 & z_1,z_2,z_3\rightarrow 0
\end{cases}$\\\\
$\mathcal{M}'_{\text{FAC-5}} = \{ 3,(23)\cap(56),(34)\cap(56) \} = \mathcal{M}'_{\Pi_{\{2,3,5\}}}$
\end{tabular}\nonumber
\ee
\be
\text{FAC-5}&=& \frac{\delta^{0|4}(\eta_1\lb 2356\rb+\text{cyclic})}
{\lb 1256\rb\lb 1236\rb\lb AB35\rb\lb AB56\rb\lb AB23\rb\lb AB(23)\cap(156) 1\rb}\nonumber\\
\text{FAC-5}_\text{Y space}&=& \frac{\lb 12356\rb^4}{\lb Y1256\rb\lb Y1236\rb\lb YAB35\rb\lb YAB56\rb\lb YAB23\rb\lb YAB(23)\cap(156) 1\rb}\nonumber
\ee\\

\be
\vcenter{\hbox{\includegraphics[width=6cm]{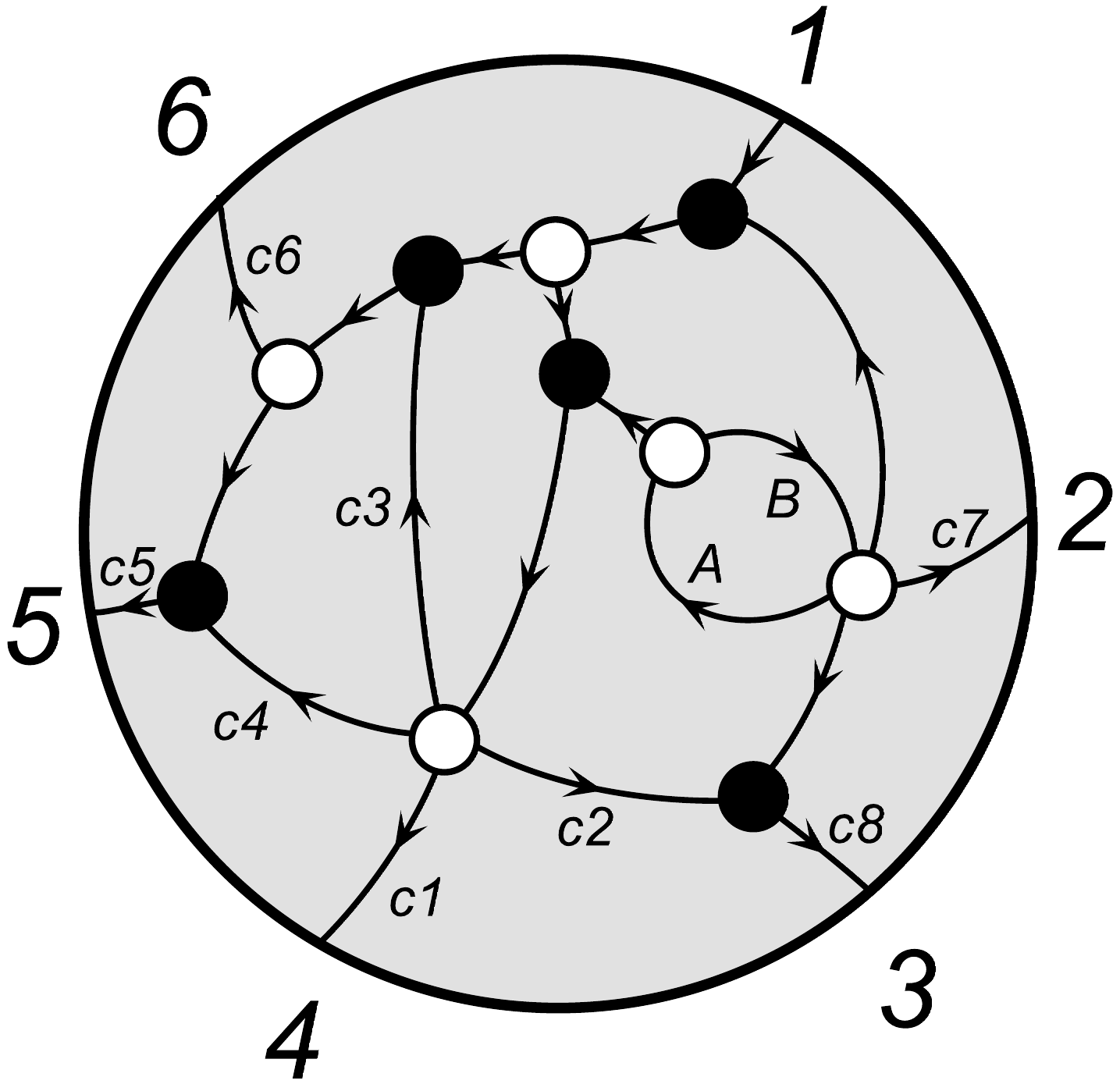}}}\;\;\;\;\;
\begin{tabular}{ll}
$\begin{pmatrix}
0 & 0 & c_2 c_8 & c_1 & c_5(c_3+c_4) & c_3 c_6\\
0 & -c_7 & c_8(-1+c_2) & c_1 & c_5(1+c_3+c_4) & c_6(1+c_3)\\
1 & 0 & c_2 c_8 & c_1 & c_5(1+c_3+c_4) & c_6(1+c_3)
\end{pmatrix}$\\ \\
$\text{Vanishing minors: } (234),(456),[2]$\\\\
$\text{Residues: }\begin{cases}
P_3 = \alpha_1 P_2 + \alpha_2 P_4 + z_1 P_6 & z_1 \rightarrow 0\\
P_5 = \alpha_3 P_4 + \alpha_4 P_6 + z_2 P_1 & z_2 \rightarrow 0\\
C_2 = z_3 & z_3\rightarrow 0
\end{cases}$\\\\
$\mathcal{M}'_{\text{FL-1}} = \{1,2,3,4,5,6\}$
\end{tabular}\nonumber
\ee
\be
\text{FL-1}&=&\frac{\delta^{0|4}(\eta_1\lb 3456\rb + \text{cyclic})\lb AB1(23)\cap(156)\rb^2}
{\parbox{5in}{$\lb 1356\rb\lb 1456\rb\lb 3456\rb\lb AB12\rb\lb AB13\rb\lb AB16\rb\lb AB23\rb\lb AB5(34)\cap(156)\rb$ \hspace*{8cm} $\lb AB(34)\cap(156)1\rb$}}\nonumber\\
\text{FL-1}_\text{Y space}&=&\frac{\lb 13456\rb^4\lb YAB1(23)\cap(156)\rb^2}
{\parbox{5in}{$\lb Y1356\rb\lb Y1456\rb\lb Y3456\rb\lb YAB12\rb\lb YAB13\rb\lb YAB16\rb\lb YAB23\rb$ \hspace*{6cm} $\lb YAB5(34)\cap(156)\rb\lb YAB(34)\cap(156)1\rb$}}\nonumber
\ee\\

\be
\vcenter{\hbox{\includegraphics[width=6cm]{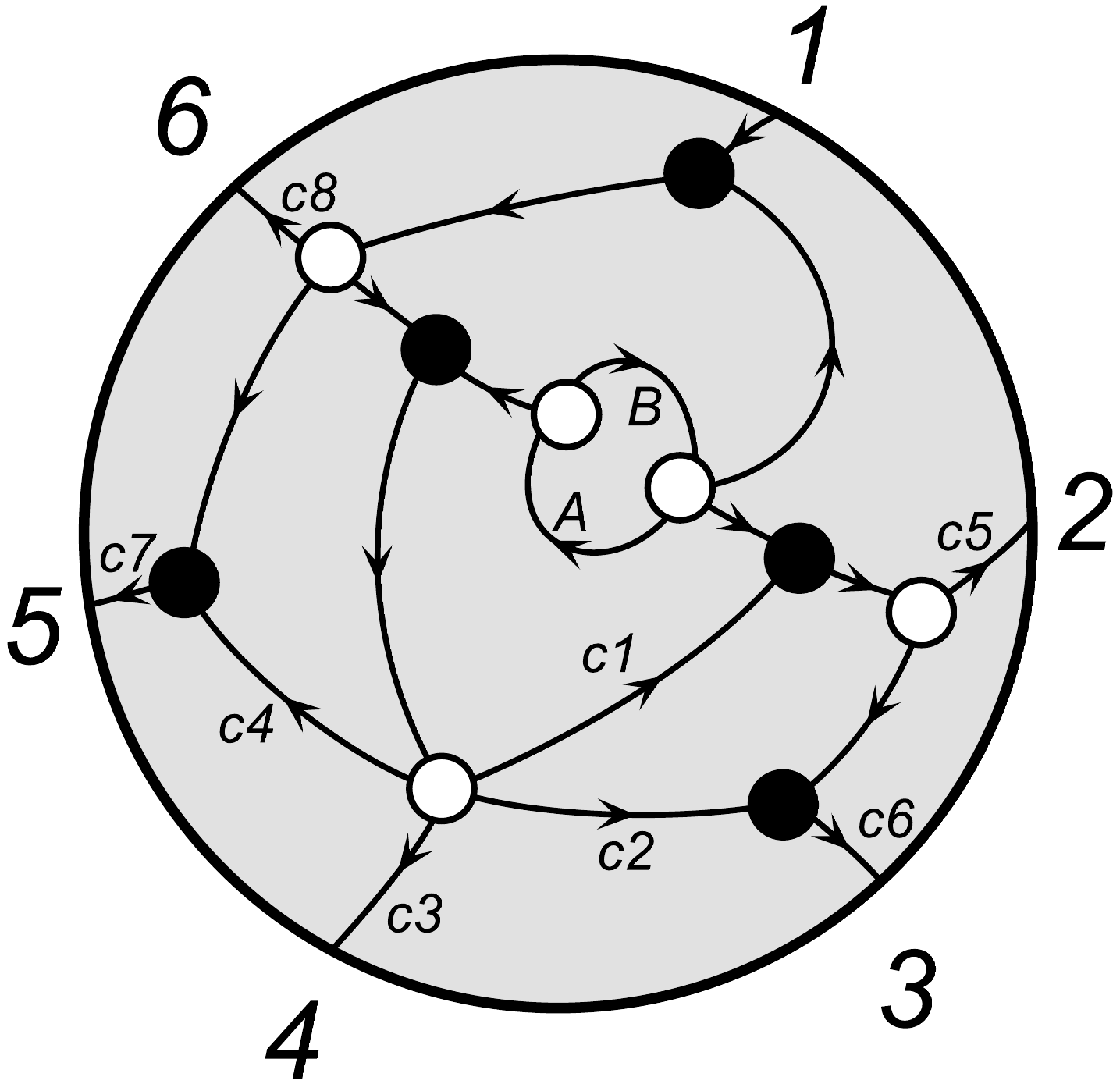}}}\;\;\;\;\;
\begin{tabular}{ll}
$\begin{pmatrix}
0 & c_1 c_5 & c_6(c_1+c_2) & c_3 & c_4 c_7 & 0\\
0 & c_5(-1+c_1) & c_6(-1+c_1+c_2) & c_3 & c_7(1+c_4) & c_8\\
1 & c_1 c_5 & c_6(c_1+c_2) & c_3 & c_7(1+c_4) & c_8
\end{pmatrix}$\\ \\
$\text{Vanishing minors: } (234),(456),[(16)\cap(24)]$\\\\
$\text{Residues: }\begin{cases}
P_3 = \alpha_1 P_2 + \alpha_2 P_4 + z_1 P_1 & z_1\rightarrow 0\\
P_5 = \alpha_3 P_4 + \alpha_4 P_6 + z_2 P_1 & z_2 \rightarrow 0\\
[(16)\cap(24)] = z_3 & z_3\rightarrow 0
\end{cases}$\\\\
$\mathcal{M}'_{\text{FL-2}} = \{(45)\cap(16),(16)\cap(23),(23)\cap(45)\} = \mathcal{M}'_{\Pi_{\{2,4,6\}}}$
\end{tabular}\nonumber
\ee
\be
\text{FL-2}&=&\frac{\delta^{0|4}(\eta_2\lb 345(AB)\cap(156)\rb+\text{cyclic})\lb AB1(23)\cap(156)\rb^2}
{\parbox{5in}{$\lb 2345\rb\lb AB12\rb\lb AB23\rb\lb AB15\rb\lb AB16\rb\lb AB56\rb
\lb AB5(23)\cap(156)\rb$ \hspace*{1cm}$\lb AB5(34)\cap(156)\rb\lb AB(156)\cap(234)\rb\lb 45(AB)\cap(156) (23)\cap(1AB)\rb$}}\nonumber
\\
\text{FL-2}_\text{Y space}&=&\frac{\lb 2345(AB)\cap(156)\rb^4\lb YAB1(23)\cap(156)\rb^2}
{\parbox{5in}{$\lb Y2345\rb\lb YAB12\rb\lb YAB23\rb\lb YAB15\rb\lb YAB16\rb\lb YAB56\rb\lb YAB5(23)\cap(156)\rb
$ \hspace*{0.5cm}$\lb YAB5(34)\cap(156)\rb\lb YAB(156)\cap(234)\rb\lb Y45(AB)\cap(156) (23)\cap(1AB)\rb$}}\nonumber
\ee

\be
\vcenter{\hbox{\includegraphics[width=6cm]{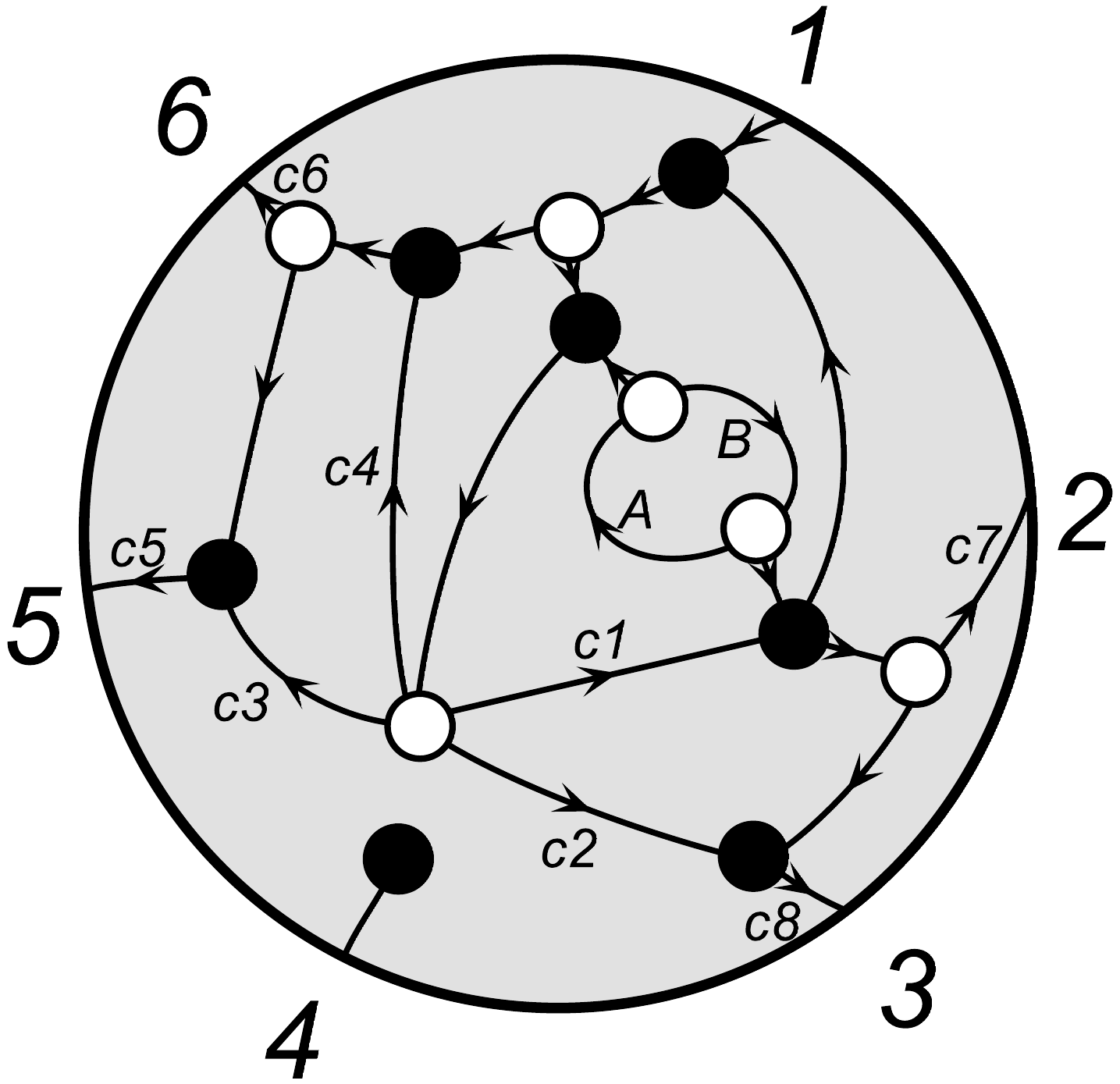}}}\;\;\;\;\;
\begin{tabular}{ll}
$\begin{pmatrix}
0 & c_1 c_7 & c_8(c_1{+}c_2) & 0 & c_5(c_3{+}c_4) & c_4 c_6\\
0 & c_7({-}1{+}c_1) & c_8(-{}1{+}c_1{+}c_2) & 0 & c_5(1{+}c_3{+}c_4) & c_6(1{+}c_4)\\
1 & c_1 c_7 & c_8(c_1{+}c_2) & 0 & c_5(1{+}c_3{+}c_4) & c_6(1{+}c_4)
\end{pmatrix}$\\ \\
$\text{Vanishing minors: } (4\bullet\bullet),[4]$\\\\
$\text{Residues: }\begin{cases}
P_4 = z_1 P_3+z_2 P_5+z_3 P_6 & z_1,z_2,z_3\rightarrow 0
\end{cases}$\\\\
$\mathcal{M}'_{\text{FL-3}} = \{1,2,(16)\cap(23)\} = \mathcal{M}'_{\Pi_{\{1,2,6\}}}$
\end{tabular}\nonumber
\ee
\be
\text{FL-3}&=&\frac{\delta^{0|4}(\eta_1\lb 2356\rb+\text{cyclic})}{\lb 2356\rb\lb 1356\rb\lb AB12\rb\lb AB16\rb\lb AB23\rb\lb AB(23)\cap(156)5\rb}\nonumber
\\
\text{FL-3}_\text{Y space}&=&\frac{\lb 12356\rb^4}{\lb Y2356\rb\lb Y1356\rb\lb YAB12\rb\lb YAB16\rb\lb YAB23\rb\lb YAB(23)\cap(156)5\rb}\nonumber
\ee

\be
\vcenter{\hbox{\includegraphics[width=6cm]{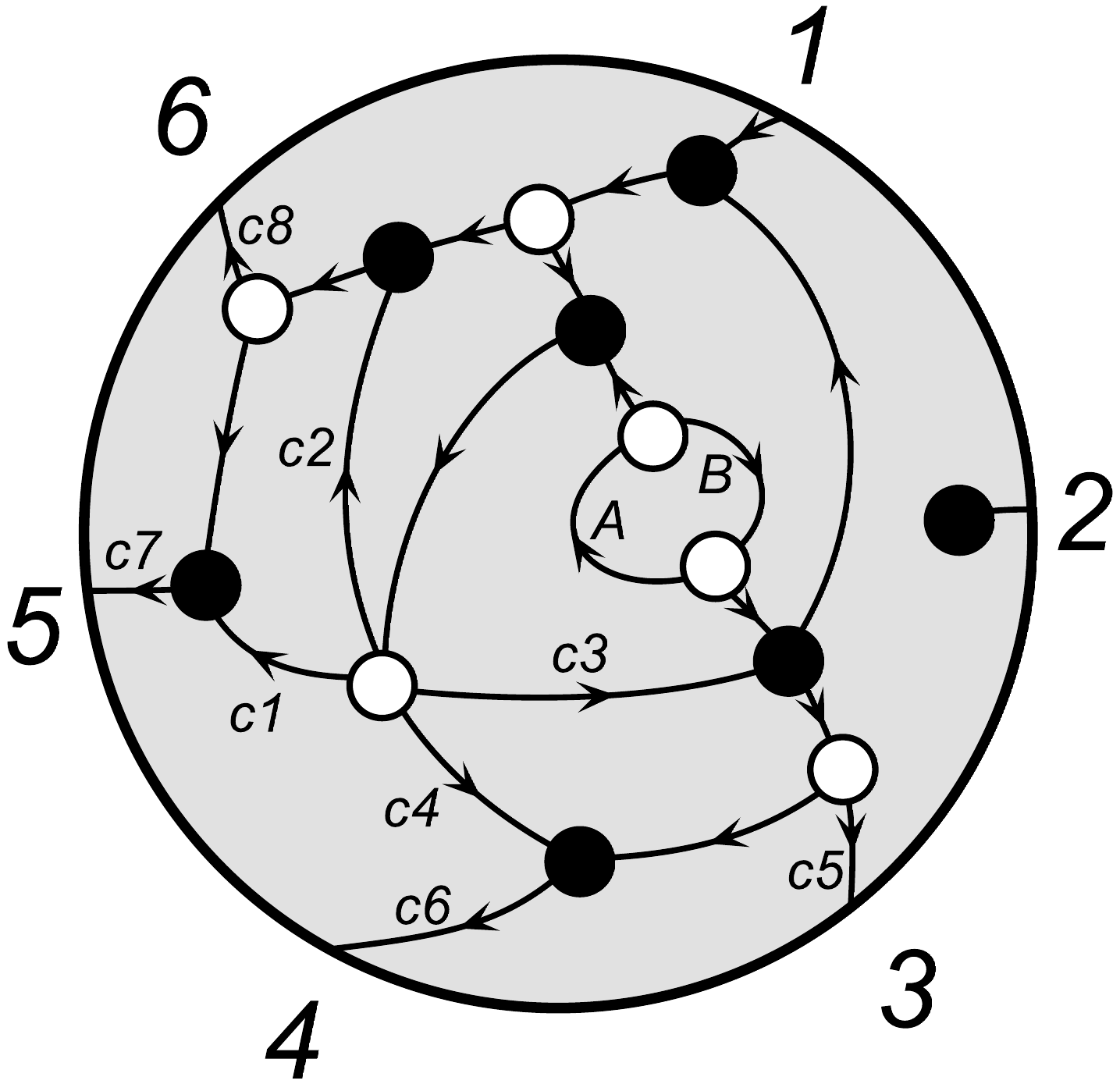}}}\;\;\;\;\;
\begin{tabular}{ll}
$\begin{pmatrix}
0 & 0 & c_3 c_5 & c_6(c_3{+}c_4) & c_7(c_1{+}c_2) & c_2 c_8\\
0 & 0 & c_5(-1{+}c_3) & c_6(-1{+}c_3{+}c_4) & c_7(1{+}c_1{+}c_2) & c_8(1{+}c_2)\\
1 & 0 & c_3 c_5 & c_6(c_3{+}c_4) & c_7(1{+}c_1{+}c_2) & c_8(1{+}c_2)
\end{pmatrix}$\\ \\
$\text{Vanishing minors: } (2\bullet\bullet), [2]$\\\\
$\text{Residues: }\begin{cases}
P_2 = z_1 P_1 + z_2 P_3 + z_3 P_4 & z_1,z_2,z_3\rightarrow 0
\end{cases}$\\\\
$\mathcal{M}'_{\text{FL-4}} = \{1,(12)\cap(34),(16)\cap(34)\} = \mathcal{M}'_{\Pi_{\{1,3,6\}}}$
\end{tabular}\nonumber
\ee
\be
\text{FL-4}&=&\frac{\delta^{0|4}(\eta_1\lb 3456\rb+\text{cyclic})}
{\lb 3456\rb\lb 1456\rb\lb AB13\rb\lb AB16\rb\lb AB34\rb\lb AB(34)\cap(156)5\rb}\nonumber
\\
\text{FL-4}_\text{Y space}&=&\frac{\lb 13456\rb^4}
{\lb Y3456\rb\lb Y1456\rb\lb YAB13\rb\lb YAB16\rb\lb YAB34\rb\lb YAB(34)\cap(156)5\rb}\nonumber
\ee

\be
\vcenter{\hbox{\includegraphics[width=6cm]{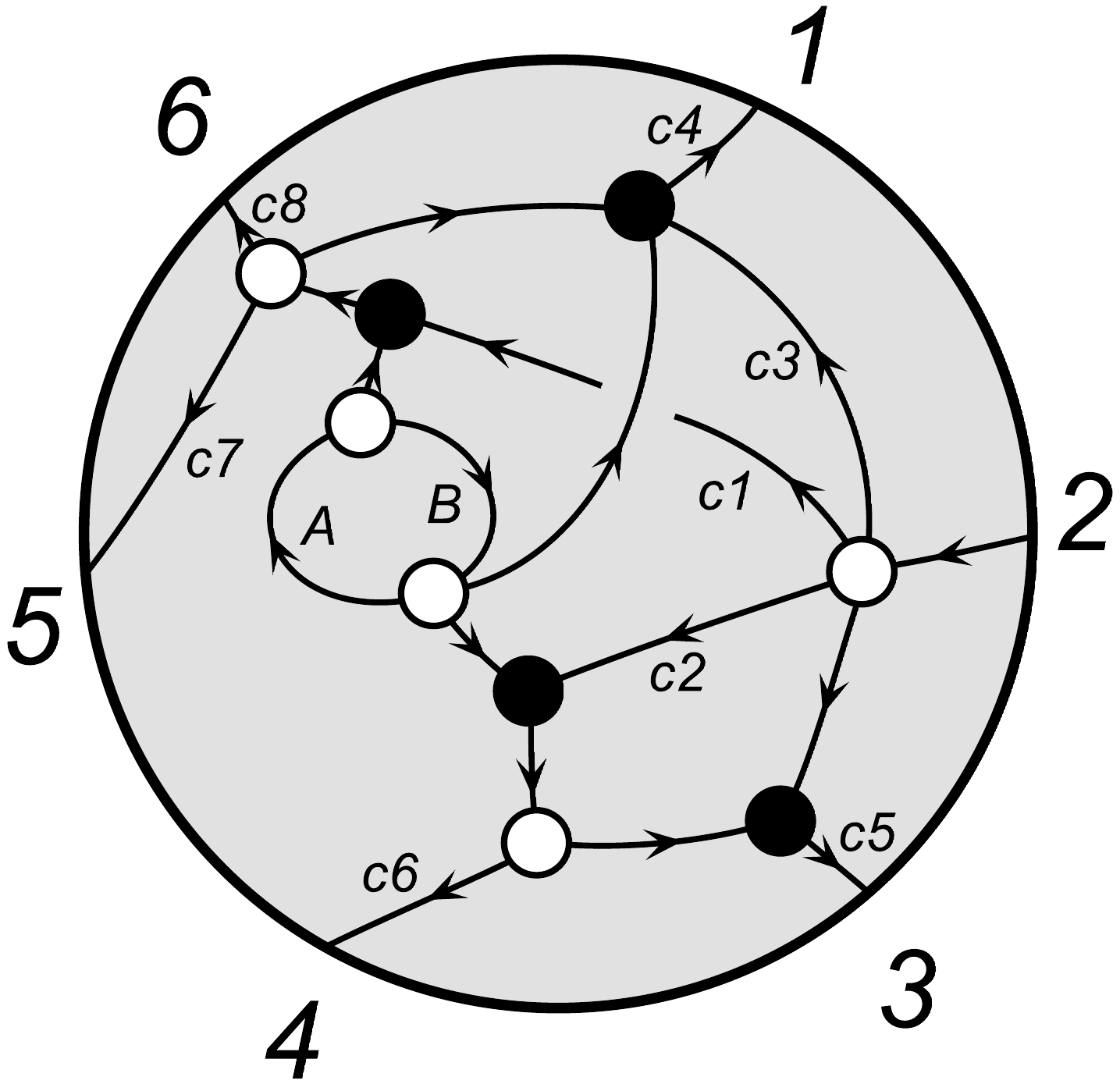}}}\;\;\;\;\;
\begin{tabular}{ll}
$\begin{pmatrix}
-c_4 & 0 & 0 & 0 & -c_7 & -c_8 \\
-c_4 & 0 & c_5 & c_6 & 0 & 0 \\
c_4(c_1+c_3) & 1 & c_5(1+c_2) & c_2 c_6 & c_1 c_7 & c_1 c_8
\end{pmatrix}$\\ \\
$\text{Vanishing minors: } (56\bullet),(234)$\\\\
$\text{Residues: }\begin{cases}
P_5 = \alpha_1 P_6 + z_1 P_4 + z_2 P_1 & z_1,z_2\rightarrow 0\\
P_3 = \alpha_2 P_2 + \alpha_3 P_4 + z_3 P_6 & z_3 \rightarrow 0
\end{cases}$\\\\
$\mathcal{M}'_{\text{FL-5}} = \{4,(16)\cap(45),(16)\cap(34)\} = \mathcal{M}'_{\Pi_{\{3,4,6\}}}$
\end{tabular}\nonumber
\ee
\be
\text{FL-5}&=&\frac{\delta^{0|4}(\eta_1\lb 234(AB)\cap(156)\rb+\text{cyclic})\lb AB1(34)\cap(156)\rb}
{\parbox{5in}{$\lb 1234\rb\lb AB14\rb\lb AB15\rb\lb AB16\rb\lb AB34\rb\lb AB56\rb\lb AB(234)\cap(156)\rb $ \hspace*{3.5cm} $\lb AB1(23)\cap(156)\rb\lb 12(AB)\cap(134) (AB)\cap(156)\rb$}}\nonumber
\\
\text{FL-5}_\text{Y space}&=&\frac{\lb 1234(AB)\cap(156)\rb^4\lb YAB1(34)\cap(156)\rb}
{\parbox{5in}{$\lb Y1234\rb\lb YAB14\rb\lb YAB15\rb\lb YAB16\rb\lb YAB34\rb\lb YAB56\rb\lb YAB(234)\cap(156)\rb $ \hspace*{4cm} $\lb YAB1(23)\cap(156)\rb\lb Y12(AB)\cap(134) (AB)\cap(156)\rb$}}\nonumber
\ee

\be
\vcenter{\hbox{\includegraphics[width=6cm]{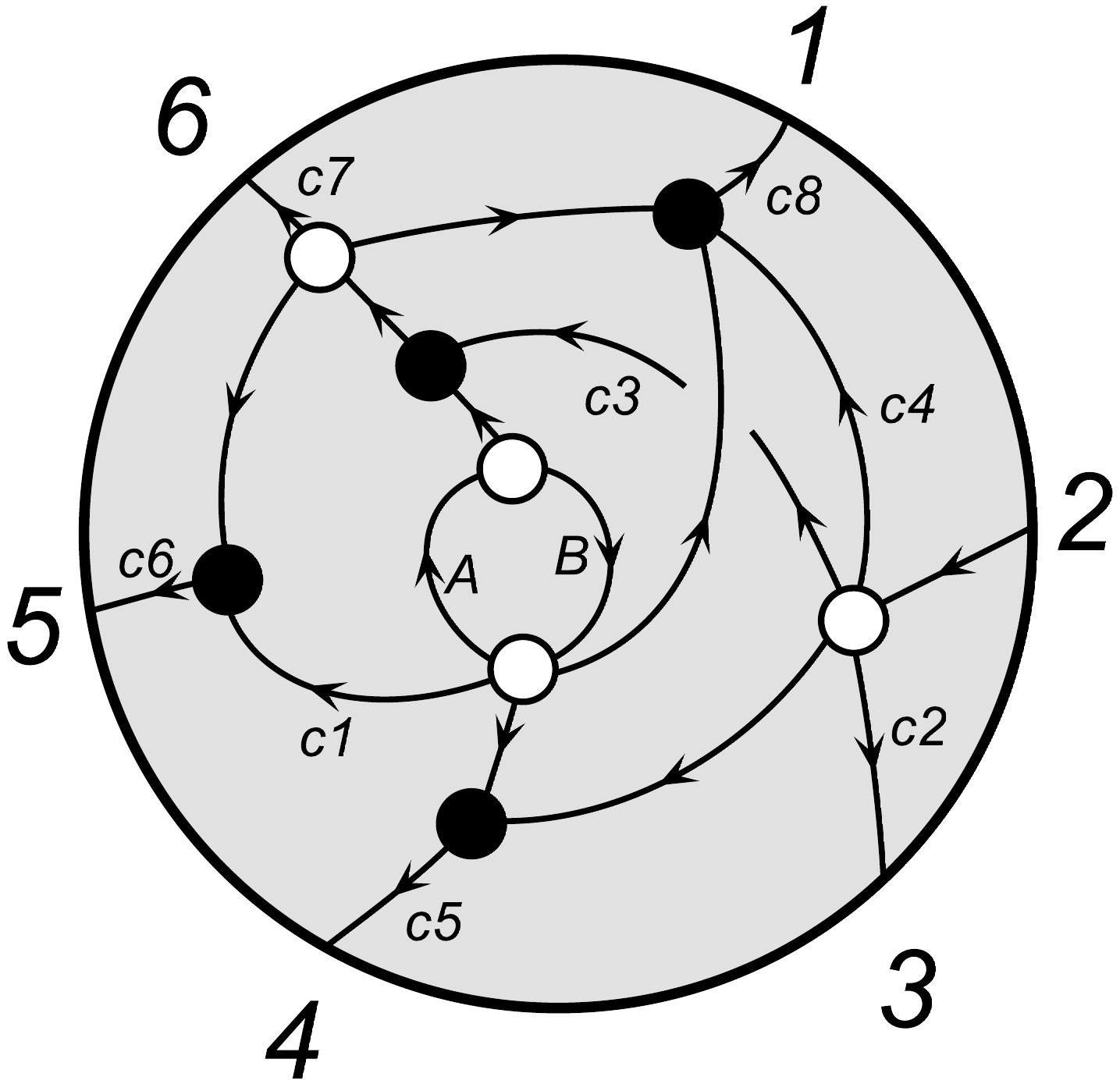}}}\;\;\;\;\;
\begin{tabular}{ll}
$\begin{pmatrix}
-c_8 & 0 & 0 & 0 & -c_6 & -c_7 \\
-c_8 & 0 & 0 & c_5 & c_1 c_6 & 0 \\
c_8(c_3+c_4) & 1 & c_2 & c_5 & c_3 c_6 & c_3 c_7
\end{pmatrix}$\\ \\
$\text{Vanishing minors: } (23\bullet),[(24)\cap(56)]$\\\\
$\text{Residues: }\begin{cases}
P_2 = \alpha_1 P_3 + z_1 P_4 + z_2 P_5 & z_1,z_2\rightarrow 0\\
[(24)\cap(56)] = z_3 & z_3\rightarrow 0
\end{cases}$\\\\
$\mathcal{M}'_{\text{FL-6}} = \{(12)\cap(56),(56)\cap(34),(34)\cap(12)\} = \mathcal{M}'_{\Pi_{\{1,3,5\}}}$
\end{tabular}\nonumber
\ee
\be
\text{FL-6}&=&\frac{\delta^{0|4}(\eta_1\lb 1234(AB)\cap(156)\rb+\text{cyclic})\lb 1456\rb^2}
{\parbox{5in}{$\lb 1234\rb\lb AB14\rb\lb AB16\rb\lb AB45\rb\lb AB56\rb\lb AB(234)\cap(156)\rb\lb AB1(34)\cap(156)\rb$ \hspace*{6cm}$\lb AB1(24)\cap(156)\rb\lb AB1(23)\cap(156)\rb$}}\nonumber
\\
\text{FL-6}_\text{Y space}&=&\frac{\lb 1234(AB)\cap(156)\rb^4\lb Y1456\rb^2}
{\parbox{5in}{$\lb Y1234\rb\lb YAB14\rb\lb YAB16\rb\lb YAB45\rb\lb YAB56\rb\lb YAB(234)\cap(156)\rb$ \hspace*{2cm}$\lb YAB1(34)\cap(156)\rb\lb YAB1(24)\cap(156)\rb\lb YAB1(23)\cap(156)\rb$}}\nonumber
\ee

\be
\vcenter{\hbox{\includegraphics[width=6cm]{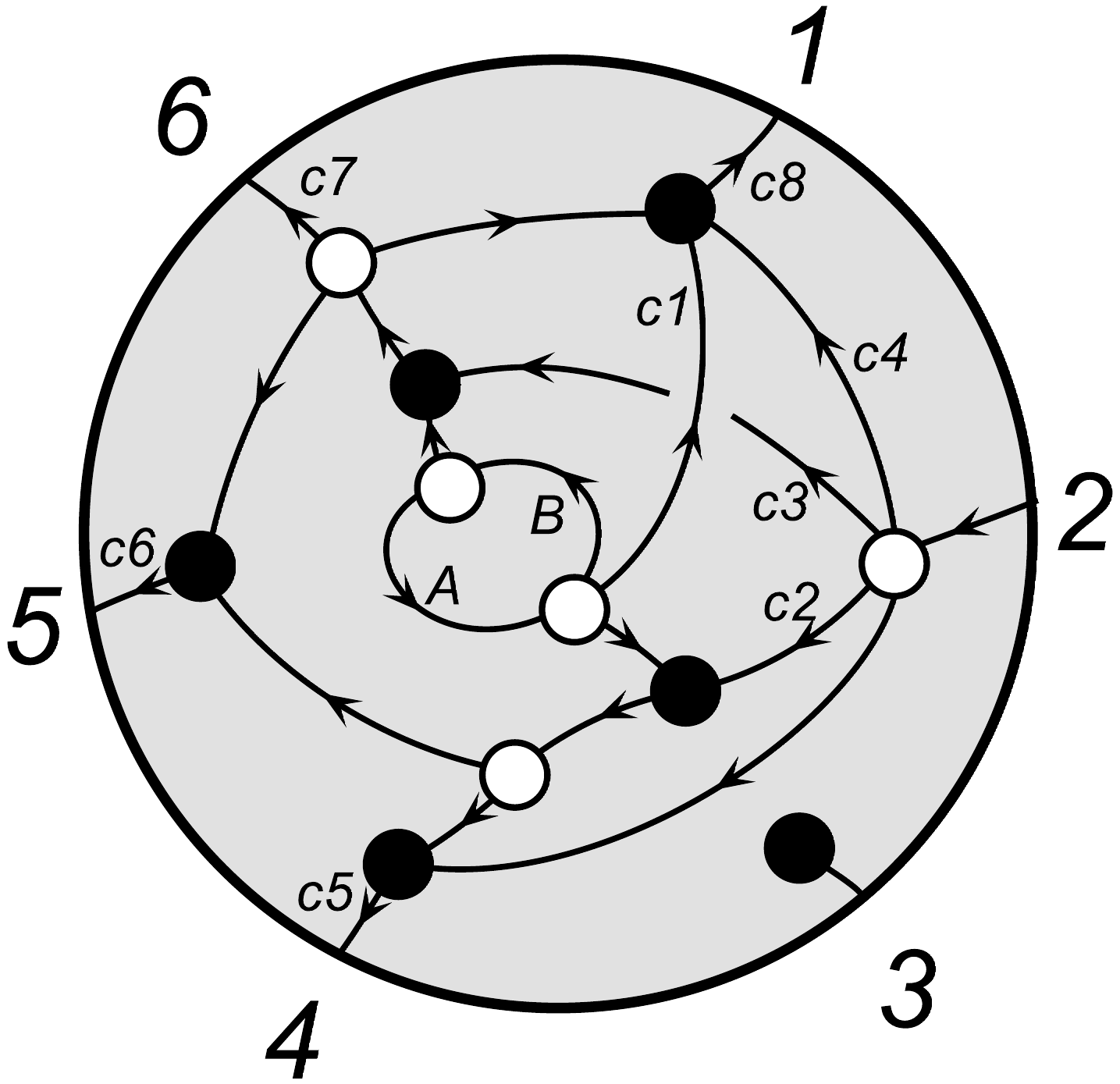}}}\;\;\;\;\;
\begin{tabular}{ll}
$\begin{pmatrix}
-c_1c_8 & 0 & 0 & c_5 & c_6 & 0\\
c_8 & 0 & 0 & 0 & c_6 & c_7 \\
c_8(c_3+c_4) & 1 & 0 & c_5(1+c_2) & c_6(c_2+c_3) & c_3 c_7
\end{pmatrix}$\\ \\
$\text{Vanishing minors: } (3\bullet\bullet),[3]$\\\\
$\text{Residues: }\begin{cases}
P_3 = z_1 P_2 + z_2 P_4 + z_3 P_5 & z_1,z_2,z_3\rightarrow 0
\end{cases}$\\\\
$\mathcal{M}'_{\text{FL-7}} = \{6,(16)\cap(23),(23)\cap(56)\} = \mathcal{M}'_{\Pi_{\{2,5,6\}}}$
\end{tabular}\nonumber
\ee
\be
\text{FL-7}&=&\frac{\delta^{0|4}(\eta_1\lb 2456\rb+\text{cyclic})\lb AB15\rb^2}
{\parbox{5in}{$\lb 1245\rb\lb AB12\rb\lb AB16\rb\lb AB45\rb\lb AB56\rb\lb AB5(24)\cap(156)\rb\lb AB1(24)\cap(156)\rb$}}\nonumber
\\
\text{FL-7}_\text{Y space}&=&\frac{\lb 12456\rb^4\lb YAB15\rb^2}
{\parbox{5in}{$\lb Y1245\rb\lb YAB12\rb\lb YAB16\rb\lb YAB45\rb\lb YAB56\rb\lb YAB5(24)\cap(156)\rb$ \hspace*{3in}$\lb YAB1(24)\cap(156)\rb$}}\nonumber
\ee

\be
\vcenter{\hbox{\includegraphics[width=6cm]{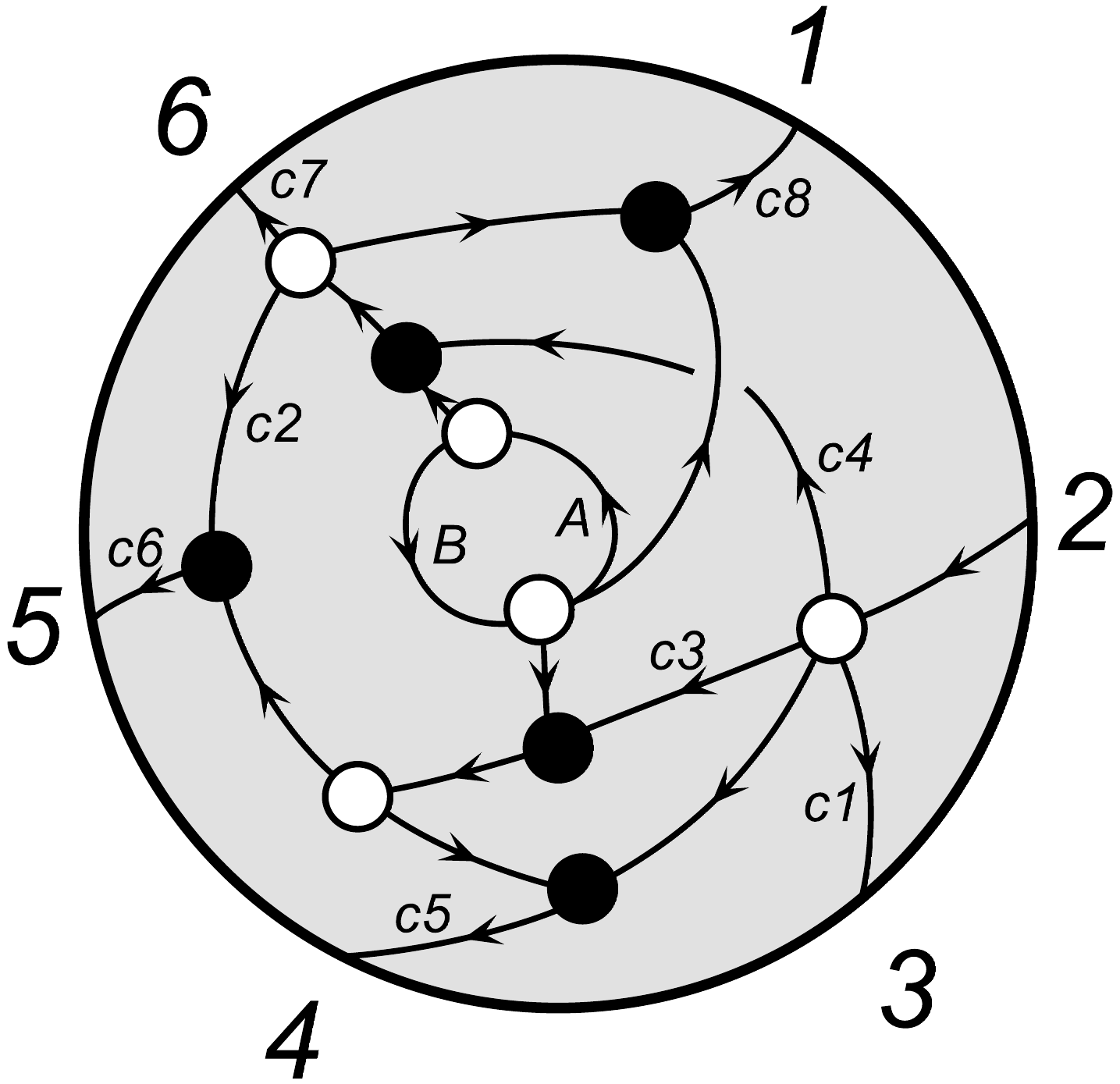}}}\;\;\;\;\;
\begin{tabular}{ll}
$\begin{pmatrix}
-c_8 & 0 & 0 & 0 & -c_2 c_6 & -c_7\\
-c_8 & 0 & 0 & c_5 & c_6 & 0\\
c_4 c_8 & 1 & c_1 & c_5(1+c_3) & c_6(c_3+c_2 c_4) & c_4 c_7
\end{pmatrix}$\\ \\
$\text{Vanishing minors: } (23\bullet), [(16)\cap(24)]$\\\\
$\text{Residues: }\begin{cases}
P_2 = \alpha_1 P_3 + z_1 P_1+ z_2 P_4 & z_1,z_2\rightarrow 0\\
[(16)\cap(24)] = z_3 & z_3\rightarrow 0
\end{cases}$\\\\
$\mathcal{M}'_{\text{FL-8}} = \{1,(12)\cap(34),(16)\cap(34)\} = \mathcal{M}'_{\Pi_{\{1,3,6\}}}$
\end{tabular}\nonumber
\ee
\be
\text{FL-8}&=&\frac{\delta^{0|4}(\lb 2345(AB)\cap(156)\rb+\text{cyclic})\lb 1456\rb^2}
{\parbox{5in}{$\lb 2345\rb\lb AB16\rb\lb AB45\rb\lb AB56\rb\lb AB5(34)\cap(156)\rb\lb AB5(24)\cap(156)\rb$ \hspace*{5cm}$\lb AB(234)\cap(156)\rb\lb 23(AB)\cap(156)(45)\cap(1AB)\rb$}}\nonumber
\\
\text{FL-8}_\text{Y space}&=&\frac{\lb 2345(AB)\cap(156)\rb^4\lb Y1456\rb^2}
{\parbox{5in}{$\lb Y2345\rb\lb YAB16\rb\lb YAB45\rb\lb YAB56\rb\lb YAB5(34)\cap(156)\rb$ \hspace*{4cm}$\lb YAB5(24)\cap(156)\rb\lb YY23(AB)\cap(156)(45)\cap(1AB)\rb$}}\nonumber
\ee

\end{document}